%% file: NE_pBC_arxiv_submission_v2.tex
\documentclass[onefignum,onetabnum]{siamart250211}

\input{commands}

\headers{Dual mechanisms of inspiratory neuron responses to norepinephrine}{S.~Venkatakrishnan, A.K.~Tryba, A.J.~Garcia III, and Y.~Wang}

\title{Dual mechanisms for heterogeneous responses of inspiratory neurons to noradrenergic modulation
\thanks{Submitted to the editors on July 25, 2025 \funding{This work was supported by NIH/NIDA R01DA057767, as part of the Collaborative Research in Computational Neuroscience Program. A.J.~Garcia 3rd is also supported by NIH/NHLBI R01HL163965 and NIH/NIDA R01DA061412.}}}

\author{Sreshta Venkatakrishnan \thanks{Department of Mathematics, Brandeis University, Waltham, MA, 02453, USA (\email{sreshtav@brandeis.edu})}
\and Andrew K. Tryba \thanks{Department of Pediatrics, Section of Neurology, The University of Chicago, Chicago, IL, 60637, USA (\email{atryba@uchicago.edu})}
\and Alfredo J. Garcia III \thanks{Institute for Integrative Physiology, The University of Chicago, Chicago, IL, 60637, USA (\email{ajgarcia3@uchicago.edu})}
\and Yangyang Wang \thanks{Department of Mathematics, Brandeis University, Waltham, MA, 02453, USA (\email{yangyangwang@brandeis.edu})}}

\ifpdf
\hypersetup{
  pdftitle={Dual mechanisms for heterogeneous responses of inspiratory neurons to noradrenergic modulation},
  pdfauthor={S. Venkatakrishnan, A. K. Tryba, A. J. Garcia III, and Y. Wang}
}
\fi

\begin{document}

\maketitle

\begin{abstract}
Respiration is an essential involuntary function necessary for survival. This poses a challenge for the control of breathing. 
The preB\"otzinger complex (preB\"otC) is a heterogeneous neuronal network responsible for driving the inspiratory rhythm. While neuromodulators such as norepinephrine (NE) allow it to be both robust and flexible for all living beings to interact with their environment, the basis for how neuromodulation impacts neuron-specific properties remains poorly understood.  
In this work, we examine how NE influences different preB\"otC neuronal subtypes by modeling its effects through modulating two key parameters: calcium-activated nonspecific cationic current gating conductance ($g_{\rm CAN}$) and inositol-triphosphate ($\rm IP_3$), guided by experimental studies. Our computational model captures the experimentally observed differential effects of NE on distinct preB\"otC bursting patterns. We show that this dual-modulation mechanism is critical for inducing conditional bursting and identify specific parameter regimes where silent neurons remain inactive in the presence of NE. Furthermore, using methods from dynamical systems theory, we uncover the mechanisms by which NE differentially modulates burst frequency and duration in NaP-dependent and CAN-dependent bursting neurons. These results align well with previously reported experimental findings and provide a deeper understanding of cell-specific neuromodulatory responses within the respiratory network. \\

\textbf{Relevance to Life Sciences.}  Experimental studies have shown that NE can stabilize inspiratory activity in healthy networks, yet destabilize it under pathological conditions. Whether NE exerts stabilizing versus destabilizing effects on inspiratory rhythmogenesis likely depends its heterogeneous actions on distinct neuronal subpopulations within the \pbc{}. Our modeling identifies a dual-modulation mechanism—through modulation of $g_{\rm CAN}$ and $\rm IP_3$—that governs neuron-specific responses to NE and helps explain its experimentally observed differential effects on \pbc{} neurons. It also reveals the emergence of mixed bursting oscillations during the induction of conditional bursting activity from spiking neurons, with initial irregular activity occurring before stabilization. This work provides a mechanistic foundation for understanding how NE specifically modulates \pbc{} neurons and offers a basis for investigating how neuron-type-specific neuromodulatory responses interact to shape respiratory network function in both healthy and diseased states. \\

\textbf{Mathematical Content.} We apply techniques from dynamical systems theory to analyze the preB\"{o}tC model neuron that involves multiple distinct timescales and exhibits differential responses to NE depending on its intrinsic state. Our approach combines phase-plane analysis, bifurcation analysis, and geometric singular perturbation theory (GSPT). These analyses not only uncover the mechanisms by which NE differentially modulates distinct types of preB\"{o}tC neurons but also generate testable predictions for future experiments.
\end{abstract}



\section{Introduction}
Breathing is a complex neurophysiological process required for survival and consists of two main phases — inspiration and expiration. In mammals, the preB{\"o}tzinger complex (\pbc) within the brainstem generates the neural rhythm that drives inspiration and is involved with regulating the phasic timing of inspiration and expiration. Rhythmic activity persists in some \pbc{} neurons even when synaptic coupling in the network is blocked and the network rhythm ceases. Such persistent activity is derived from intrinsic membrane properties among these so-called ``pacemaker neurons" \cite{smith1991pre, johnson1994pacemaker, funk1993generation, ramirez2011role}. Pacemaker neurons in the \pbc{} exhibit spontaneous rhythmic bursting activity, where a discrete barrage of action potentials is generated in a periodic manner, and are hence also referred to as intrinsic bursting neurons or bursters.
Multiple mechanisms underlie bursting, including the calcium-activated non-specific cationic current ($I_{\rm CAN}$), driven by calcium oscillations within the \pbc{}, and the voltage-sensitive persistent sodium current ($I_{\rm NaP}$) \cite{del2005sodium, koizumi2008persistent, pena2004differential, pena2007effects, ramirez2011role, pace2007inspiratory}.

Experimental observations have inspired numerous computational studies (e.g., \cite{butera1999modelI, butera1999modelII, rubin2009calcium, jasinski, TB, PR}), aimed at elucidating the cellular mechanisms underlying intrinsic rhythmic bursting activity in individual \pbc{} neurons, the dynamic interplay between intrinsic bursting conductances, and their impact at the network level. An early influential computational model of isolated \pbc{} neurons only captures bursting driven by the $I_{\rm NaP}$-dependent mechanism \cite{butera1999modelI}. Later models incorporated synaptic activation of $I_{\rm CAN}$ \cite{rubin2009calcium}, followed by studies that combined both $I_{\rm NaP}$- and $I_{\rm CAN}$- dependent bursting mechanisms (e.g., \cite{jasinski, TB, PR,phillips2022burstlet}). Beyond computational modeling, dynamical systems theory has provided valuable insights into the interplay between these two bursting mechanisms, revealing complex mixed bursting dynamics characterized by distinct bursting patterns \cite{wang2016jcompneur, wang2020complex}. Notably, while some of these mixed bursting patterns exhibit significant qualitative similarities, mathematical analysis suggests they arise from fundamentally different mechanisms \cite{wang2017timescales}, underscoring the importance of complementing computational models with dynamical systems analysis. 

Since the activity of these neurons is also constantly modulated by various neuromodulators via altering the intrinsic properties of neurons as well as properties of the network, several works have been dedicated to studying the effects of neuromodulation on the \pbc{} neurons \cite{doi2008neuromodulation, dhingra2024asymmetric, TB, zanella2014norepinephrine, VR}. Various neuromodulators such as serotonin, norepinephrine, acetylcholine, substance P have been shown to increase the frequency of the respiratory rhythmic activity \cite{doi2008neuromodulation, baertsch2019insights, VR, shao2000acetylcholine, richter2003serotonin, guiard2008functional,ben2010substance,tryba2008differential}.
These neuromodulators can have unique effects on the different types of \pbc{} neurons, altering neuronal excitability and regularity of activity patterns, which, in turn, can enhance or diminish the stability of the network rhythm. Indeed, neuromodulation of the \pbc{} by norepinephrine (NE) stimulates rhythmogenesis and differentially affects activity patterns of synaptically isolated \pbc{} neurons \cite{VR}. Specifically, NE stimulated the burst frequency without affecting burst duration in $I_{\rm NaP}$-dependent bursting neurons. Whereas, NE increased burst duration of $I_{\rm CAN}$-dependent bursters while minimally affecting their burst frequency.  Moreover, NE had differential effects on different synaptically isolated non-bursting neurons.  Silent non-bursting neurons (i.e., neurons that did not generate action potentials in synaptic isolation) remained silent in NE. Whereas, in active non-bursting neurons (i.e., neurons continued to generate action potentials but did not burst in synaptic isolation), NE induced conditional bursting properties that are $I_{\rm CAN}$-dependent \cite{VR}.

Previously, computational works such as \cite{TB, PR} have attempted to model the effects of NE on single \pbc{} neurons via increasing the conductance for the CAN-current ($g_{\rm CAN}$). Their choice in modeling the application of NE through the model parameter $g_{\rm CAN}$ was attributed to earlier findings that NE, which is an $\alpha_1$-receptor agonist, increases CAN-current conductances in various cell types. Both models successfully reproduced the effects of NE on synaptically isolated intrinsic bursting \pbc{} neurons as observed in \emph{in vitro} experiments in \cite{VR} in terms of the frequency and duration of the bursts (see \cref{modelNE} for details). However, increasing $g_{\rm CAN}$ in these \emph{in silico} models did not produce the conditional bursting properties that were observed experimentally, nor did it capture the NE-insensitivity of silent non-bursters, suggesting that additional mechanisms are needed to comprehensively capture the experimental effects of NE on single \pbc{} neurons \cite{VR}. This paper addresses this gap by integrating additional mechanisms through \RED{which} NE modulates neuronal activity. We hypothesize that NE enhances second messenger-mediated ${\rm Ca}^{2+}$ flux, a mechanism previously described in \cite{Exton,choy2018noradrenaline}. Our modeling reveals that interactions between $g_{\rm CAN}$ and inositol-triphosphate (${\rm IP_3}$) are critical for inducing conditional bursting and identifies discrete parameter regimes where silent non-bursting neurons continue to remain inactive even in the presence of NE. Furthermore, our model simulations and analyses (detailed in \cref{results}) uncover the mechanisms by which NE differentially modulates burst frequency and duration in NaP-dependent and CAN-dependent bursting neurons. These findings align closely with experimental data, supporting the model's relevance for probing the mechanisms underlying NE modulation of \pbc{} neurons.

The paper is organized as follows: \Cref{model} presents a mathematical model of a single \pbc{} neuron that builds on previous models with modifications to facilitate the modeling of NE effects. \Cref{activity_patterns} outlines the various activity patterns observed in this model. \Cref{modelNE} introduces our proposed mechanism for NE modulation of \pbc{} neurons. In \cref{gspt}, we review the mathematical tools used for analyzing the model. \Cref{results} presents our main results and detailed analysis of NE’s effects on NaP-dependent bursters (\cref{NB}), CAN-dependent bursters (\cref{CB}), Tonic Spiking neurons (\cref{TS}) and Quiescent neurons (\cref{Q}). Finally, we conclude with a discussion in \cref{discussion}.

\section{Mathematical Modeling and Preliminaries} \label{prelims}

In this section, we first present a single preB{\"o}tC neuron model, followed by an outline of the various activity patterns it exhibits. We then introduce our proposed NE modulation mechanism and conclude by reviewing the geometric singular perturbation theory used in our analysis.

\subsection{preB{\"o}tC neuron Model} \label{model}

The preB{\"o}tC neuron model considered in this paper is a single compartment dynamical system that incorporates Hodgkin-Huxley-style conductances, adapted from previously described models \cite{TB,PR,wang2020complex}. The model is described by the following equations:
\begin{subequations}
    \begin{align}
        V' &= (-I_{\rm L}-I_{\rm K}-I_{\rm Na}-I_{\rm NaP}-I_{\rm CAN}-I_{\rm Ca})/C_m \label{eq:pBC_Ica_v} \\
        n' &= (n_{\infty}(V) - n)/\tau_n(V) \label{eq:pBC_Ica_n} \\
        h' &= (h_{\infty}(V) - h)/\tau_h(V) \label{eq:pBC_Ica_h} \\
        [\rm Ca]' &= f_i (\rm{J_{ER_{in}}} - \rm{J_{ER_{out}}}) - \alpha_{\rm Ca}I_{\rm Ca} - ([\rm Ca]-[\rm Ca]_{min})/\tau_{\rm Ca} \label{eq:pBC_Ica_ca} \\
        [\rm Ca]'_{\rm Tot} &= -\alpha_{\rm Ca}I_{\rm Ca} - ([\rm Ca]-[\rm Ca]_{min})/\tau_{\rm Ca} \label{eq:pBC_Ica_catot} \\
        l' &= AK_d(1-l) - A[{\rm Ca}]l. \label{eq:pBC_Ica_l} 
    \end{align}
    \label{eq:pBC_Ica}
\end{subequations}
Equations \cref{eq:pBC_Ica_v}-\cref{eq:pBC_Ica_h}, henceforth referred to as \textit{the voltage subsystem}, describe the membrane potential dynamics $V$ and the voltage-dependent activation $n$ and inactivation $h$ variables, respectively. 
The model includes a leak current ($I_{\rm L}$), a spike-generating fast potassium ($I_{\rm K}$) current, a fast spiking sodium ($I_{\rm Na}$) current, a persistent sodium current ($I_{\rm NaP}$), and a calcium-activated nonspecific cationic current ($I_{\rm CAN}$), through which calcium dynamics $[\rm Ca]$ influence membrane potential dynamics $V$. In addition to the currents from \cite{TB,PR}, we incorporate an additional voltage-dependent calcium current \cite{wang2020complex, phillips2019biophysical, phillips2022burstlet} to account for the bidirectional interaction between membrane potential and calcium dynamics, which plays an important role in capturing the effects of NE, as discussed later in \cref{modelNE}. As a result, the neuron should no longer be considered closed to calcium flux into and out of the cell, as in \cite{TB,PR}, where the total intracellular calcium concentration within the cell ($[\rm Ca]_{\rm Tot}$) was assumed constant. Instead, model \cref{eq:pBC_Ica} now represents an open cell in which $[\rm Ca]_{\rm Tot}$ is treated as a dynamic variable. 

The ionic currents are defined as follows: 
\begin{equation}
    \begin{aligned}
        I_{\rm L} &= g_{\rm L} (V - V_{\rm L}) \\
        I_{\rm K} &= g_{\rm K} n^4 (V - V_{\rm K}) \\
        I_{\rm Na} &= g_{\rm Na} m_\infty^3(V) (1 - n) (V - V_{\rm Na}) \\
        I_{\rm NaP} &= g_{\rm NaP} mp_\infty(V) h (V - V_{\rm Na}) \\
        I_{\rm CAN} &= g_{\rm CAN} f([{\rm Ca}]) (V - V_{\rm Na}) \\
        I_{\rm Ca} &= g_{\rm Ca} mp_\infty(V) (V - V_{\rm Ca}),
    \end{aligned}
    \label{eq:pBC_ICa_currents}
\end{equation}
where $g_i$ is the maximum conductance and $V_i$ denotes the reversal potential for each current $I_i$.
To simplify the model, we followed prior studies \cite{TB,PR,toporikova2015sigh,chevalier2016development,wang2017timescales} in our treatment of CAN and calcium currents, adopting similar formulations and parameter choices.
The steady-state activation/inactivation ($x_{\infty}$) and time constant ($\tau_x$) for $I_{\rm Na}$, $I_{\rm NaP}$ and $I_{\rm Ca}$ take the following forms:
\begin{equation}
    \begin{aligned}
        x_\infty &= \frac{1}{1 + \exp{((V-V_x)/s_x)}}, \;\; x \in \{mp, m, n, h\} \\
        \tau_x &= \bar{\tau}_y / \cosh{((V-V_y)/(2s_y))}, \;\; y \in \{n, h\}.
    \end{aligned}
    \label{eq:pBC_ICa_other_eqns}
\end{equation}
$I_{\rm CAN}$ activation takes a different form and depends on the calcium concentration in the cytoplasm $\rm [Ca]$:
\begin{equation}
    \begin{aligned}
        f([{\rm Ca}]) &= \frac{1}{1 + (\frac{K_{\rm CAN}}{[{\rm Ca}]})^{n_{\rm CAN}}}.
    \end{aligned}
    \label{eq:pBC_ICAN_activation}
\end{equation}

Equations \cref{eq:pBC_Ica_ca} - \cref{eq:pBC_Ica_l}, henceforth referred to as \textit{the calcium subsystem}, describe the dynamics of intracellular calcium concentration in the cytoplasm $\rm [Ca]$, the combined intracellular and endoplasmic reticulum (ER) total calcium concentration $\rm [Ca]_{Tot}$, and the fraction of ${\rm IP_3}$ receptors that have not been inactivated. The calcium dynamics are influenced by the voltage-dynamics via the term $-\alpha_{\rm Ca}I_{\rm Ca}$ in \cref{eq:pBC_Ica_ca} and \cref{eq:pBC_Ica_catot}, which represents $\rm Ca^{2+}$ influx from the extracellular space through voltage-gated calcium channels. In \cref{eq:pBC_Ica_ca}, $\rm{J_{ER_{in}}}$ represents the calcium flux per unit volume from the endoplasmic reticulum (ER) into the cell cytoplasm, and this depends on $l$. ${\rm{J_{ER_{out}}}}$ represents the calcium flux that flows out of the cytoplasm into the ER. These fluxes are modeled by \cref{eq:pBC_jerin} - \cref{eq:pBC_caer}:
\begin{subequations}
    \begin{align}
        {\rm{J_{ER_{in}}}} &= \left( L_{\rm IP_3} + P_{\rm IP_3} \left[ \frac{[{\rm IP_3}][{\rm Ca}]l}{([{\rm IP_3}] + K_I)([{\rm Ca}] + K_a)} \right]^3 \right) ([\rm Ca]_{\rm ER} - [\rm Ca]) \label{eq:pBC_jerin} \\
        \rm{J_{ER_{out}}} &= {V_{\rm SERCA}} \frac{[\rm Ca]^2}{K_{\rm SERCA}^2 + [\rm Ca]^2} \label{eq:pBC_jerout} \\    [\rm Ca]_{\rm ER} &= \frac{[\rm Ca]_{\rm Tot} - [\rm Ca]}{\sigma}. \label{eq:pBC_caer}
    \end{align}
\end{subequations}

The last term in \cref{eq:pBC_Ica_ca} and \cref{eq:pBC_Ica_catot} represents the membrane $\rm Ca^{2+}$ pump, which expels free intracellular $\rm Ca^{2+}$ from the cytoplasm, where $\rm [Ca]_{\rm min}$ sets a minimal baseline calcium concentration and $\tau_{\rm Ca}$ is the time constant for the $\rm Ca^{2+}$ pump \cite{jasinski,phillips2022burstlet,rubin2009calcium}. 

The parameter values associated with this model are listed in \cref{tab:par_val}. Additional details about the model can be found in \cite{PR,wang2020complex,jasinski}.

\begin{table}[!ht]
    \centering
    \begin{tabular}{|c|c|}
        \hline 
        \textbf{Currents} & \textbf{Associated Parameter Values} \\ [1ex]
        \hline 
        $I_{\rm L}$ & $g_{\rm L} = 2.3$nS, $V_{\rm L} = -58$mV. \\ [1ex]
        \hline
        $I_{\rm K}$ & $g_{\rm K} = 11.2$nS, $V_{\rm K} = -85$mV. \\ [1ex]
        \hline
        \multirow{2}{1.5em}{$I_{\rm Na}$} & $g_{\rm Na} = 28$nS, $V_m = -34$mV, $V_n = -29$mV, \\ [1ex]
         & $s_m = -5$mV, $s_n = -4$mV, $\bar{\tau}_n = 10$ms, $V_{\rm Na} = 50$mV. \\ [1ex]
        \hline
        \multirow{2}{2em}{$I_{\rm NaP}$} & $g_{\rm NaP} \in [0, 5]$nS, $V_h = -48$mV, $V_{mp} = -40$mV, \\ [1ex]
         & $s_h = 5$mV, $s_{mp} = -6$mV, $\bar{\tau}_h = 10000$ms. \\ [1ex]
        \hline
        $I_{\rm CAN}$ & $g_{\rm CAN} \in [0, 4]$nS, $K_{\rm CAN} = 0.74 \mu$M, $n_{\rm CAN} = 0.97$. \\ [1ex]
        \hline
        $I_{\rm Ca}$ & \RED{$g_{\rm Ca} \in [0, 0.0008]$nS}, $V_{\rm Ca} = 150$mV. \\ [1ex]
        \hline
        \multirow{4}{4em}{ER $[{\rm Ca}]$} & $[{\rm IP_3}] \in [0, 1] \mu$M, $L_{\rm IP_3} = 0.37/$ms, $P_{\rm IP_3} = 31000/$ms, \\ [1ex]
         & $K_I = 1 \mu$M, $K_a = 0.4 \mu$M, $\sigma = 0.185$, $f_i = 0.000025$, \\ [1ex]
         & $V_{\rm SERCA} = 400 \mu$M/ms, $K_{\rm SERCA} = 0.2 \mu$M, $\alpha_{\rm Ca} = 0.025$mM/fC, \\ [1ex]
         & $[\rm Ca]_{min} = 0.005 \mu$M, $\tau_{\rm Ca} = 500$ms, $A = 0.005 \mu$M/ms, $K_d = 0.4 \mu$M. \\ [1ex]
        \hline
        Other & $C_m = 21$pF. \\ [1ex]
        \hline
    \end{tabular}
    \caption{The parameter values for model \cref{eq:pBC_Ica}.}
    \label{tab:par_val}
\end{table}

\subsection{Activity Patterns in the Model}\label{activity_patterns}

\begin{figure}[!h]
    \begin{center}
    \begin{tabular}
    {@{}p{0.5\linewidth}@{\quad}p{0.5\linewidth}@{}}
    \subfigimg[width=1\linewidth]{\bfseries{{(i)}}}{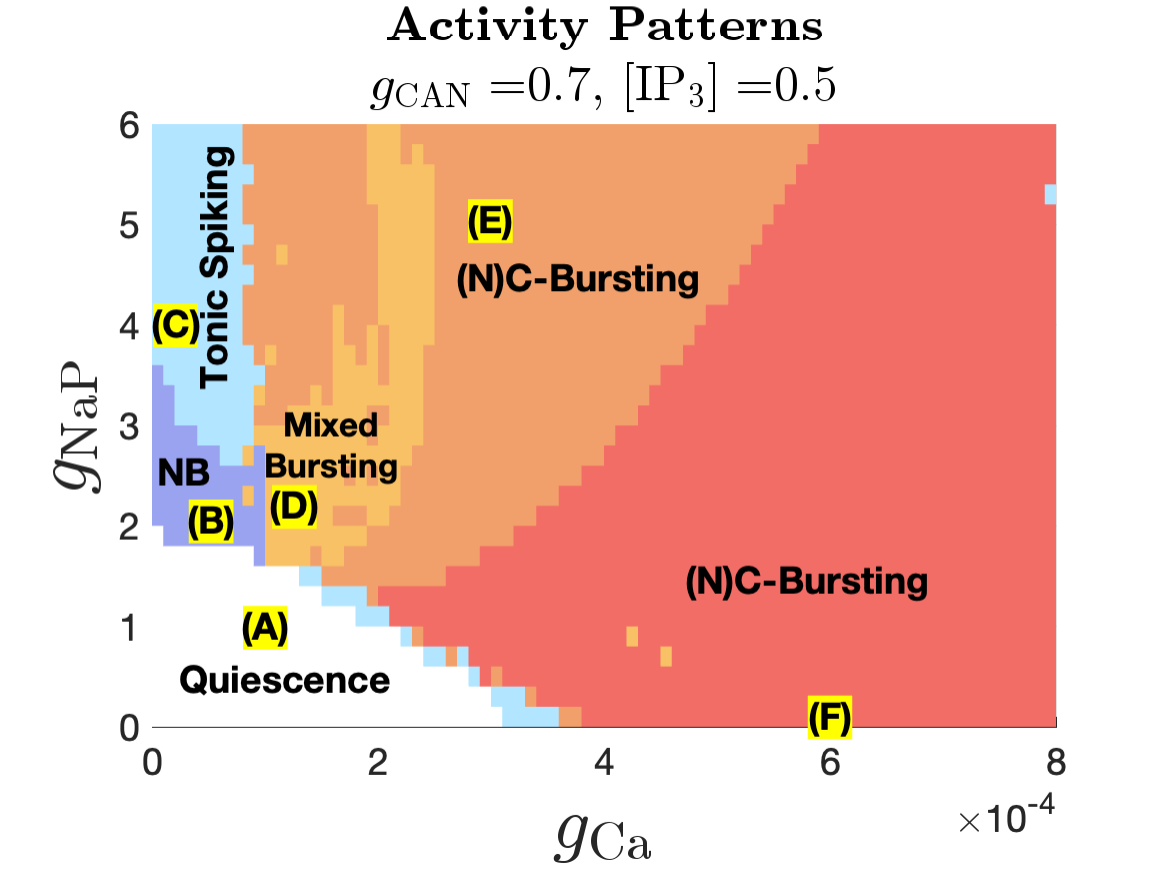} & 
    \subfigimg[width=0.9\linewidth]{}{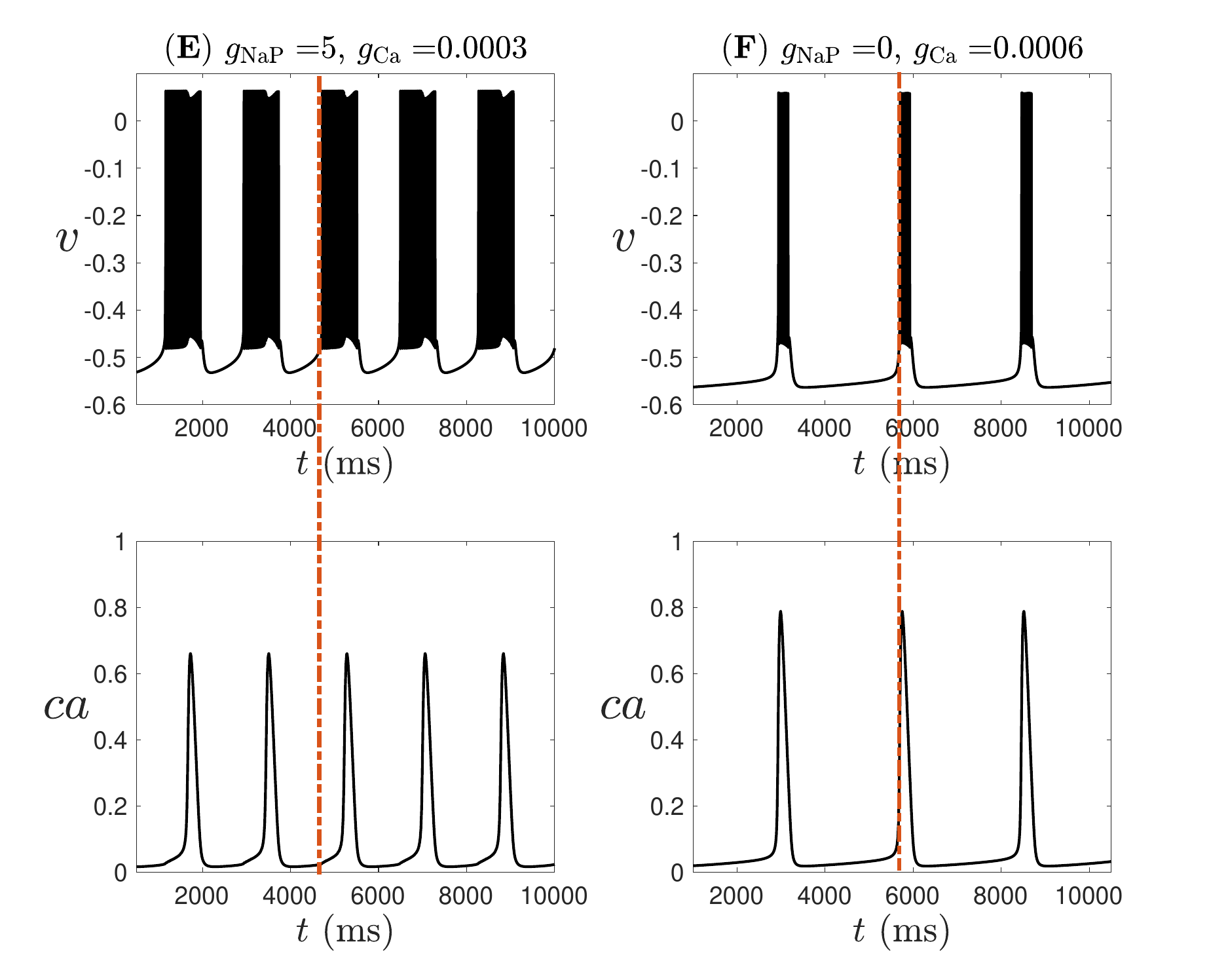} \\
    \multicolumn{2}{c}{
    \subfigimg[width=\linewidth]{}{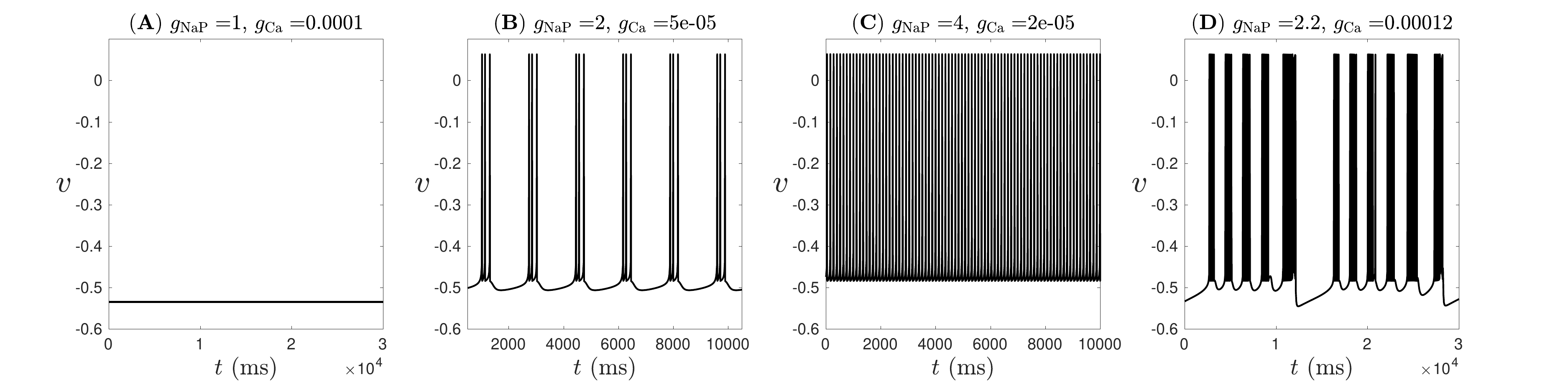}}
    \end{tabular}
    \end{center}
    \caption{Activity patterns of the dimensionless model \cref{eq:pBC_ICa_slow}, derived from model \cref{eq:pBC_Ica}, depend on parameter values $g_{\rm NaP}$ and $g_{\rm Ca}$, for $g_{\rm CAN}=0.7$ and $[\rm IP_3]=0.5$. (i) The range of $g_{\rm Ca}$ and $g_{\rm NaP}$ for different solution patterns. White color denotes quiescence and other colors are tonic spiking (light blue) or bursting (dark blue, orange or red). Panels (A) through (F) illustrate sample traces for each activity pattern corresponding to labeled parameter values in panel (i). The red dashed line in panels (E) and (F) marks burst onset, highlighting a key difference between the two (N)C-Bursting regions (orange and red): in the orange region, the burst initiates before $ca$ spikes, whereas in the red region, $ca$ spikes before burst onset. 
    \label{fig:activity_patterns}}
\end{figure}

The \pbc{} model \cref{eq:pBC_Ica} exhibits a diverse range of dynamically different activity patterns depending on variations in $g_{\rm NaP}$ and $g_{\rm Ca}$, the two critical parameters that govern transitions between different bursting mechanisms (see \cref{fig:activity_patterns}(i)). The voltage and calcium time series of a dimensionless version of model \cref{eq:pBC_Ica} (see \cref{eq:pBC_ICa_slow} in \cref{gspt}) are shown for parameters corresponding to different regions in \cref{fig:activity_patterns}(i). Specifically, panels illustrate (A) Quiescence, (B) $I_{\rm NaP}$-dependent bursting, (C) Tonic Spiking, (D) Mixed Bursting, and (E, F) two different types of $I_{\rm CAN}$-dependent bursting behaviors. 

Increasing $g_{\rm NaP}$ at low $g_{\rm Ca}$ values causes transitions from quiescence (\cref{fig:activity_patterns}(i), white region; \cref{fig:activity_patterns}A) to NB (\cref{fig:activity_patterns}(i), dark blue region; \cref{fig:activity_patterns}B) and eventually to tonic spiking (\cref{fig:activity_patterns}(i), pale blue region; \cref{fig:activity_patterns}C). Bursting in the NB region occurs when the persistent sodium current is in the bursting range and there are no large $\rm Ca^{2+}$ oscillations. Following nomenclature established in prior studies \cite{TB,PR,wang2020complex}, we refer to $I_{\rm NaP}$-dependent bursting as \emph{N-bursting (NB)}. For relatively large $g_{\rm Ca}$ values, the model \cref{eq:pBC_Ica} generates a distinct type of bursting that depends on the activation of $I_{\rm CAN}$ through $\rm Ca^{2+}$ oscillations. We refer to this bursting as \emph{(N)C-bursting}. The parentheses around `N' indicate that the dependence of this bursting on $I_{\rm NaP}$ varies depending on $g_{\rm NaP}$ and $g_{\rm Ca}$. Specifically, for $g_{\rm Ca} \geq 0.0004$, and either $0 \leq g_{\rm NaP} \leq 1.8$ or $g_{\rm NaP} \geq 3.5$, the bursts are driven by $I_{\rm CAN}$ in the sense that they persist when $I_{\rm NaP}$ is blocked but disappear when $I_{\rm CAN}$ is blocked. This bursting type, known as \emph{C-bursting}, represents a regime where the rhythmic bursting activity is maintained exclusively by $I_{\rm CAN}$. Within (N)C-bursting regime, there is also a subclass of bursting that depends on both $I_{\rm NaP}$ and $I_{\rm CAN}$, which we refer to as \emph{NC-bursting}. This bursting exhibits two distinct subtypes: (Type 1) Both NaP and CAN mechanisms are essential for the NC bursting -  blocking either channel abolishes the bursts. 
These bursts occur for $0.0001 \leq g_{\rm Ca} \leq 0.00035$ and $0 \leq g_{\rm NaP} \leq 1.8$ or $g_{\rm NaP} \geq 3.5$ within the red and orange regions in \cref{fig:activity_patterns}(i); (Type 2) Bursting involves both $I_{\rm NaP}$ and $I_{\rm CAN}$, but the bursting continues if only one of the two currents is blocked. Bursting stops only when both channels are blocked. This type of (N)C-bursting occurs for $g_{\rm Ca} \geq 0.0004$ and $2 \leq g_{\rm NaP} \leq 3$ in \cref{fig:activity_patterns}(i). 
In this paper, we focus on exploring the effects of NE on N- and C-bursters, excluding NC-bursters due to the limited experimental data on NE's impact on those neurons, as a result of practical difficulties in isolating such pacemakers in experiments. 
For simplicity, we do not differentiate between C-bursting and the subtypes of NC-bursting in \cref{fig:activity_patterns}(i), but instead distinguish (N)C-bursting types based on the timing of the elevation in cytoplasmic $\rm Ca^{2+}$ (i.e., $ca$ jump-up) relative to burst initiation: the orange region corresponds to cases where bursting begins before the $ca$ jump-up (\cref{fig:activity_patterns}E), whereas the red region represents cases where $ca$ jumps up before or near the onset of the burst (\cref{fig:activity_patterns}F). 

Finally, for intermediate values of $g_{\rm Ca}$, we observe \emph{mixed bursting} (MB) patterns in the light orange region of \cref{fig:activity_patterns}(i) (see \cref{fig:activity_patterns}D for a representative trace).
Such MB patterns, characterized by long bursts separated by sequences of short bursts, have been previously observed in \pbc{} models (e.g., \cite{PR,jasinski,wang2016jcompneur,wang2017timescales}), as well as in \emph{in vitro} experiments, where the long burst can be associated with sighing \cite{dunmyre2011interactions,lieske2000reconfiguration,tryba2008differential,pena2004differential}. It remains unclear whether the MB solution shown in \cref{fig:activity_patterns}D shares similar underlying mechanisms as those described in previous studies \cite{wang2016jcompneur,wang2017timescales}. Moreover, in \RED{Appendix \hyperlink{appE}{E}}, we demonstrate that model \cref{eq:pBC_Ica} can generate a diverse range of MB patterns, some of which differ qualitatively from previously reported MB solutions and may exhibit quasi-periodic or chaotic dynamics (see \cref{fig:MB}). 

\subsection{Modeling effects of norepinephrine (NE)} \label{modelNE}

Neuromodulation is a process by which various substances influence neuronal activity. Many experimental and computational studies have made directed efforts into understanding the principles of neuromodulation, as well as how neuromodulators regulate respiratory rhythmic activity \cite{doi2008neuromodulation, dhingra2024asymmetric, toporikova2013dynamics, zanella2014norepinephrine, VR, baertsch2019insights, shao2000acetylcholine, richter2003serotonin, guiard2008functional}. Experimental evidence has shown that neuromodulators like acetylcholine, serotonin, norepinephrine (NE), histamine, dopamine, and substance P can affect the frequency, amplitude, and regularity of respiratory activity (see \cite{doi2008neuromodulation} for review).  

In mice, noradrenergic effects on respiration are considered to be primarily excitatory and mediated by $\alpha_1$-receptors \cite{viemari2004phox2a}. In \cite{VR}, NE was applied to synaptically isolated \pbc{} neurons in mice medullary brain slices, and the authors similarly found that the observed effects were mainly mediated by $\alpha_1$-receptor activation, despite NE also acting on other noradrenergic receptor subtypes. 
Further, the effects of NE on intrinsic bursting neurons (i.e., ``pacemakers") critically depended on whether their underlying bursting mechanisms were cadmium-insensitive (i.e., not dependent on calcium currents) or cadmium-sensitive (i.e., reliant on calcium currents). Specifically, NE increased the burst frequency, but did not change the burst duration in cadmium-insensitive bursters (e.g., N-bursting neurons). In contrast, in cadmium-sensitive bursters (e.g., (N)C-bursting neurons), the burst frequency did not change with NE, but the burst duration increased. Nonpacemakers also exhibited differential effects under the application of NE. In an ``active nonpacemaker", which continues to spike tonically when isolated from the network (e.g., a tonic spiking neuron), NE induced bursting. This induced pacemaking behavior was further found to be $I_{\rm CAN}$-dependent, as the NE-induced bursting activity continued in the presence of NaP-blocker riluzole, but was lost when $I_{\rm CAN}$ was blocked. In contrast, a ``silent nonpacemaker", which does not produce spikes when isolated from the network, continued to remain silent in the presence of NE.

\RED{Since $\alpha_1$-receptor activation increases the CAN-current conductance in different cell types (\cite{hilletal}, \cite{HL}), prior computational studies \cite{TB,PR} modeled the effects of NE, an $\alpha_1$-receptor agonist, on the \pbc{} neurons via an increase in $g_{\rm CAN}$.} 
These studies effectively captured several NE-mediated effects on \pbc{} neurons, such as the increase in burst frequency with unchanged burst duration in N-bursters and the increase in burst duration with unchanged burst frequency in C-bursters. However, none of these models account for the effects of NE on non-bursting neurons. A natural question arises: Can an increase in $g_{\rm CAN}$ also produce NE-induced effects on silent and tonic spiking neurons? The answer is no. We demonstrate in \cref{results} that increasing $g_{\rm CAN}$ alone does not induce a transition from spiking to bursting in single \pbc{} neuron models \cite{TB,PR,jasinski,phillips2022burstlet}. In fact, bifurcation analysis in \cite{wang2016jcompneur} suggests that increasing $g_{\rm CAN}$ can shrink the bursting region, favoring tonic spiking neurons - a result that is contrary to what has been observed in the experiments in \cite{VR}. Moreover, \RED{simulations using the models from \cite{TB,PR}} 
indicate that with a sufficient increase in $g_{\rm CAN}$, silent neurons eventually transition to either a bursting or spiking behavior. Thus, while increasing $g_{\rm CAN}$ captures certain aspects of NE effects on bursting \pbc{} neurons, it does not fully capture all experimentally observed effects, particularly in non-bursting neurons. 

In the review by \cite{Exton}, experiments in hepatocytes showed that ${\rm IP_3}$ functions as a second messenger for $\alpha_1$-adrenergic agonists and other calcium-mediated agonists. This was further supported by experiments in neurons \cite{choy2018noradrenaline}. Building on this, in addition to the previously described increases in $g_{\rm CAN}$, we propose that 
NE application in our model also leads to an increase in $[{\rm IP_3}]$. 
Incorporating this mechanism, we observe an overall increase in burst frequency in N-bursters as $g_{\rm CAN}$ and $[\rm IP_3]$ increase (see \cref{fig:NB_gcan_ip3} in the Results section), consistent with the experimental findings in \cite{VR}. This increase is primarily driven by the enhanced depolarizing potential due to $g_{\rm CAN}$, while $[\rm IP_3]$ has minimal impact on N-bursting dynamics. In C-bursters, burst frequency also increases with $g_{\rm CAN}$ (\cref{fig:CB_gcan_ip3}) and we show in \cref{CB} that this is mainly due to the involvement of the voltage-gated calcium current $I_{\rm Ca}$. However, we also demonstrate in \cref{CB} that an increase in $[\rm IP_3]$ may counteract the $g_{\rm CAN}$-induced frequency increase in the model \cref{eq:pBC_Ica}, resulting in a constant burst frequency in the presence of NE, consistent with experimental observations. Nonetheless, there exist parameter regimes where increasing $[\rm IP_3]$ and $g_{\rm CAN}$ leads to increases in both burst frequency and duration, which contradicts experimental results \cite{VR}. 

Different from previous NE modeling studies that overlook silent and tonic spiking neurons, a key novelty of our proposed mechanism is its ability to successfully capture NE-induced conditional pacemaking in tonic spiking neurons while also preserving the inactivity of silent neurons in the presence of NE. We show in \cref{TS} that a tonic-spiking neuron can transition to bursting with increases in both $g_{\rm CAN}$ and $[{\rm IP_3}]$. Moreover, both mechanisms are essential: increasing $g_{\rm CAN}$ alone results in continued tonic spiking; while increasing $[{\rm IP_3}]$ alone can induce bursting, these bursts are not C-bursts since they are lost when either $I_{\rm NaP}$ or $I_{\rm CAN}$ is blocked. Only bursts induced by simultaneous increases in $g_{\rm CAN}$ and $[{\rm IP_3}]$ are truly CAN-dependent, as they persist when $I_{\rm NaP}$ is blocked but disappear when $I_{\rm CAN}$ is blocked - consistent with experimental data described above. This further confirms the necessity of incorporating both $g_{\rm CAN}$ and $[{\rm IP_3}]$ increases to accurately model NE effects on \pbc{} neurons. Finally, we also identify a subset of silent neurons in our model that remains inactive despite increased $g_{\rm CAN}$ and $[{\rm IP_3}]$ (see \cref{Q}), aligning with observations in~\cite{VR}.

In the following subsection, we provide a brief overview of geometric singular perturbation theory (GSPT) and the derivation of relevant subsystems and important geometric structures, which will be used in \cref{results} to analyze the distinct effects of NE on the different solution patterns of \cref{eq:pBC_Ica} as described above.

\subsection{Geometric Singular Perturbation Theory} \label{gspt}

In addition to our modeling efforts, we apply techniques from dynamical systems theory, including phase plane analysis, fast-slow decomposition, and bifurcation analysis, to uncover mechanisms underlying different effects of NE on various types of intrinsic bursting \pbc{} neurons as well as the conditional bursting activity induced by NE. In this section, we use geometric singular perturbation theory (GSPT) \cite{fenichel1979, rinzel1987formal} to identify the underlying geometric structures that organize the dynamics of \cref{eq:pBC_Ica}, which will be used for our analysis in \cref{results} of the mechanisms underlying NE modulation in individual \pbc{} neurons. 
For readers less familiar with GSPT and bifurcation analysis, we include a basic introduction to these mathematical approaches in Appendix \hyperlink{appA}{A}. 


GSPT has been widely applied to study \pbc{} neurons, approached as either a two-timescale or a three-timescale problem \cite{PR,wang2016jcompneur,wang2017timescales,wang2020complex}. Traditionally, three-timescale problems were often simplified to two-timescale problems, which aligns with the natural setting of the GSPT approach \cite{PR, baldemir2020pseudo}. However, a two-timescale decomposition can fail to capture crucial dynamical aspects of a three-timescale system and is hence an insufficient approach for modeling and analysis of such problems \cite{Nan2015,phan2024mixed}. Our approach is to analyze model \cref{eq:pBC_Ica} as a three-timescale problem, based on our dimensional analysis detailed in Appendix \hyperlink{appB}{B}. This analysis transforms \cref{eq:pBC_Ica} to the following dimensionless system of equations:
\RED{
\begin{equation}
    \begin{aligned}
        \varepsilon \frac{dv}{dt_s} &= f_1 (v, n, h, ca) \\
        \varepsilon \frac{dn}{dt_s} &= f_2 (v, n) \\
        \frac{dh}{dt_s} &= g_1 (v, h) \\
        \frac{d ca}{dt_s} &= R_{ca}\cdot g (v, ca, ca_{tot}, l) \\
        \frac{d ca_{tot}}{dt_s} &= \delta h (v, ca) \\
        \frac{dl}{dt_s} &= g_2 (ca,l),
    \end{aligned}
    \label{eq:pBC_ICa_slow}
\end{equation}}
where $\varepsilon, \delta \ll 1$ are independent timescale parameters, $t_s$ is the slow dimensionless time variable, \RED{$R_{ca}$ is a dimensionless coefficient of order $O(10)$,} $f_1, \allowbreak f_2, \allowbreak g_1, \allowbreak g, \allowbreak h, \allowbreak g_2$ are $O(1)$ functions specified in Appendix \hyperlink{appB}{B}.
The lowercase variables $v$, $ca$, $ca_{tot}$ represent the dimensionless forms of $V$, $\rm [Ca]$ and $\rm [Ca]_{Tot}$, respectively, while $n,h$ and $l$ are already dimensionless. Throughout the paper, all model variables are plotted in their dimensionless forms. Time traces, however, are shown using the original time unit (ms). 
\RED{From our analysis in Appendix \hyperlink{appB}{B}, we conclude that $v$ and $n$ evolve on a fast timescale of $O(\varepsilon^{-1})$, $h$ and $l$ evolve on a slow timescale of $O(1)$, and $ca_{tot}$ evolves on a superslow timescale of $O(\delta)$. Although $ca$ evolves on an intermediate timescale, slower than the fast variables but faster than the slow variables, we group it with the slow variables for convenience in the GSPT analysis (Appendix \hyperlink{appC}{C}). This choice does not change the resulting critical manifold $M_s$ and the superslow manifold $M_{ss}$, which are defined below.} We call system \cref{eq:pBC_ICa_slow} that evolves over the \emph{slow timescale} $t_s$ the \textit{slow system}, which is the same as \cref{eq:nondim} in Appendix \hyperlink{appB}{B}. 

The existence of two independent singular perturbation parameters, $\varepsilon$ and $\delta$, implies there are various ways to implement GSPT, each yielding distinct singular limit predictions \cite{Nan2015,phan2024mixed}, as detailed in Appendix \hyperlink{appC}{C}. In brief, applying GSPT yields the \emph{fast layer problem} (see \cref{eq:pBC_ICa_fastlayer} in Appendix \hyperlink{appC}{C}), whose set of equilibrium points defines the \textit{critical manifold} $M_s$:
\RED{
\begin{equation}
    M_s := \{(v,n,h,ca,ca_{tot},l) : f_1 (v,n,h,ca)  = f_2 (v,n) = 0\}.
    \label{eq:pBC_ICa_Ms}
\end{equation}}
The slow evolution of trajectories of the \emph{slow reduced problem} (see \cref{eq:pBC_ICa_slowreduced} in Appendix \hyperlink{appC}{C}) is slaved to $M_s$ until nonhyperbolic points - such as the fold points $L_s$ (defined in  \cref{eq:pBC_ICa_fold_app}) - are encountered. 

Moreover, GSPT analysis also yields the \emph{slow layer problem} (see \cref{eq:pBC_ICa_slowlayer} in Appendix \hyperlink{appC}{C}), whose equilibrium points form a one-dimensional subset of $M_s$ called the \textit{superslow manifold} $M_{ss}$:
\RED{
\begin{equation}
    M_{ss} := \{(v,n,h,ca,ca_{tot},l) \in M_s: g_1 (v,h)  = g(v, ca, ca_{tot}, l) =  
     g_2 (ca,l)= 0\}.
    \label{eq:pBC_ICa_Mss}
\end{equation}}
Near the singular limit $\delta\to 0$, the full solution trajectories travel near $M_{ss}$ on the superslow timescale until they reach the nonhyperbolic points on $M_{ss}$ - such as Hopf or saddle-node bifurcations (denoted as $L_{ss}$) of the slow layer problem \cref{eq:pBC_ICa_slowlayer} - where Fenichel's theory (GSPT) breaks down \cite{fenichel1979}. For further details of these subsystems and their derivations, please refer to Appendix \hyperlink{appC}{C}. For brevity, we do not show the detailed constructions of singular orbits, which are solution segments of singular limit systems. Instead, we present orbits of the full system \cref{eq:pBC_ICa_slow} and refer to different segments as being governed by various subsystems derived from the GSPT analysis, evolving under the fast, slow, or superslow flow. For further details on constructing singular orbit and using singular limit systems to understand the nature of the oscillations in the full system, see, e.g., \cite{Desroches2012,phan2024mixed}.

\RED{
In addition to using the reduced subsystems as described above, we also exploit the full four-timescale structure of the model through a complementary viewpoint within the framework of GSPT \cite{Nan2015}. Specifically, we separate the system into the two-timescale voltage subsystem $(v, n, h)$ and the three-timescale calcium subsystem $(ca, ca_{tot},l)$, analyze each using its natural timescale separation, and couple the results via two-parameter bifurcation analysis, following \cite{wang2017timescales,wang2020complex}.  }




\section{Results and Bifurcation Analysis} \label{results}

In this section, we use GSPT and bifurcation analysis to examine the dynamic mechanisms underlying the various activity patterns of model \cref{eq:pBC_ICa_slow} as summarized in \cref{fig:activity_patterns}(i) and analyze how changes in $g_{\rm CAN}$ and $[{\rm IP_3}]$ replicate NE effects on \pbc{} neurons, as discussed in \cref{modelNE}. 
Specifically, we analyze the effects of NE on three distinct activity patterns of \cref{eq:pBC_ICa_slow}, using representative parameter values for each case: the $I_{\rm NaP}$-dependent N-bursting pattern with $g_{\rm NaP}=2$, $g_{\rm CAN}=0.7$, $g_{\rm Ca}=0.00002$ and $[{\rm IP_3}] = 0.5$ (\cref{NB}); the C-bursting pattern with $g_{\rm NaP}=0$, $g_{\rm CAN} = 0.7$, $g_{\rm Ca}=0.0005$ and $[{\rm IP_3}] = 0.5$ (\cref{CB}); and the tonic spiking pattern with $g_{\rm NaP}=4$, $g_{\rm CAN} = 0.7$ , $g_{\rm Ca}=0.0002$ and $[{\rm IP_3}] = 0.1$ (\cref{TS}). We also identify a subset of silent \pbc{} neurons that remain inactive despite increased $g_{\rm CAN}$ and $[{\rm IP_3}]$ in \cref{Q}. In the following subsections, we fix $g_{\rm NaP}$ and $g_{\rm Ca}$ for each solution pattern while varying $g_{\rm CAN}$ and $[\rm IP_3]$ to investigate NE effects. All other parameters remain fixed as listed in \cref{tab:par_val}.

\subsection{Effects of NE on N-bursting neurons} \label{NB}

\begin{figure}[!t]
    \begin{center}
    \begin{tabular}
    {@{}p{0.3\linewidth}@{\quad}p{0.3\linewidth}@{\quad}p{0.3\linewidth}@{}}
    \subfigimg[width=1.1\linewidth]{\bfseries{\footnotesize{(A)}}}{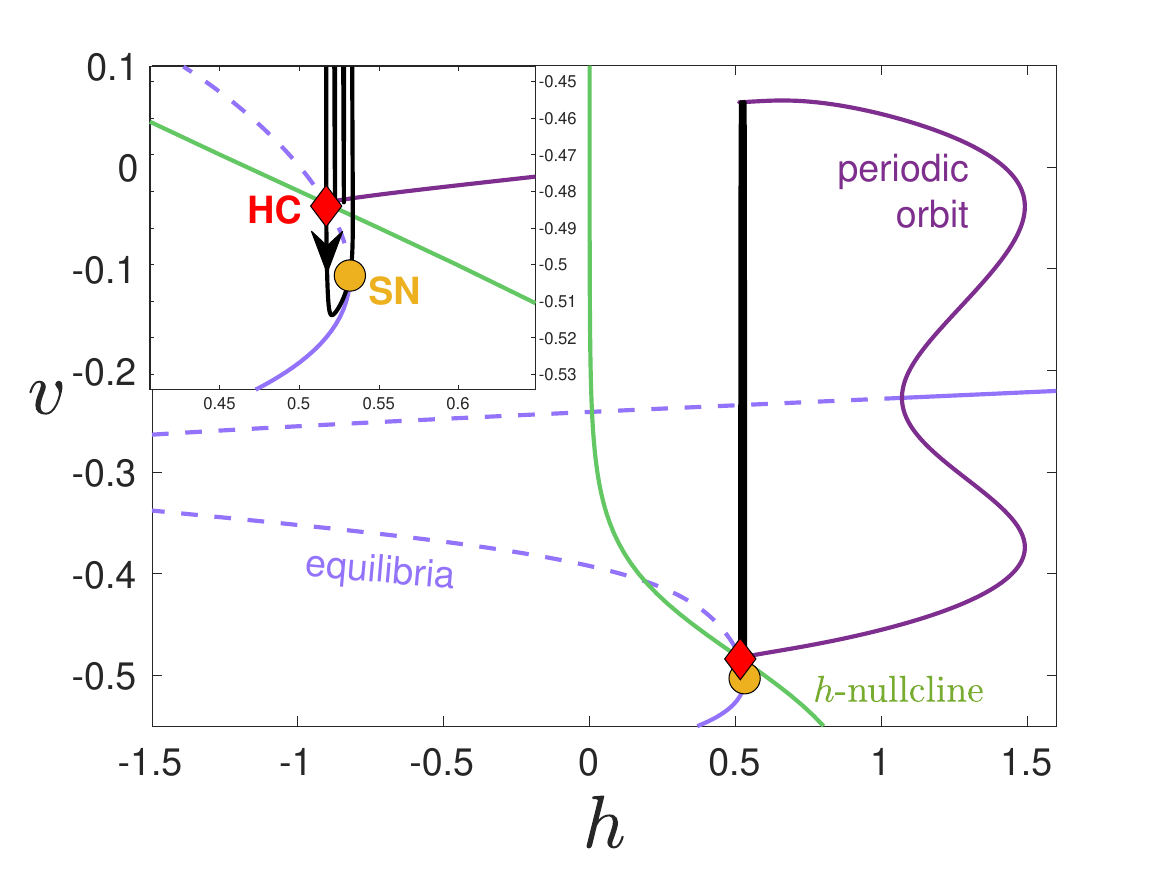} &
    \subfigimg[width=1.1\linewidth]{\bfseries{\footnotesize{(B)}}}{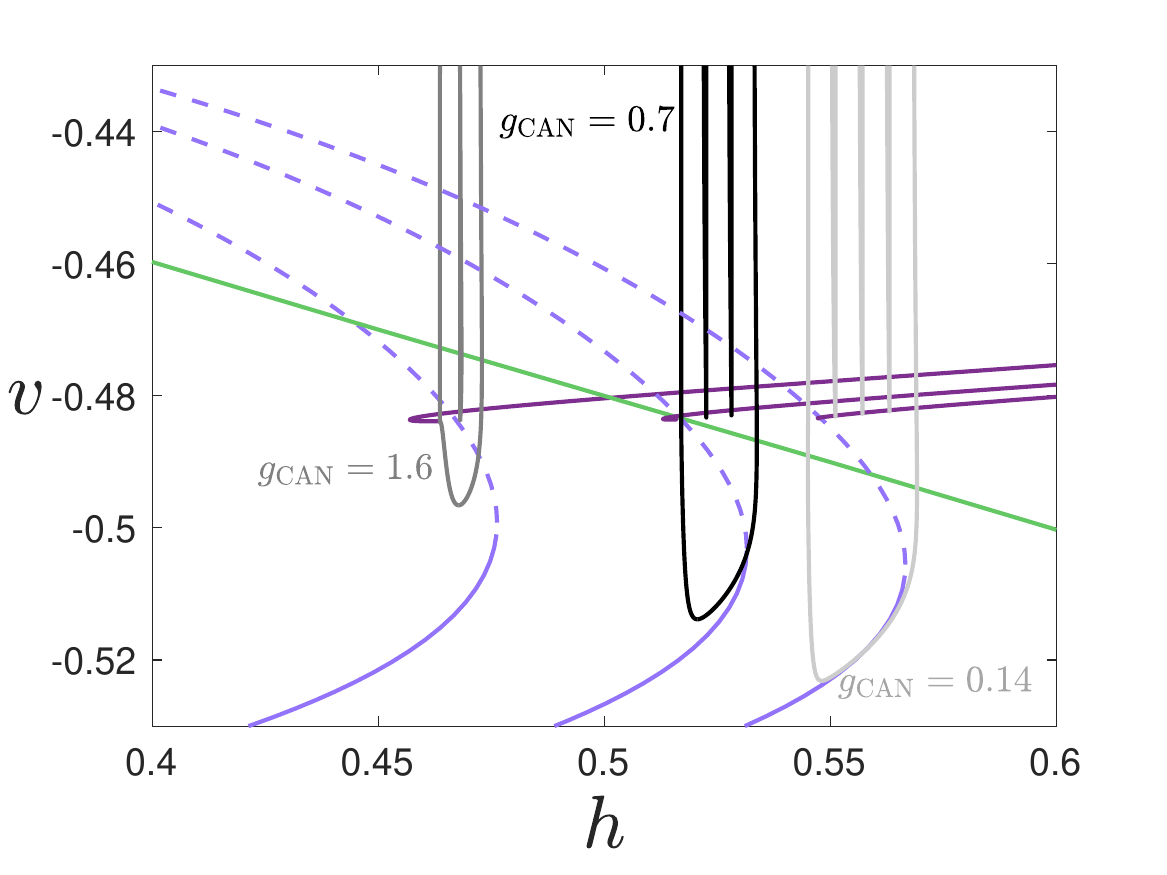} &
    \subfigimg[width=1.1\linewidth]{\bfseries{\footnotesize{(C)}}}{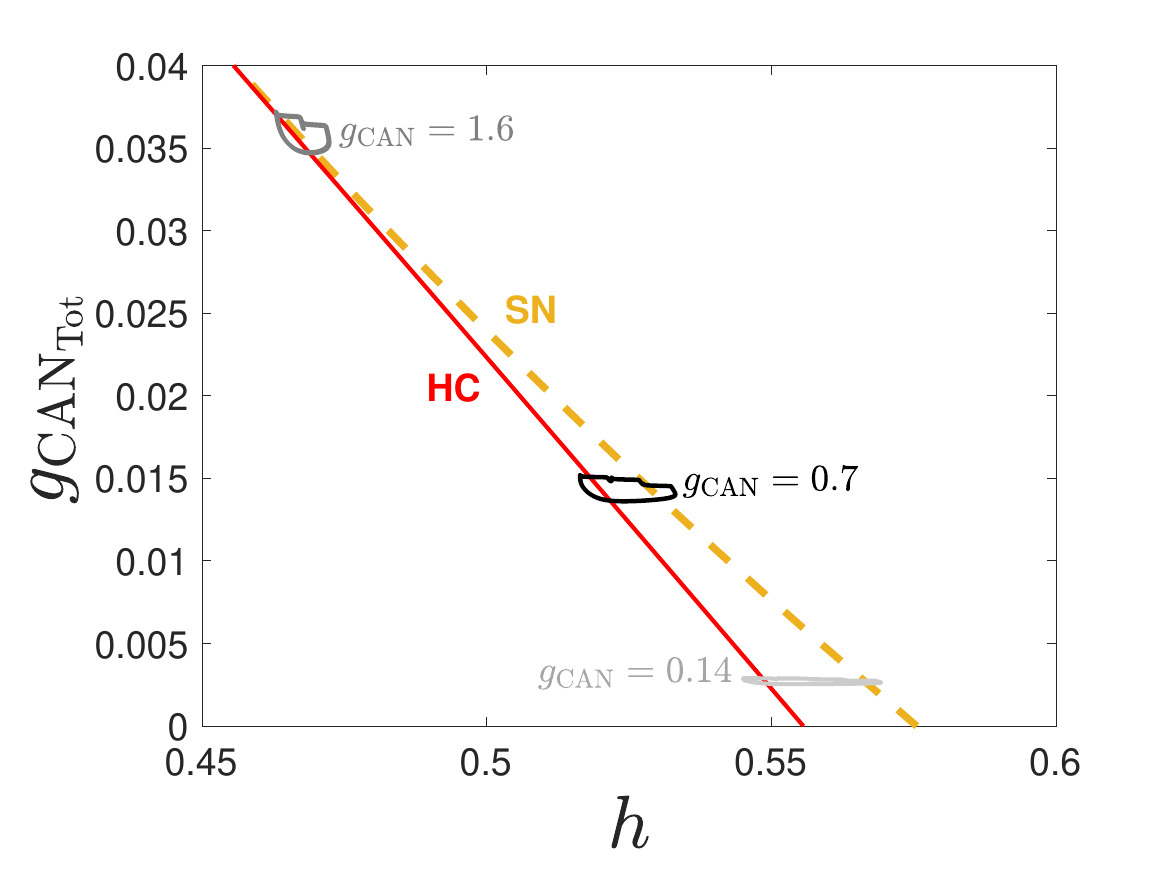}
    \end{tabular}
    \end{center}
        \caption{Projection of the $I_{\rm NaP}$-dependent N-burst solution of \cref{eq:pBC_ICa_slow} along with the bifurcation diagrams for $g_{\rm NaP}=2$, $g_{\rm Ca}=0.00002$, $[\rm IP_3]=0.5$, and varying values of $g_{\rm CAN}$. (A) $g_{\rm CAN}=0.7$. Projection of the solution (black) and the bifurcation diagram for the fast $(v,n)$ subsystem with respect to $h$, along with the $h$-nullcline (green). The S-shaped light \RED{blue} curve (solid where attracting, dashed otherwise) denotes the equilibria of the $v,n$ equations and represents the projection of the critical manifold $M_s$. The dark purple curves show the maximum and minimum $v$ along two families of periodic orbits born at the subcritical Andronov-Hopf bifurcation (HB). The yellow circle and red diamond respectively denote the lower saddle-node (SN) bifurcation of $M_s$ and the homoclinic (HC) bifurcation in which the outer periodic orbit branch terminates. (B): The effect of $g_{\rm CAN}$ on the solution trajectory and bifurcation diagram. From right to left, $g_{\rm CAN}=0.14, 0.7, 1.6$. (C): Projection of bursting solutions from panel (B) onto the 2-parameter bifurcation diagram in the $(h, g_{\rm CAN_{Tot}})$-space, where $g_{\rm CAN_{Tot}}=g_{\rm CAN}f([\rm Ca])$ is given in \cref{eq:pBC_ICAN_activation}.  
        }
    \label{fig:NB_analysis}
\end{figure}

The $I_{\rm NaP}$-dependent N-bursting dynamics can be generated by the voltage subsystem $(v,n,h)$ without any calcium dynamics by setting $g_{\rm CAN}=0$. This type of bursting has been extensively studied using the fast-slow decomposition approach \cite{rinzel1987formal}, by treating $(v,n)$ as fast variables and $h$ as a slow variable \cite{PR,wang2016jcompneur,wang2020complex} (see our dimensional analysis in \cref{gspt}). In the parameter regime for N-bursting in model \cref{eq:pBC_Ica}, both $g_{\rm Ca}$ and $[\rm IP_3]$ are low (see \cref{fig:activity_patterns}(i)), keeping the calcium concentration $ca$ at a low level. Although voltage dynamics influence the calcium compartment through $I_{\rm Ca}$, leading to small oscillations in $ca$ around a low baseline, this effect remains negligible due to the low calcium levels. Similarly, the impact of $[\rm IP_3]$ on the full dynamics via its effect on $ca$ can also be neglected. As a result, $I_{\rm CAN}$, which is partially activated by $ca$, is functioning as a depolarizing leak current for $v$. In the following, we briefly review the bifurcation analysis for the N-burst and analyze the effects of NE on N-bursting dynamics by only considering the effect of increasing $g_{\rm CAN}$ on the voltage subsystem by examining its effect on the bifurcation diagram (see \cref{fig:NB_analysis}).

\Cref{fig:NB_analysis}A shows the one-parameter bifurcation of the $(v,n)$ subsystem with respect to $h$, together with the projection of the N-bursting solution (black trajectory) when $g_{\rm NaP}=2$, $g_{\rm CAN}=0.7$, $g_{\rm Ca}=0.00002$, $[\rm IP_3]=0.5$ onto the $(h,v)$-space. The bifurcation diagram consists of an S-shaped curve of equilibria (S) and a family of stable periodic orbits (PO). The PO branch is born at a subcritical Andronov-Hopf bifurcation (HB) and terminates at a homoclinic bifurcation (HC). The equilibria curve S, which represents the projection of the critical manifold $M_s$ onto the \RED{$(h,v)$}-space, folds at two saddle-node (SN) bifurcations or fold points $L_s$. The N-bursting solution is a square-wave burst that consists of a silent phase (i.e., interburst interval) along the lower branch of S and an active spiking phase along PO, initiating at SN and terminating at HC (\cref{fig:NB_analysis}A inset). 

\Cref{fig:NB_analysis}B shows the effect of increasing $g_{\rm CAN}$ on the bifurcation diagram from \cref{fig:NB_analysis}A. As $g_{\rm CAN}$ increases, S and PO shift toward lower $h$, while the $h$-nullcline remains unchanged. The corresponding bursting trajectories at various $g_{\rm CAN}$ values are also superimposed onto $(h,v)$-space. \Cref{fig:NB_analysis}C summarizes how the key bifurcation points SN (yellow curve) and HC (red curve) move to lower $h$ values with a continuous increase of $g_{\rm CAN}$ in the $(h, g_{\rm CAN_{Tot}})$-space, where $g_{\rm CAN_{Tot}}=g_{\rm CAN} f(ca)$ is a monotonically increasing function of $ca$ (see \cref{eq:pBC_ICAN_activation}). 

\begin{figure}[!t]
    \begin{center}
    \begin{tabular}{@{}p{0.45\linewidth}@{\quad}p{0.45\linewidth}@{}} 
    \subfigimg[width=0.9\linewidth]{\bfseries{{(i)}}}{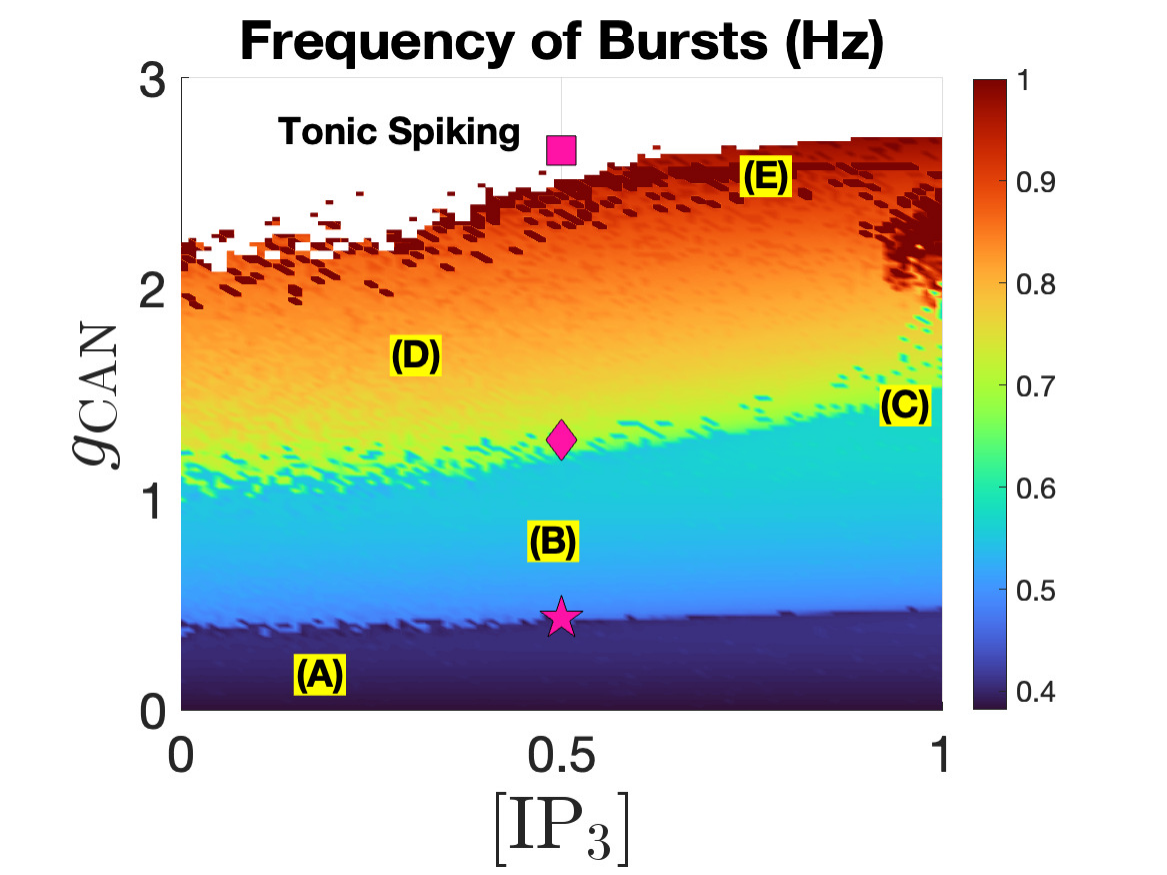} & 
    \subfigimg[width=0.9\linewidth]{\bfseries{{(ii)}}}{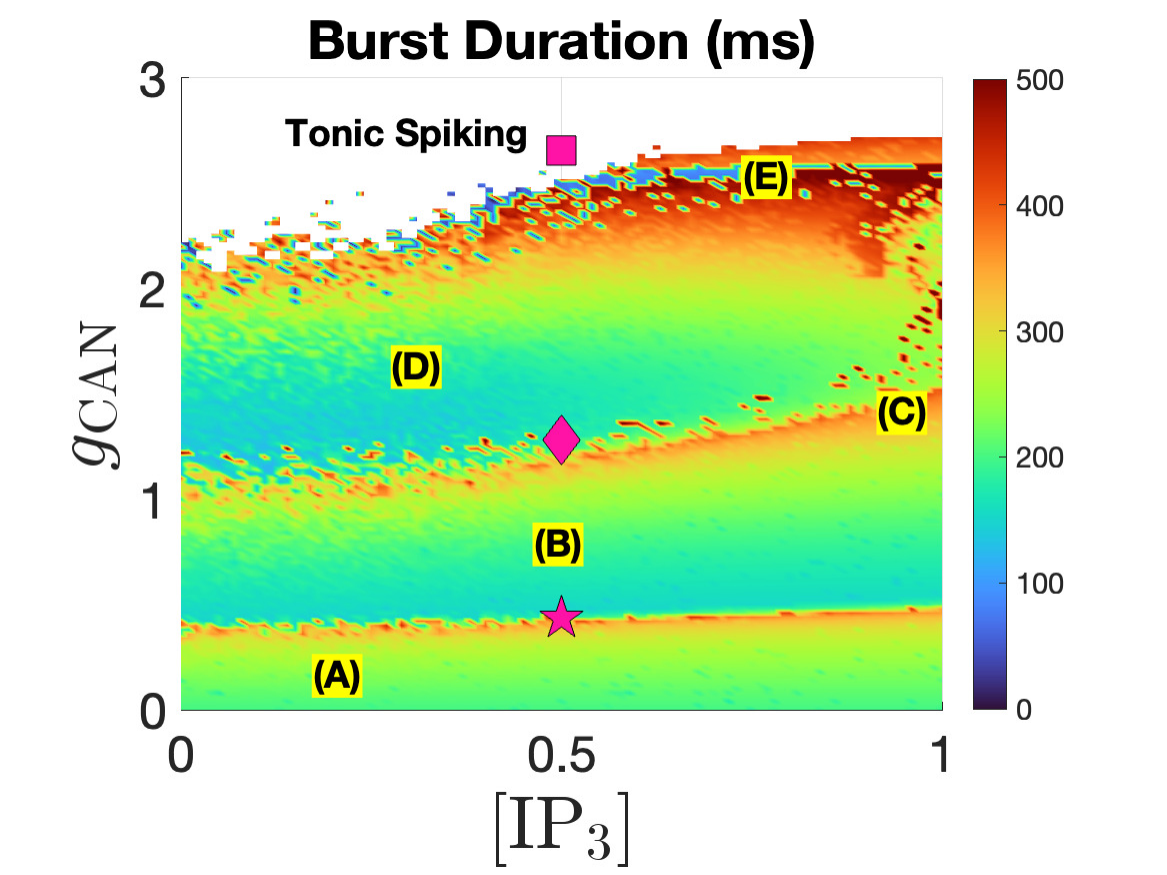} \\
    \multicolumn{2}{c}{\subfigimg[width=0.8\linewidth]{}{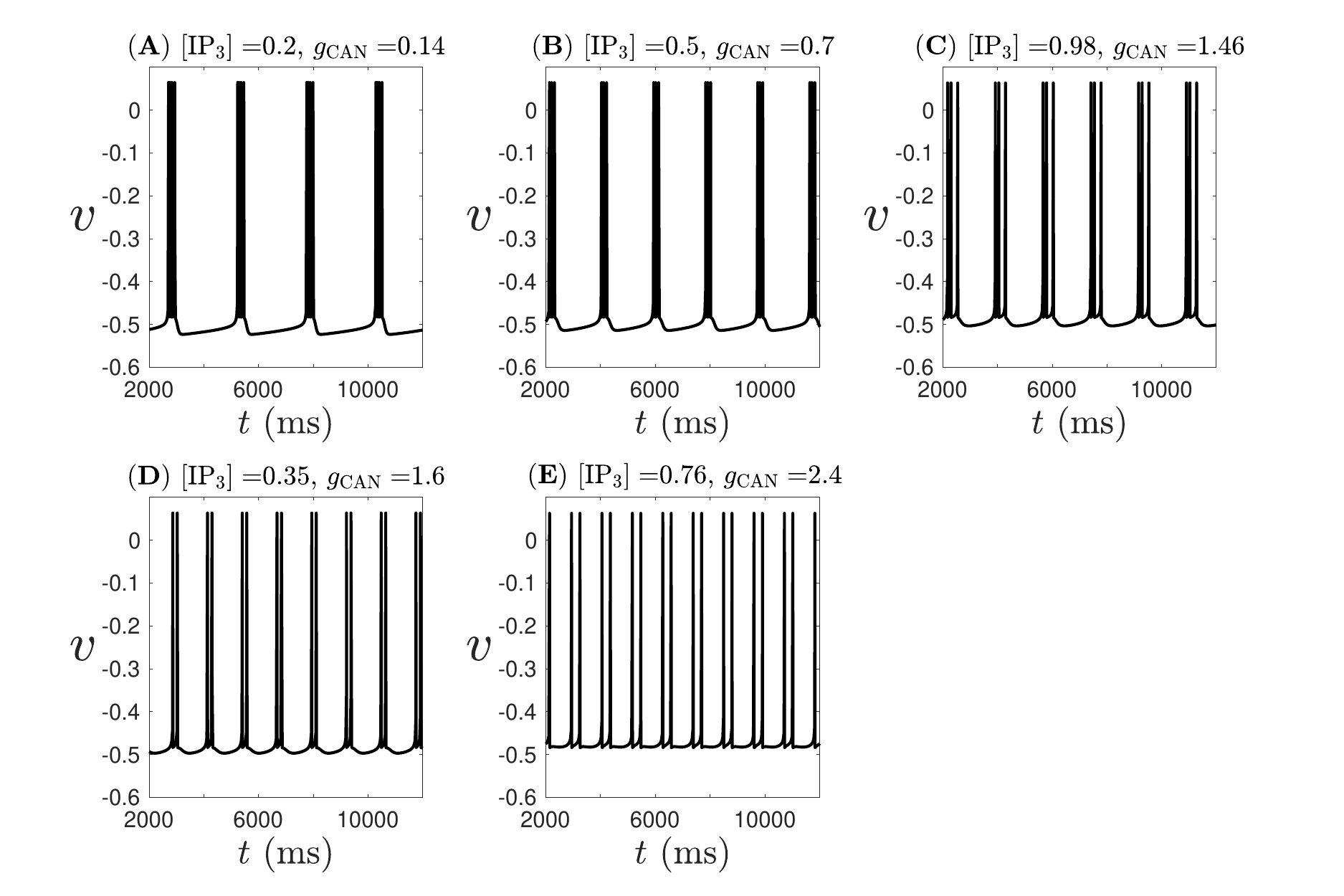}}
    \end{tabular}
    \end{center}
    \caption{Two-parameter diagrams showing the effect of $([{\rm IP_3}], g_{\rm CAN})$ on (i) N-burst frequency and (ii) N-burst duration for $g_{\rm NaP} = 2$ and $g_{\rm Ca} = 0.00002$. Pink symbols correspond to the $g_{\rm CAN}$ values at the saddle-node (SN) bifurcation points of the bursting branches in \cref{fig:NB_freq_analysis} for fixed $[{\rm IP_3}] = 0.5$. Sample voltage traces corresponding to labeled parameter values (A) through (E) are displayed below.}
    \label{fig:NB_gcan_ip3}
\end{figure}

We first analyze the effect of $g_{\rm CAN}$ on the burst frequency by examining its influence on the interburst interval - the time between the end of one burst and the start of the next - which primarily determines the burst period \cite{phillips2019biophysical}. For an N-bursting solution, this time interval can be approximated as 
\[(h_{\rm SN} - h_{\rm HC})/\langle \dot{h}_{sp} \rangle,\] 
where $\langle \dot{h}_{sp} \rangle$ denotes the average speed $h$ during the silent phase. As $g_{\rm CAN}$ increases, the lower fold of S shifts further away from the $h$-nullcline, leading to an increase in $\langle \dot{h}_{sp} \rangle$. In the meantime, $(h_{\rm SN} - h_{\rm HC})$ decreases with increasing $g_{\rm CAN}$, as the SN and HC curves move closer together and eventually merge in a saddle-node on an invariant circle (SNIC) bifurcation for $g_{\rm CAN}$ large enough. Together, the decrease in the numerator $(h_{\rm SN} - h_{\rm HC})$ and the increase in the denominator $\langle \dot{h}_{sp} \rangle$ shorten the interburst interval, ultimately resulting in a higher burst frequency as $g_{\rm CAN}$ increases. This trend is confirmed by our numerical simulations in \cref{fig:NB_gcan_ip3}(i), where $g_{\rm CAN}=0.14$, $0.7$ and $1.6$ correspond to the dark blue, light blue and orange regions. See also \cref{fig:NB_gcan_ip3}(A), (B) and (D) for representative voltage traces, despite different $[\rm IP_3]$ values.  

 Notably, the burst duration remains similar across these three regions. To understand this, we switch to examining the active spiking phase within the burst. During this phase, the trajectory oscillates between the maximum and minimum $v$ branches along the family of periodic orbits (PO). As $g_{\rm CAN}$ increases, the lower $v$ branch shifts to lower $h$ values while the $h$-nullcline remains static (see \cref{fig:NB_analysis}B). As a result, the trajectory transitions from oscillating entirely above the $h$-nullcline (e.g., $g_{\rm CAN}=0.14$) to a state where part of the trajectory during the active phase falls below the $h$-nullcline (e.g., $g_{\rm CAN}=1.6$). In other words, as $g_{\rm CAN}$ increases, a larger portion of the full system trajectory's projection in the $(h, v)$-space during the active phase lies close to or below the $h$-nullcline. Consequently, the average rate of change of $h$ during the active phase becomes less negative (i.e., the spiking frequency within the burst decreases). Meanwhile, as discussed above, the distance $(h_{\rm SN}-h_{\rm HC})$ decreases with increasing $g_{\rm CAN}$. These two effects counterbalance each other, maintaining a nearly constant burst duration, as confirmed by our numerical simulations in \cref{fig:NB_gcan_ip3}(ii), where the burst duration remains constant in most green regions. However, in certain regions, where the reduced $h$ distance can no longer fully offset the slower $h$ rate, the burst duration increases as indicated by the red regions (also compare \cref{fig:NB_gcan_ip3}(B) and (C)). 

The effect of $[{\rm IP_3}]$ on calcium dynamics has been analyzed previously in \cite{wang2016jcompneur} in the absence of $I_{\rm Ca}$. Specifically, increasing $[\rm IP_3]$ triggers greater release of stored $\rm Ca^{2+}$ from the ER into the cytoplasm, thereby raising intracellular calcium levels. This phenomenon largely holds in our model, despite the presence of $I_{\rm Ca}$ which remains small in the NB parameter regime. During the active phase, $\rm Ca^{2+}$ enters the cell through $I_{\rm Ca}$, contributing to neuronal depolarization via $I_{\rm CAN}$. However, during the interburst interval, calcium quickly returns back to its low baseline level by calcium pumps and thus has minimal impact on voltage dynamics compared to the effect of $g_{\rm CAN}$ (see \cref{fig:NB_gcan_ip3}). 


\begin{figure}[!ht]
    \begin{center}
    \begin{tabular}
    {@{}p{0.5\linewidth}@{}}
    \subfigimg[width=1\linewidth]{}{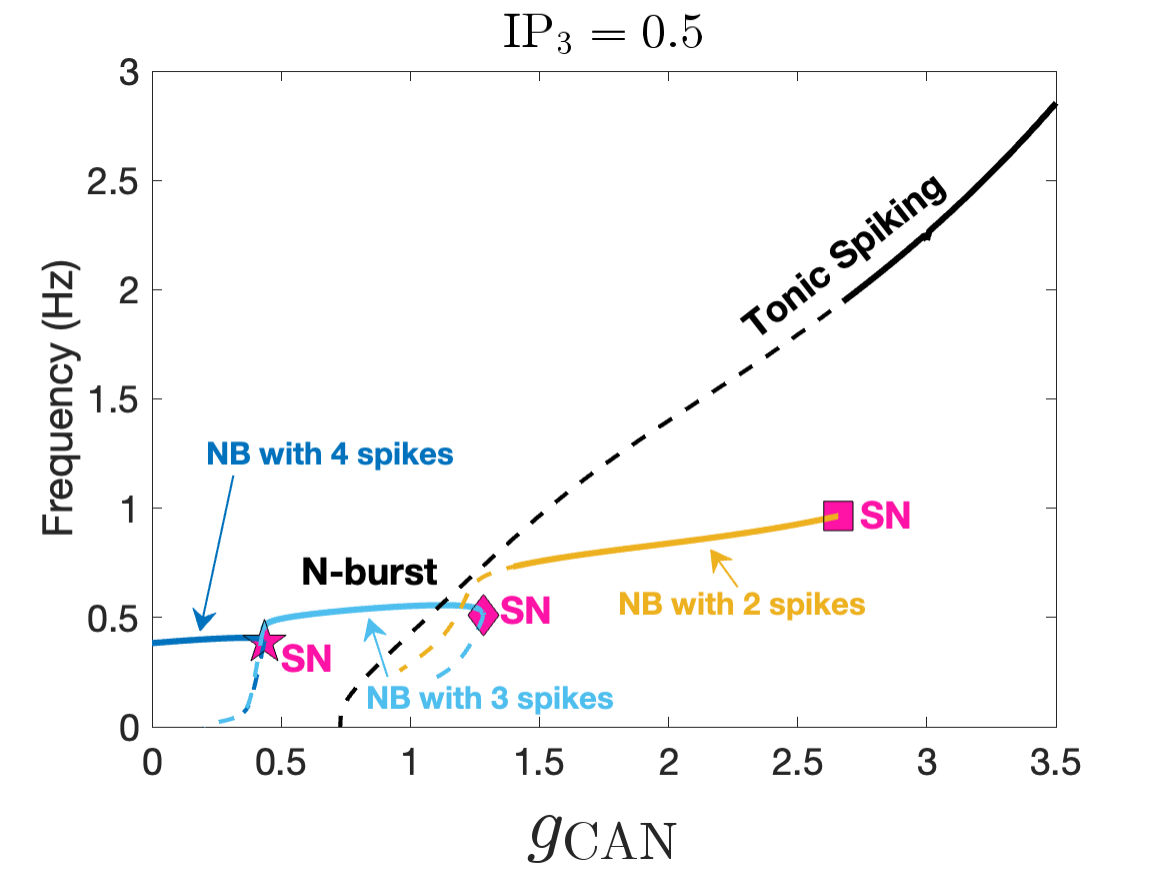}
    \end{tabular}
    \end{center}
        \caption{Bifurcation diagram of \cref{eq:pBC_ICa_slow}, showing the frequency of different N-burst solutions (colored curves) and the tonic spiking branch (black curve) as functions of $g_{\rm CAN}$. As $g_{\rm CAN}$ increases, the number of spikes per burst decreases from 4 (dark blue) to 3 (light blue), then to 2 (yellow), before eventually transitioning to tonic spiking. Solid lines denote stable branches; dashed lines denote unstable branches. 
        All parameter values are as in \cref{fig:NB_analysis}A.}
    \label{fig:NB_freq_analysis}
\end{figure}

Furthermore, our analysis suggests that the increase in burst frequency is neither continuous nor strictly monotonic but instead occurs in a discrete fashion, exhibiting a phasic pattern. Specifically, \cref{fig:NB_gcan_ip3}(i) shows that increasing $g_{\rm CAN}$ has minimal effects on burst frequency over a relatively broad parameter range, but can lead to a sudden, significant increase in burst frequency upon crossing a threshold. Similarly, discrete changes in burst duration are observed across the same threshold curves (\cref{fig:NB_gcan_ip3}(ii)). To better understand these thresholds, we analyze bifurcations of the full system \cref{eq:pBC_ICa_slow} with respect to $g_{\rm CAN}$ at fixed $[\rm IP_3]=0.5$. \Cref{fig:NB_freq_analysis} shows the effect of $g_{\rm CAN}$ on the frequency of different bursting branches (dark blue, light blue and yellow curves — color-coded to match the burst frequency values in \cref{fig:NB_gcan_ip3}(i)) and the tonic spiking branch (black). As $g_{\rm CAN}$ increases, the frequency of each N-burst branch remains nearly constant until its saddle-node (SN) bifurcation is reached. After crossing this point, the solution transitions to the next branch, resulting in a discrete jump in burst frequency. After crossing the SN of the yellow burst branch at the pink square, the solution jumps to the tonic spiking branch, leading to tonic spiking solutions for $g_{\rm CAN}$ large enough. These three SN bifurcation points are overlaid on the $(\mathrm{[IP_3]}, g_{\rm CAN})$-space in \cref{fig:NB_gcan_ip3} and align with the three threshold curves, suggesting that the discrete changes in the N-burst frequency are driven by the saddle-node bifurcations of the N-burst solution branches.  
Another interesting observation in \cref{fig:NB_freq_analysis} is the loss of a spike within each burst as the solution crosses each SN bifurcation threshold with increasing $g_{\rm CAN}$ (see also \cref{fig:NB_analysis}B). This is because, as $g_{\rm CAN}$ increases, the burst duration remains almost constant while the intra-burst spike frequency decreases. As a result, the system can no longer sustain the original number of spikes per burst, leading to a gradual loss of spikes as SN points are crossed. 

To summarize, our analysis and simulations of \cref{eq:pBC_ICa_slow} predict that increasing $g_{\rm CAN}$ increases the burst frequency of N-bursting solutions through a sequence of SN bifurcations, while having minimal impact on burst duration. Additionally, $[\rm IP_3]$ has little effect on both burst frequency and duration. The agreement with experimental data described before in \cref{modelNE} supports our proposed mechanism for NE, which involves increasing $g_{\rm CAN}$ and $[\rm IP_3]$. Although it seems that ${\rm IP_3}$ is not a prerequisite for modeling the effect of NE on N-bursts, we show later that it is critical to capture the effects of NE on other cell types of the \pbc{}.

\subsection{Effect of NE on C-bursting neurons} \label{CB}

\begin{figure}[!htp]
\begin{center}
\begin{tabular}
{@{}p{0.3\linewidth}@{\quad}p{0.3\linewidth}@{\quad}p{0.3\linewidth}@{}}
\subfigimg[width=1.1\linewidth]{\bfseries{\footnotesize{(A)}}}{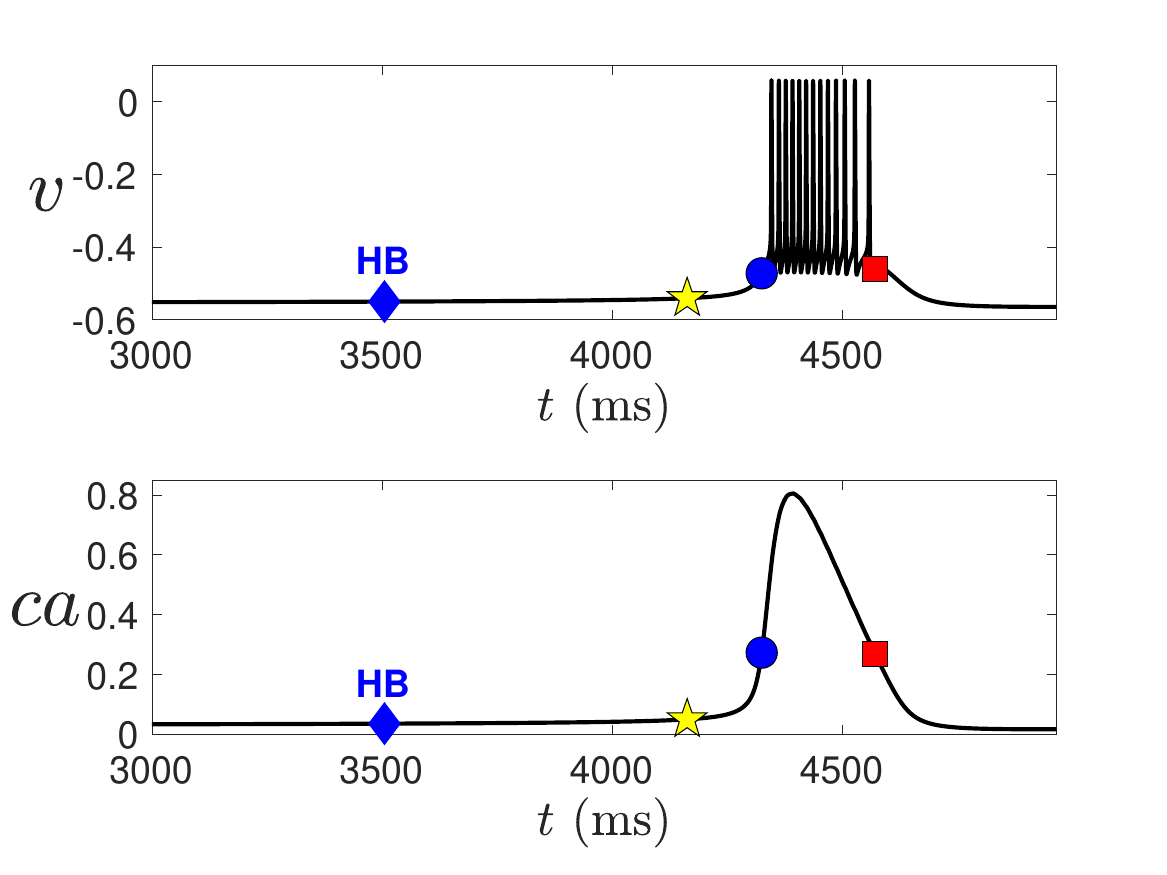} & \subfigimg[width=1.1\linewidth]{\bfseries{\footnotesize{(B)}}}{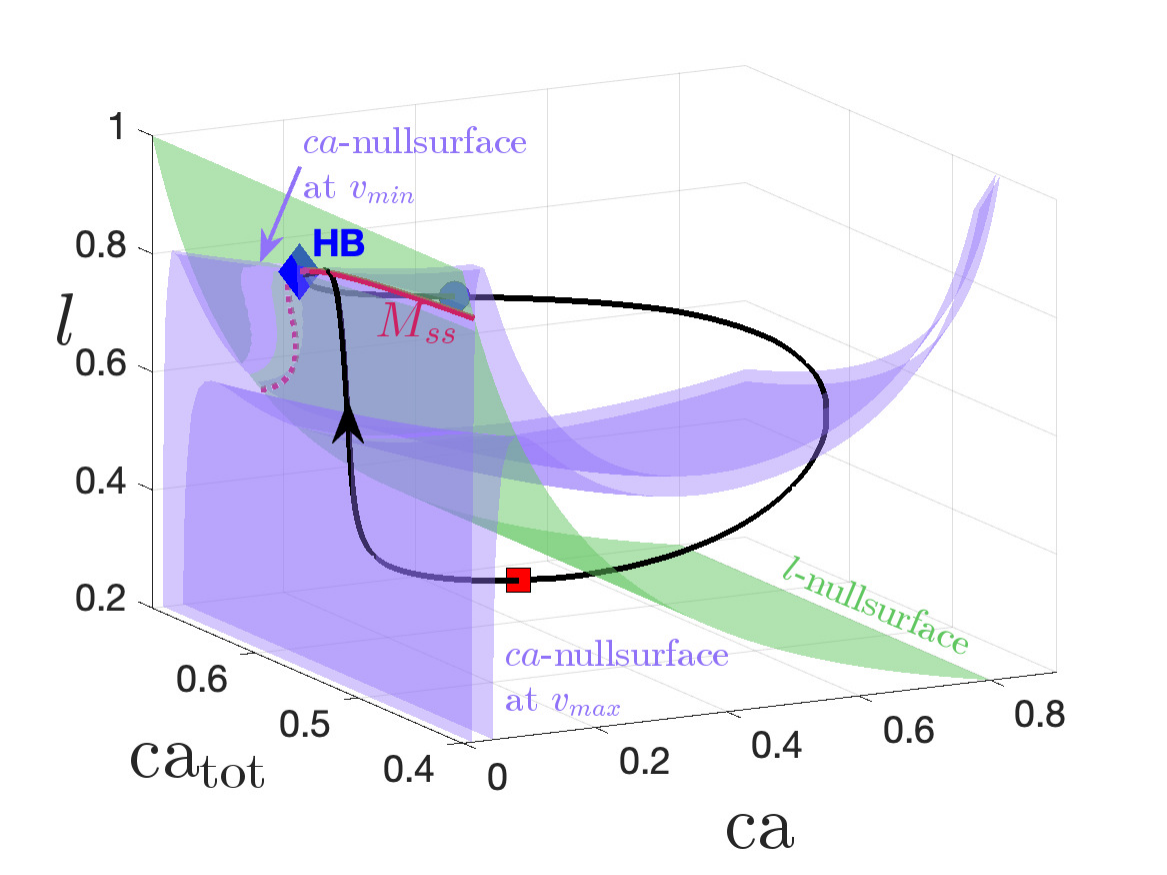} &
\subfigimg[width=1.1\linewidth]{\bfseries{\footnotesize{(C)}}}{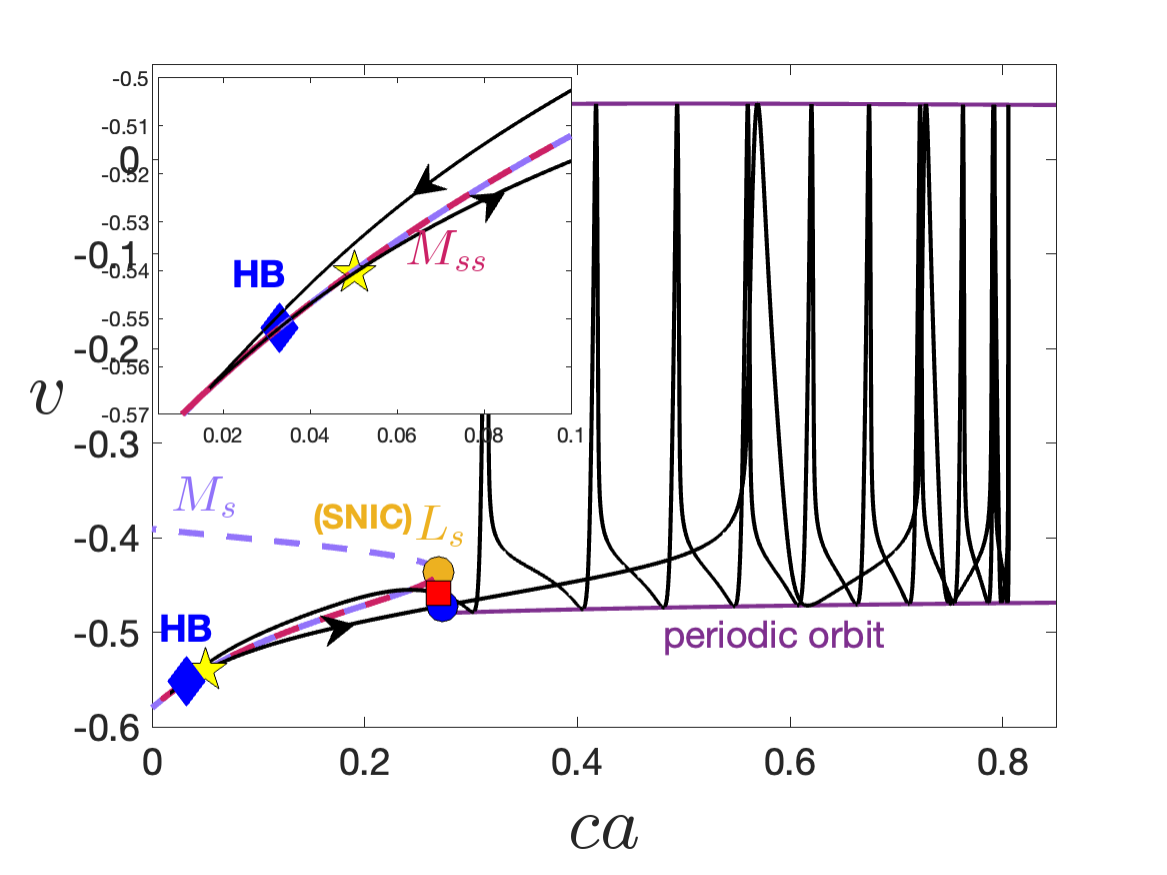}
\end{tabular}
\end{center}
\caption{Simulation of one cycle of the C-bursting solution generated by \cref{eq:pBC_ICa_slow} for $g_{\rm NaP} = 0$, $g_{\rm CAN} = 0.7$, $g_{\rm Ca} = 0.0005$ and $[\rm IP_3]=0.5$, and other parameters at default values. The blue diamond denotes the HB point. The yellow star denotes the jump up point of $ca$, defined as crossing the threshold of $ca = 0.05$ from below, whereas the blue circle and red square mark the initiation and termination of the burst. 
(A) Temporal evolution of $v$ and $ca$.
(B) \RED{ Projections of $ca$-nullsurfaces (denoted by blue surfaces)} with $v$ at its minimum and maximum in the $(ca, ca_{tot},l)$-space, together with the black solution trajectory from panel (A), the superslow manifold $M_{ss}$ (red curve). The green surface denotes the $l$-nullsurface.
 \RED{(C) Projections of the critical manifold $M_s$ (blue curve; solid for stable and dashed for unstable) and $M_{ss}$ (red curve from panel (B)) in the $(ca, v)$-space, along with the black solution trajectory from panels (A) and (B). The yellow circle denotes the SNIC bifurcation where the fast subsystem homoclinic bifurcation occurs at the fold $L_s$ of $M_s$. The purple curve shows the maximum and minimum $v$ along the family of periodics of the fast layer problem. }}
\label{fig:CB_burst_mechanism}
\end{figure}

\begin{figure}[!htp]
    \begin{center}
    \begin{tabular}    
{@{}p{0.45\linewidth}@{\quad}p{0.45\linewidth}@{}}
    \subfigimg[width=\linewidth]{\bfseries{\small{(A)}}}{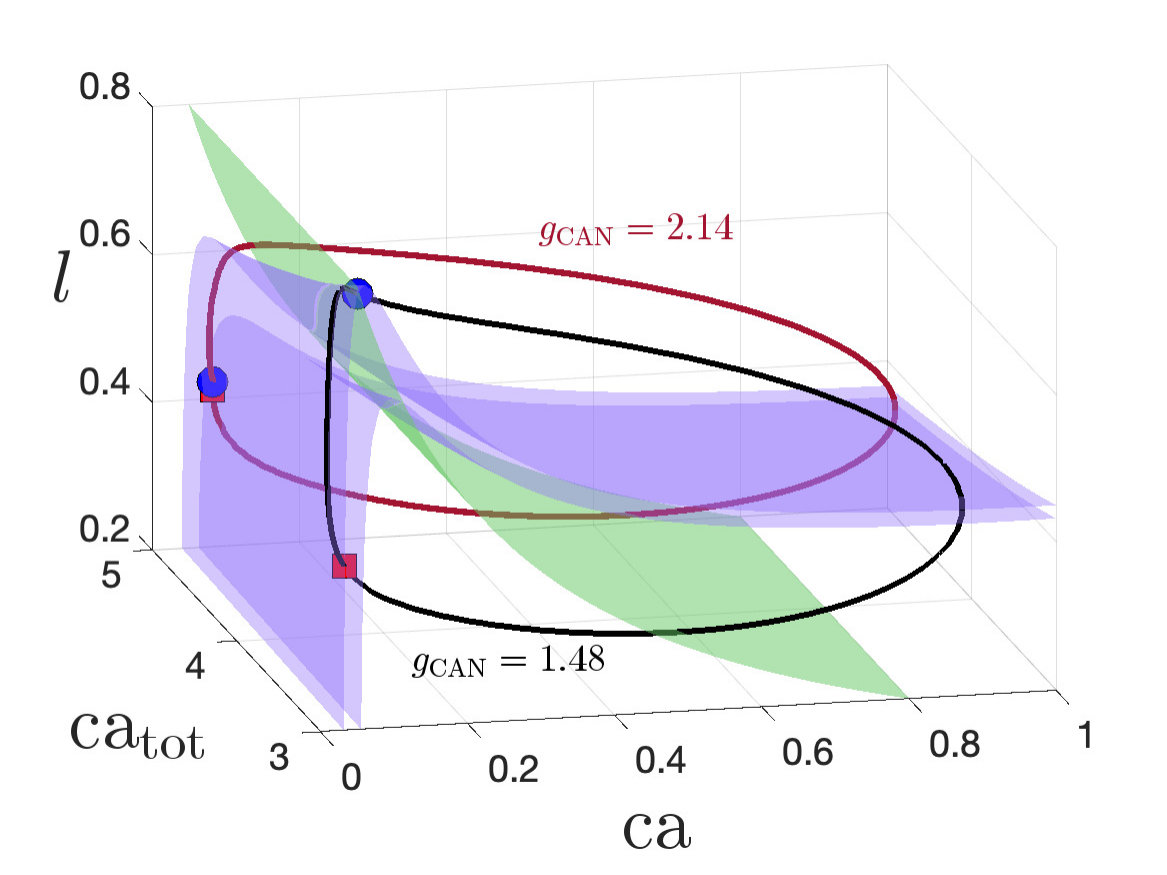} & \subfigimg[width=\linewidth]{\bfseries{\small{(B)}}}{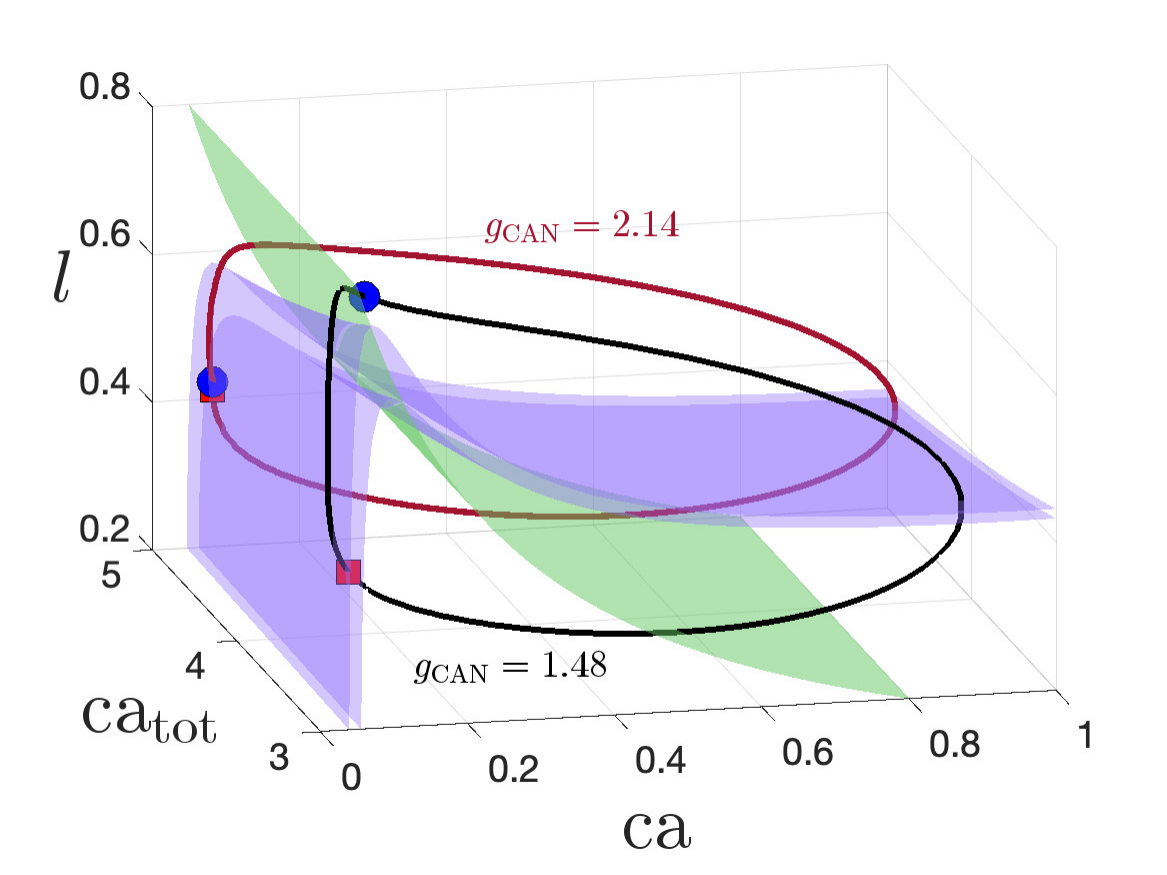} 
    \end{tabular}
    \end{center}
    \caption{$ca$-nullsurfaces (blue surface) and $l$-nullsurface (green surface) for $[\rm IP_3]=0.2$ and (A) $g_{\rm CAN}=1.48$ and (B) $g_{\rm CAN}=2.14$, projected onto $(ca, ca_{tot},l)$-space.     
    Solution trajectories of \cref{eq:pBC_ICa_slow} with $g_{\rm CAN}=1.48$ and $g_{\rm CAN}=2.14$ are shown in both panels by the black and red curves, respectively. Other parameters, color coding and symbols are the same as \cref{fig:CB_burst_mechanism}B.
 }
    \label{fig:CB_analysis_gcan}
\end{figure}

\begin{figure}[!ht]
    \begin{center}
    \begin{tabular}    
    {@{}p{0.45\linewidth}@{\quad}p{0.45\linewidth}@{}}
    \subfigimg[width=\linewidth]{\bfseries{\small{(A)}}}{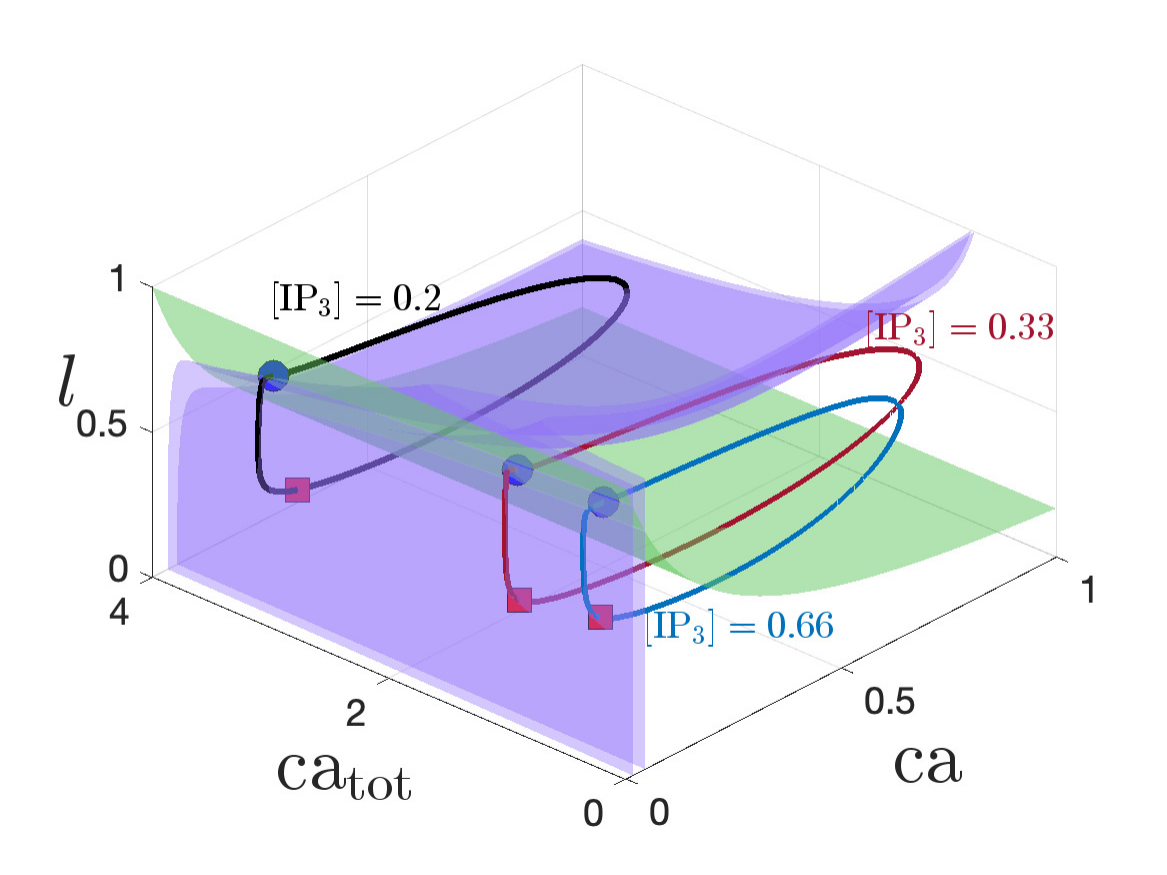} & \subfigimg[width=\linewidth]{\bfseries{\small{(B)}}}{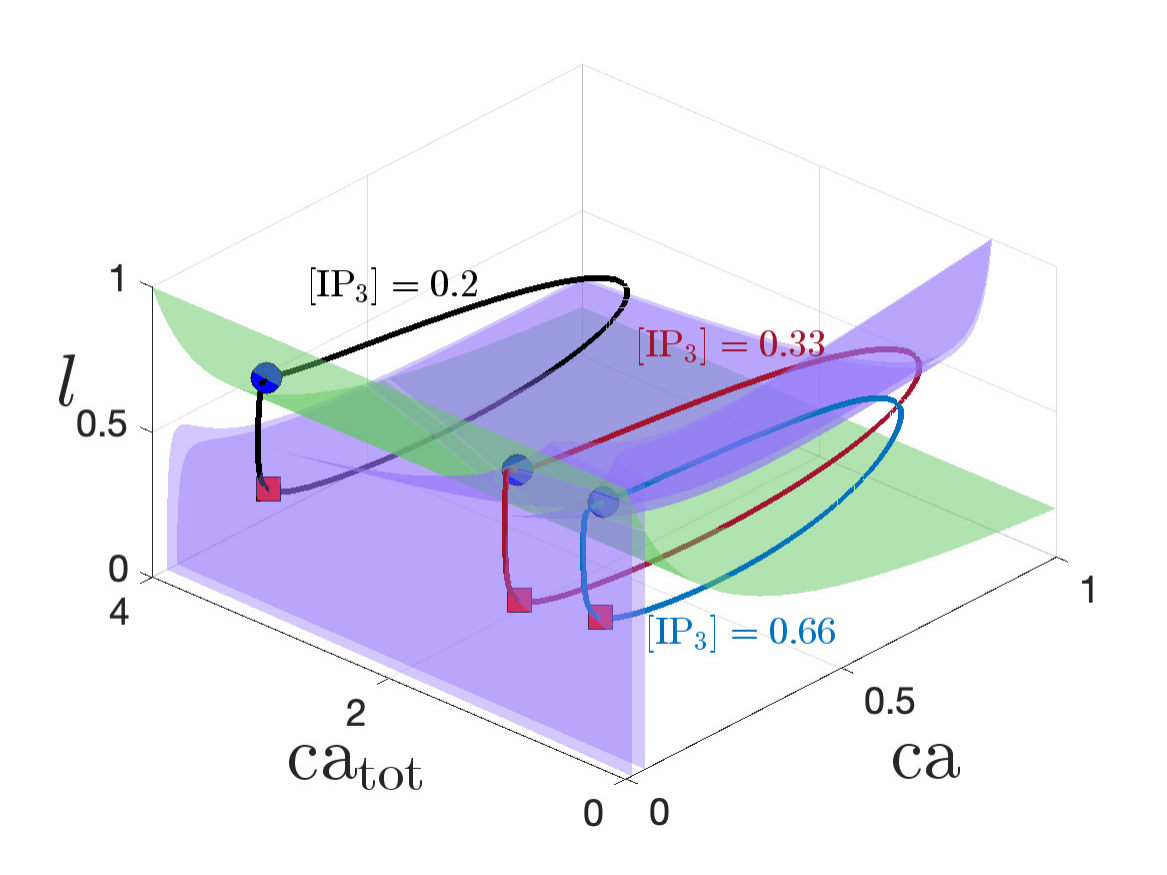} \\
    \subfigimg[width=\linewidth]{\bfseries{\small{(C)}}}{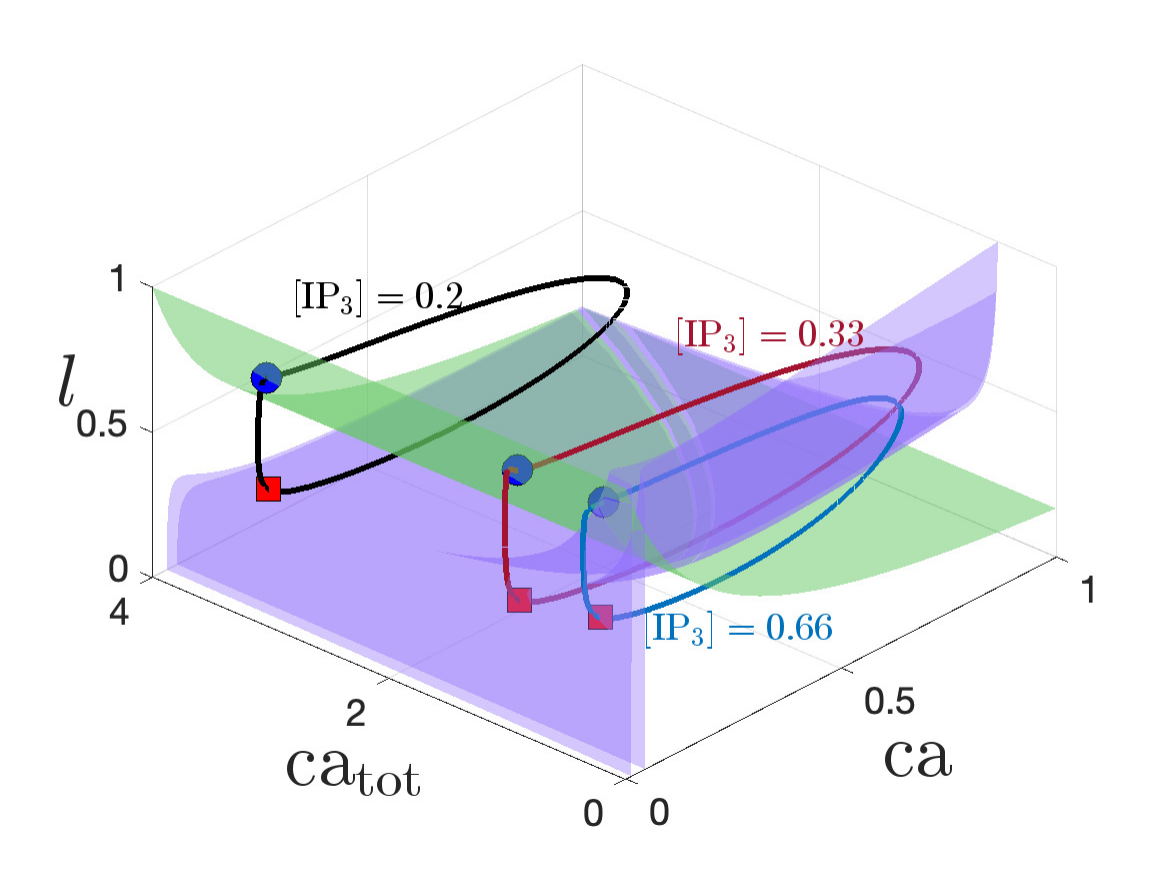} & \subfigimg[width=\linewidth]{\bfseries{\small{(D)}}}{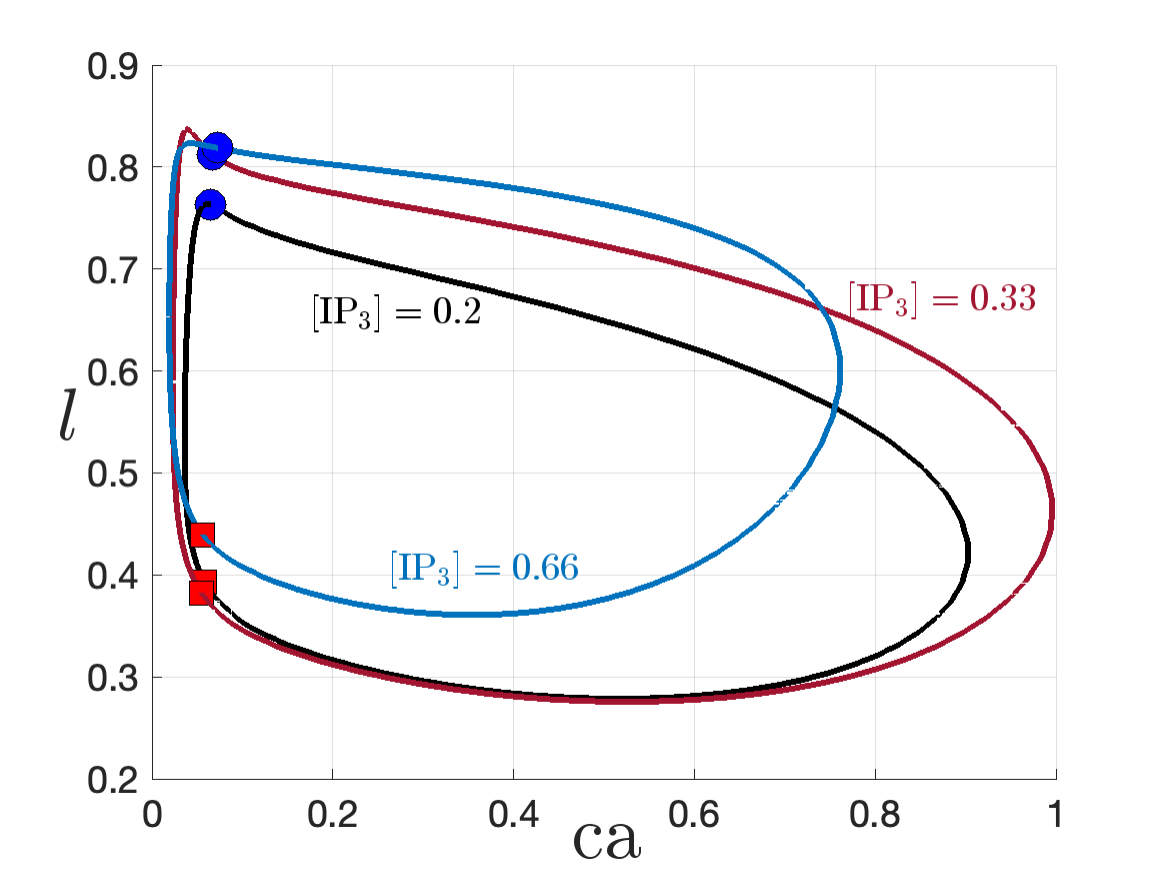}
    \end{tabular}
    \end{center}
    \caption{$ca$-nullsurfaces and $l$-nullsurface for $g_{\rm CAN}=1.48$ and (A) $[{\rm IP_3}] = 0.2$, (B) $[{\rm IP_3}] = 0.33$ and (C) $[{\rm IP_3}] = 0.66$, projected onto $(ca, ca_{tot},l)$-space, and (D) onto $(ca,l)$-space.  
    Solution trajectories of \cref{eq:pBC_ICa_slow} with $[{\rm IP_3}] = 0.2$, $[{\rm IP_3}] = 0.33$ and $[{\rm IP_3}] = 0.66$ are shown in all panels by the black, red and blue curves, respectively, in all panels. Other parameters, color coding and symbols are the same as in  \cref{fig:CB_burst_mechanism}B.}
    \label{fig:CB_analysis_ip3}
\end{figure}

When $g_{\rm NaP} = 0$, $g_{\rm CAN} = 0.7$, $g_{\rm Ca} = 0.0005$ and $[\rm IP_3]=0.5$, model \cref{eq:pBC_ICa_slow} produces C-bursting dynamics (see \cref{fig:activity_patterns}(i), lower right red region; also \RED{shown} in \cref{fig:CB_burst_mechanism}A). 
\RED{Below, we first apply GSPT analysis (see \cref{gspt}) to elucidate the underlying mechanism for the C-burst. We analyze the calcium subsystem $(ca, ca_{tot},l)$ using the projections of relevant nullsurfaces and the superslow manifold $M_{ss}$ onto the calcium phase space (\cref{fig:CB_burst_mechanism}B) and then examine its interaction with the voltage dynamics using the bifurcation diagram in which the intermediate variable $ca$ is treated as a bifurcation parameter (\cref{fig:CB_burst_mechanism}C).} We finally investigate the effect of $g_{\rm CAN}$ and $[\rm IP_3]$ on the C-bursting dynamics by considering their influence on the geometric structures of the calcium subsystem (see \cref{fig:CB_analysis_gcan,fig:CB_analysis_ip3}).  


\RED{Past studies \cite{rubin2009calcium,PR,wang2016jcompneur} have shown that C-bursting is driven by $ca$-oscillations. These oscillations begin at the yellow star (defined as crossing $ca = 0.05$ from below) and precede the burst onset at the blue circle (see \cref{fig:CB_burst_mechanism}A).} Since $g_{\rm NaP}$ remains small in the C-bursting parameter regime, the bifurcation structure of the fast voltage subsystem discussed in the previous section is no longer the primary determinant of the dynamics. \RED{Instead, following \cite{wang2017timescales}, we analyze the calcium subsystem by 
examining the full-system trajectory and $M_{ss}$ in $(ca,ca_{tot},l)$-space, together with the $ca$-nullsurface evaluated at the minimum and maximum values of $v$ to represent the influence of the fast voltage dynamics (\cref{fig:CB_burst_mechanism}B). A projection of the fast subsystem bifurcation structure onto the $(ca,v)$-plane then reveals how $ca$-oscillations drive the full-system C-bursting dynamics via $I_{\rm CAN}$ activation (\cref{fig:CB_burst_mechanism}C).} 

\RED{The $ca$-nullsurfaces, each defined for $v$ fixed, are shown as the two blue surfaces in \cref{fig:CB_burst_mechanism}B.} Also shown are the projections of the solution trajectory (black), the superslow manifold $M_{ss}$ (red), and the $l$-nullsurface (green). After the burst ends at the red square in \cref{fig:CB_burst_mechanism}B, the trajectory jumps to the left branch of the $ca$-nullsurface at $v_{min}$, entering the silent phase. It then evolves on the slow timescale under the slow reduced layer problem, \RED{where $ca$ and $l$ evolve slowly (with $ca$ slaved to $l$) while the superslow variable $ca_{tot}$ remains approximately constant}, until reaching the stable portion of the superslow manifold $M_{ss}$ (red curve). At this point, the dynamics transitions to the superslow timescale under the superslow reduced problem \cref{eq:pBC_ICa_superslowreduced}. The trajectory closely follows the attracting side of $M_{ss}$, passes over the HB (blue diamond) to the repelling side, and undergoes a delay in which it traces the repelling branch before making a jump at the yellow star to large $ca$ (see \cref{fig:CB_burst_mechanism}A for the delay between blue diamond and yellow star). \RED{Afterward, the trajectory moves near the right branch of the $v$-dependent family of $ca$-nullsurfaces until passing the curve of lower folds and jumping back to the left branch. This completes a full calcium oscillation cycle.} \RED{We make two remarks. First, the three-timescale structure of the calcium subsystem, as revealed by our dimensional analysis in Appendix \hyperlink{appB}{B}, is visible in the projection shown in \cref{fig:CB_burst_mechanism}B: the trajectory exhibits $ca$ jumps, slow drift along the $ca$-nullsurface, and superslow drift along $M_{ss}$. Second, the trajectory only loosely follows the right branch of the $ca$-nullsurface, reflecting the fact that the timescale separation between $ca$ and $l$ is moderate and therefore relatively far from the singular limit. } 

\RED{To illustrate how the $ca$ oscillation drives the onset and termination of bursting in $v$ at the blue circle and red square, respectively (\cref{fig:CB_burst_mechanism}A), we now turn to the bifurcation structure of the fast layer problem $(v,n)$, projected onto $(ca, v)$-space, in \cref{fig:CB_burst_mechanism}C. 
In this projection, $M_{ss}$ (red curve) overlaps with the critical manifold $M_s$ (blue curve), which undergoes a homoclinic bifurcation at its fold $L_s$ (also known as a saddle-node on invariant circle (SNIC) bifurcation \cite{shilnikov2001, ermentrout2010mathematical}). After this SNIC bifurcation, a family of stable periodic orbits emerges. When the trajectory jumps to larger $ca$ values at the yellow star, it crosses the SNIC at the blue circle, thereby initiating the fast spiking phase of the C-burst. When $ca$ jumps back to small values, the trajectory in $(ca, v)$-space crosses the SNIC again at the red square, terminating the burst and returning the system to the silent phase, completing the full cycle.} 

\begin{figure}[!t]
    \begin{center}
    \begin{tabular}{@{}p{0.45\linewidth}@{\quad}p{0.45\linewidth}@{}} 
    \subfigimg[width=0.9\linewidth]{\bfseries{{(i)}}}{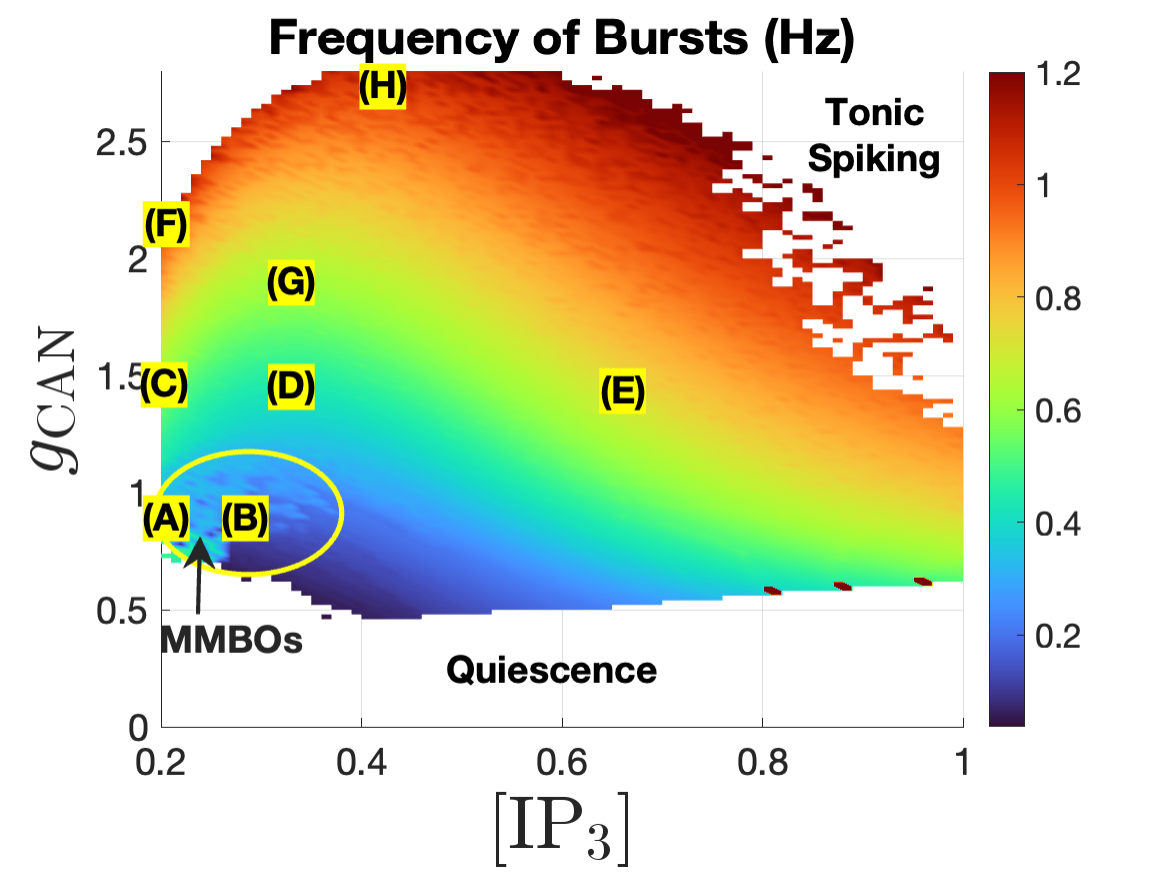} & 
    \subfigimg[width=0.9\linewidth]{\bfseries{{(ii)}}}{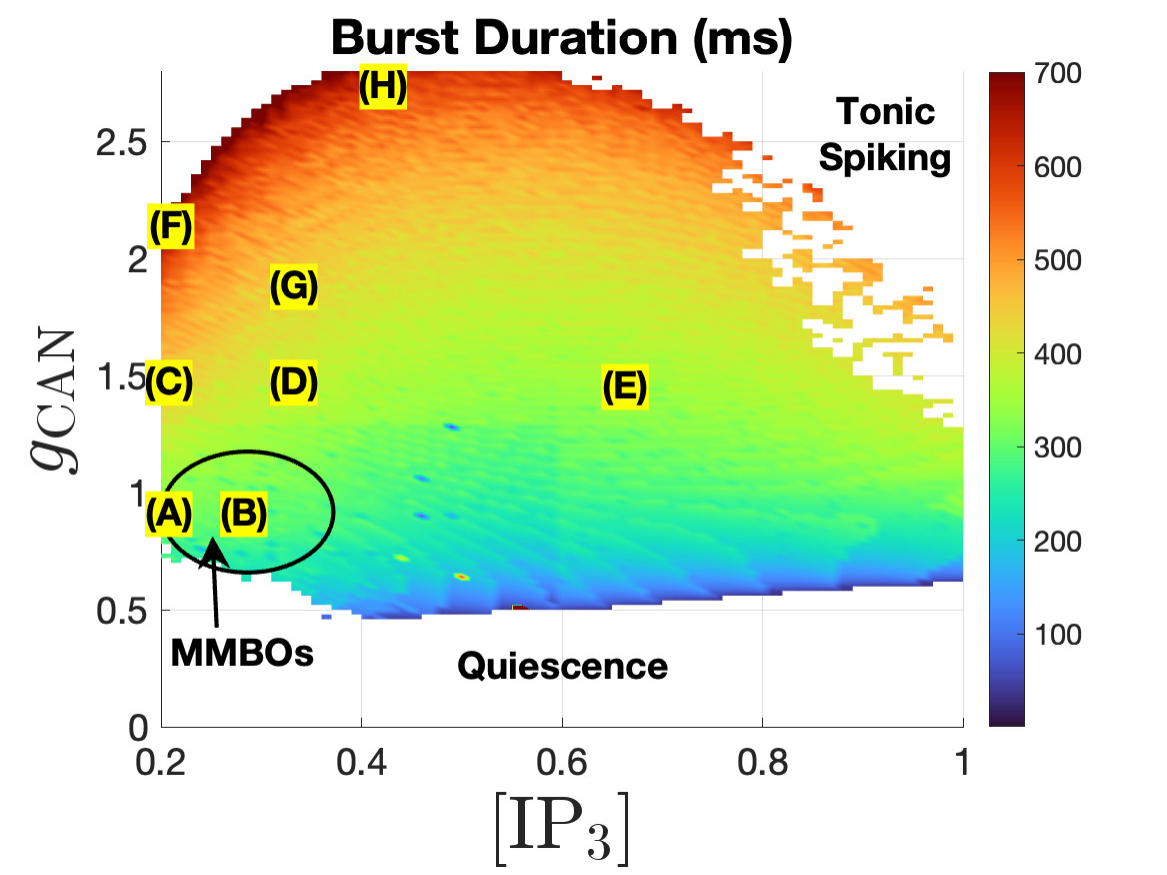} \\
    \multicolumn{2}{c}{\subfigimg[width=0.8\linewidth]{}{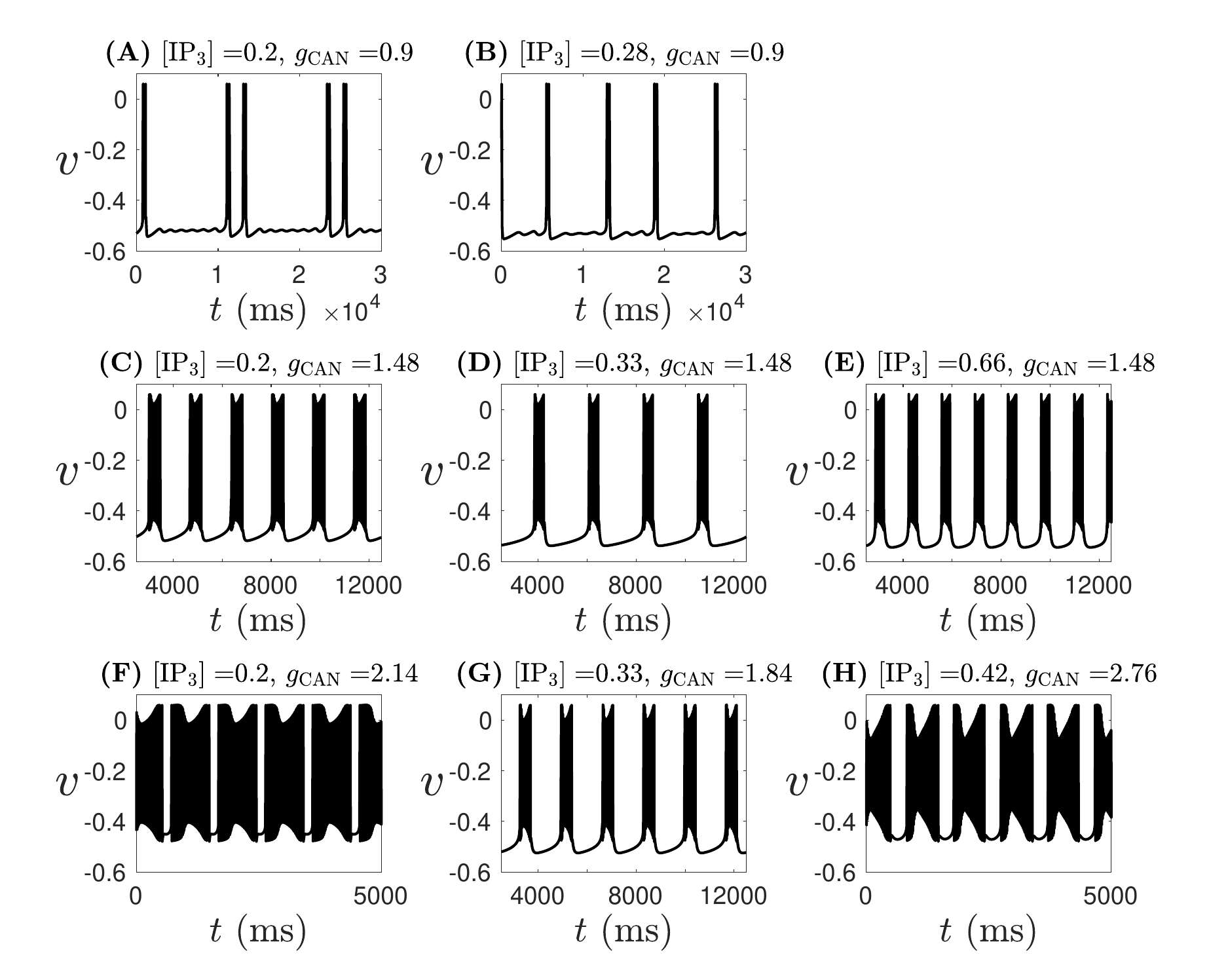}}
    \end{tabular}
    \end{center}
    \caption{Two-parameter diagrams showing the effects of $([{\rm IP_3}], g_{\rm CAN})$ on (i) C-burst frequency and (ii) C-burst duration. Sample voltage traces corresponding to labeled parameter values (A) through (H) are displayed below. 
    }
    \label{fig:CB_gcan_ip3}
\end{figure}

Next we investigate how increasing $g_{\rm CAN}$ and $[{\rm IP_3}]$ affects the C-burst frequency and duration (see \cref{fig:CB_gcan_ip3}(i) and (ii)) by examining the projections of solution trajectories onto $(ca, ca_{tot},l)$ (see \cref{fig:CB_analysis_gcan,fig:CB_analysis_ip3}). \RED{For simplicity, only the burst initiation (blue circle) and burst termination (red square) points are shown.}

\RED{In \cref{fig:CB_analysis_gcan}, we analyze the effect of $g_{\rm CAN}$ on C-bursting dynamics by fixing $\rm [IP_3]$ at $0.2$, which more clearly reveals the differences in the underlying dynamics. The same effects also occur at other $[\rm IP_3]$ values, including the control value of $0.5$ used in \cref{fig:CB_burst_mechanism}, as demonstrated in  \cref{fig:CB_gcan_ip3}.}
Under fixed $[{\rm IP_3}]$, increasing $g_{\rm CAN}$ depolarizes the cell, which increases the total calcium concentration within the cell through $I_{\rm Ca}$. As a result, the solution for higher $g_{\rm CAN}$ lies at higher $ca_{tot}$ values than a solution for low $g_{\rm CAN}$ (see \cref{fig:CB_analysis_gcan}). The projections of $ca$-nullsurfaces for $g_{\rm CAN}=1.48$ and $2.14$ are shown respectively in panels (A) and (B). 
\RED{From our analysis, the burst period is primarily determined by the interburst interval, which corresponds to the time the trajectory spends along the left branch of the $ca$-nullsurface near $v_{min}$, between the burst termination point (red square) and the burst initiation point (blue circle) (see \cref{fig:CB_analysis_gcan}, upper blue surface).} In contrast, the burst duration is determined by the duration of the active phase when $ca$ is relatively large (\cref{fig:CB_analysis_gcan}, the right segment of the trajectory between blue circle and red square). Therefore, to examine the effect of NE on C-burst frequency and duration, we analyze different phases of the $ca$-oscillation.

Beginning at the red square in \cref{fig:CB_analysis_gcan}A, the trajectory for $g_{\rm CAN}=1.48$ (black curve) follows the left branch of the $ca$-nullsurface at $v_{min}$ (upper blue surface) until it reaches the fold, then undergoes a jump to the right branch, initiating burst onset of the black trajectory at the blue circle. 
As $g_{\rm CAN}$ increases, the trajectory shifts to higher $ca_{tot}$ and moves further away from the $l$-nullsurface (red curve). At the same time, the $ca$-nullsurface (blue surfaces in \cref{fig:CB_analysis_gcan}B) shifts slightly toward lower $l$-values. As a result, the rate of increase of $l$ during the silent phase becomes slightly higher and the red trajectory crosses the fold of the $ca$-nullsurface at a lower $l$-value, leading to faster ca-oscillations. 
\RED{Additionally}, a higher $g_{\rm CAN}$ lowers the $ca$-threshold required to trigger a $v$-spike in the voltage subsystem \RED{(compare the burst initiation points marked by blue circles for the two trajectories)}, which advances the onset of the C-burst and significantly shortens the silent phase duration. \RED{Together, these effects increase C-burst frequency. On the other hand, the trajectory with larger $g_{\rm CAN}$ remains in the active spiking phase for a longer portion of the cycle, because the $ca$-threshold required to sustain spiking is lower, resulting in a longer burst duration.} \RED{This} analysis is consistent with our numerical simulations, which show that increasing $g_{\rm CAN}$ for fixed $[\rm IP_3]$ consistently increases both burst frequency and duration (\cref{fig:CB_gcan_ip3}).


\Cref{fig:CB_analysis_ip3} shows the effect of increasing $[\rm IP_3]$ on the C-bursting dynamics for fixed $g_{\rm CAN}=1.48$. The projections of the $ca$-nullsurfaces for $[{\rm IP_3}] = 0.2$, $[{\rm IP_3}] = 0.33$ and $[{\rm IP_3}] = 0.66$ are shown respectively in panels (A), (B) and (C). The solution trajectories for all three $[\rm IP_3]$ values are shown in all panels. Comparing panels (A-C) indicates that increasing ${[\rm IP_3]}$ shifts the $ca$-nullsurface towards lower $l$-values while pushing the trajectory to lower $ca_{tot}$ values. This occurs because higher $[\rm IP_3]$ enhances calcium release from the ER into the cytosol, which in turn increases calcium extrusion out of the cell through plasma membrane $\rm Ca^{2+}$ pumps, reducing the total intracellular calcium.   

We first compare $[\rm IP_3]=0.2$ and $0.33$, with their trajectories also projected onto $(ca, l)$-space (\cref{fig:CB_analysis_ip3}D). This projection clearly shows that the C-burst for $[\rm IP_3]=0.33$ initiates at a higher $l$ value than for $[\rm IP_3]=0.2$, while both bursts terminate at similar $l$ values (red square). To understand why, we examine the projection onto $(ca, ca_{tot},l)$-space in panels (A) and (B). For both $[\rm IP_3]$ values, burst onset (blue circle) occurs approximately along the fold curve of the $ca$-nullsurface. While increasing $[\rm IP_3]$ shifts the $ca$-nullsurface to lower $l$ values, it also significantly lowers the trajectory's $ca_{tot}$, which in turn increases the $l$ value at the $ca$-nullsurface's fold. As a result, the jump up of $ca$ for $[\rm IP_3]=0.33$ occurs at a higher value of $l$ than for $[\rm IP_3]=0.2$ (see \cref{fig:CB_analysis_ip3}D, blue circles). Similarly, if the $ca$-nullsurface remained fixed while the trajectory alone shifted to lower $ca_{tot}$, one would expect a higher $l$ value at burst termination (red square) as the right branch of the $ca$-nullsurface moves to higher $l$ as $ca_{tot}$ decreases. Nonetheless, since the increased $[\rm IP_3]$ also shifts the surface to lower $l$, these two effects counteract each other, leading to similar $l$ values at burst termination. Hence, as a result of the evolution rate of $l$ remaining similar, the longer distance between the blue circle and red square along the left branch of $M_s$ at $[\rm IP_3]=0.33$ results in a longer interburst interval, thereby reducing burst frequency compared to $[\rm IP_3]=0.2$ (see \cref{fig:CB_gcan_ip3}C and D).  

A similar argument also holds when comparing $[{\rm IP_3}] = 0.33$ and $[{\rm IP_3}] = 0.66$. While increasing $[\rm IP_3]$ continues to push the left branch of $ca$-nullsurface to lower $l$ (compare \cref{fig:CB_analysis_ip3}B and C), the effect is less pronounced than in the previous comparison. As a result, the two opposing effects described above largely cancel out near the left branch, resulting in a similar $l$ value at burst onset (blue circles). However, the jump-back value of $l$ (red square) is higher for $[{\rm IP_3}] = 0.66$ than for $[{\rm IP_3}] = 0.33$, as the right branch is only minimally affected by the increased [${\rm IP_3}$] but shifts to higher $l$ with larger $ca_{tot}$ along the blue trajectory. 
Consequently, the C-burst for $[{\rm IP_3}] = 0.66$ has a higher burst frequency than for $[{\rm IP_3}] = 0.33$ (see \cref{fig:CB_gcan_ip3}D and E). This suggests a non-monotonic effect of $[{\rm IP_3}]$ on the C-burst frequency, where an initial increase in $[\rm IP_3]$ slows down burst frequency, but a further increase accelerates it (see \cref{fig:CB_gcan_ip3}(i)). \RED{Moreover, this phenomenon arises from the differential effects of $[\rm IP_3]$ on the $l$-values at burst initiation and termination (see \cref{fig:CB_analysis_ip3}D). To further support this, we simulated the full model over a broader range of $\rm [IP_3]\in [0.2, 0.94]$ and examined the $l$ values at the beginning and end of each burst (see \cref{fig:CB_lval_SP} in Appendix \hyperlink{appD}{D}). Consistent with the analysis above, the $l$ values at burst initiation exhibits a non-monotonic dependence on $[\rm IP_3]$, first increasing and then decreasing, whereas the $l$ values at burst termination remain nearly constant at lower $[\rm IP_3]$ before increasing at higher $[\rm IP_3]$.} 
We omit a detailed analysis of the effect of $[{\rm IP_3}]$ on burst duration, which requires a closer examination of the trajectory along the right branch, its shape variations, and the average rate of $ca$ during the burst. It is worth noting that the effect of $[\rm IP_3]$ on the size of the $(ca,l)$-projection of the trajectory during the burst is also non-monotonic (see \cref{fig:CB_analysis_ip3}D). 

Having analyzed the effects of $g_{\rm CAN}$ and $\rm [IP_3]$ separately, we now examine their combined influence. \cref{fig:CB_gcan_ip3} suggests that increasing both $g_{\rm CAN}$ and $[{\rm IP_3}]$ generally leads to an increase in the C-burst frequency. However, as previously analyzed, while increasing $g_{\rm CAN}$ consistently raises burst frequency, an increase of $[\rm IP_3]$ from a relatively low level initially reduces it. 
Thus, in theory, there should exist a parameter region where these two opposing effects counterbalance, resulting in a constant burst frequency, as observed in experiments.
Indeed, \cref{fig:CB_gcan_ip3}C and G show that the C-burst frequency remains nearly unchanged despite a $20\%$ increase in $g_{\rm CAN}$, due to the opposing effect of simultaneously increasing $[{\rm IP_3}]$. A similar example is seen between \cref{fig:CB_gcan_ip3}F and H. 
Similarly, as a result of the opposing effects of $g_{\rm CAN}$ and $[{\rm IP_3}]$ on the burst duration, as seen in \cref{fig:CB_gcan_ip3}(ii), the combined influence leads to one of the two outcomes: (1) the increase in C-burst duration caused by $g_{\rm CAN}$ is counterbalanced by the decrease induced by $[{\rm IP_3}]$, resulting in little change or a slight decrease in duration (see \cref{fig:CB_gcan_ip3}C-G for example); or (2) C-burst duration increases (see \cref{fig:CB_gcan_ip3}G-H, note the different time scale). 

Finally, we note that model \cref{eq:pBC_ICa_slow} exhibits \emph{mixed-mode bursting oscillations} (MMBOs) in a small region of parameter space (\cref{fig:CB_gcan_ip3}(i), yellow circle; see also, \cref{fig:CB_gcan_ip3}A,B). These oscillations exhibit complex dynamics characterized by alternating large-amplitude bursts and small-amplitude oscillations \cite{Desroches2013, Desroches2022}. While not central to our present modeling study, the emergence of MMBOs highlights the richness of the model's dynamics, and investigating their underlying mechanism represents an interesting direction for future work. 



\subsection{NE-induced Bursting in a Tonic Spiking neuron} \label{TS}


\begin{figure}[!t]
    \begin{center}
    \begin{tabular}
    {@{}p{0.45\linewidth}@{\quad}p{0.45\linewidth}@{}}
    \subfigimg[width=0.9\linewidth]{\bfseries{\small{(i)}}}{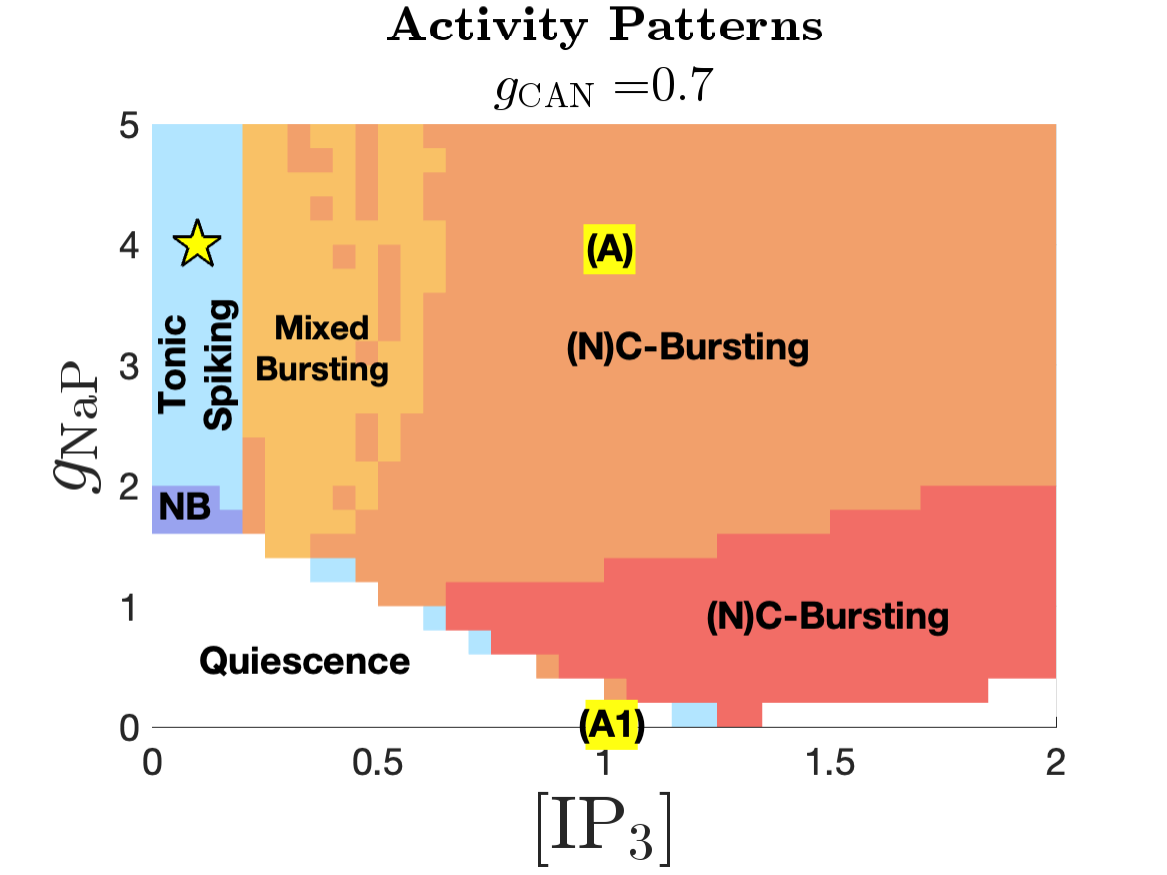} &
    \subfigimg[width=0.9\linewidth]{\bfseries{\small{(ii)}}}{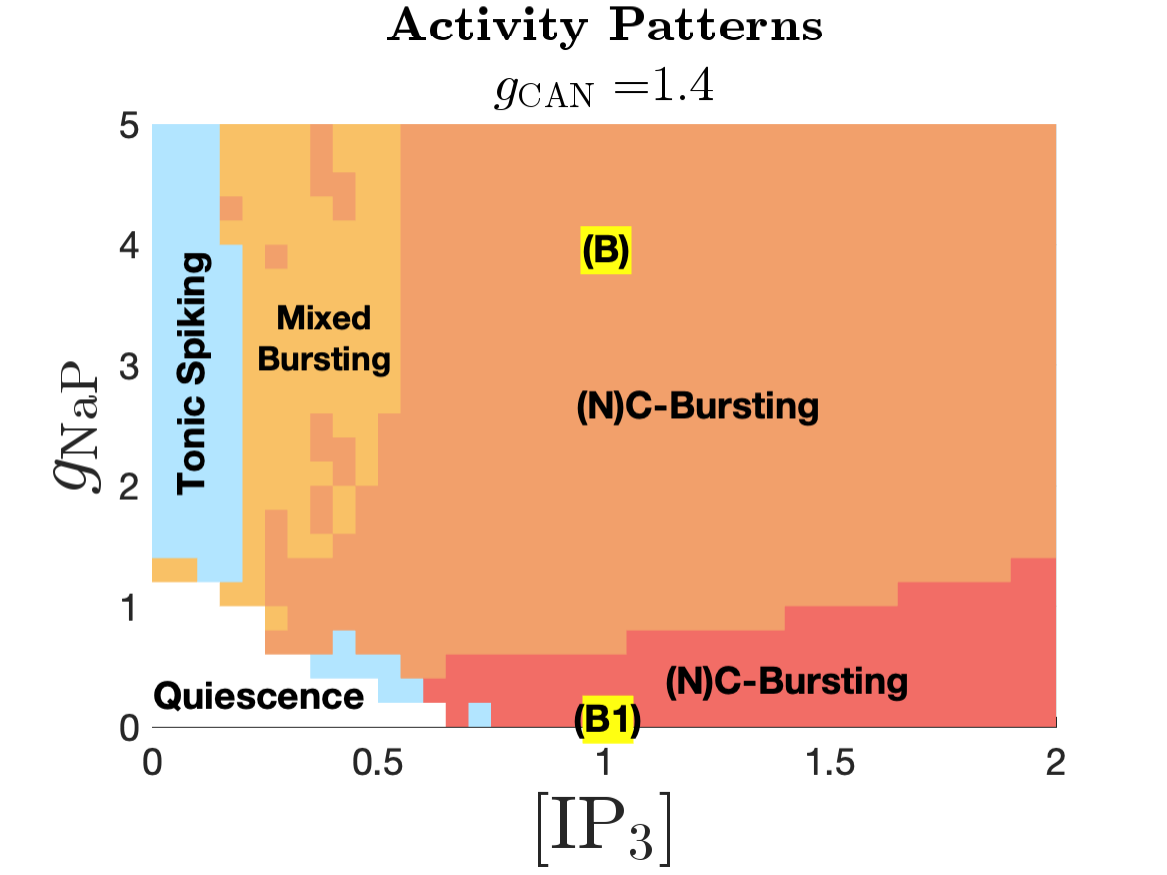} \\
    \multicolumn{2}{c}{\subfigimg[width=0.8\linewidth]{}{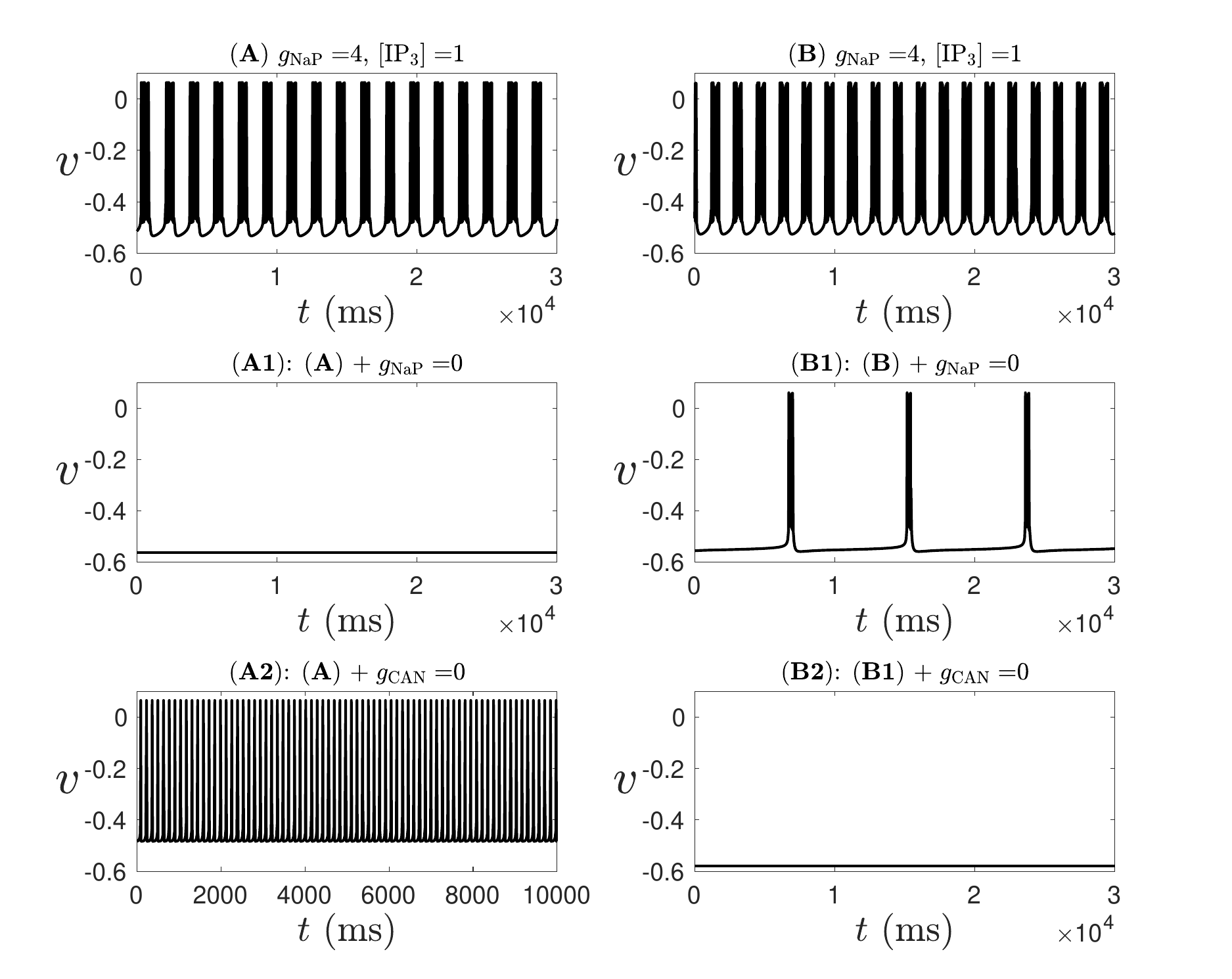}} 
    \end{tabular}
    \end{center}
    \caption{Activity patterns of model \cref{eq:pBC_ICa_slow} for $g_{\rm Ca} = 0.0002$, shown as a function of $g_{\rm NaP}$ and $[\rm IP_3]$, with (i) $g_{\rm CAN} = 0.7$ and (ii) $g_{\rm CAN} = 1.4$.  Color coding follows that in \cref{fig:activity_patterns}(i). (A): A sample voltage trace demonstrates that increasing $[\rm IP_3]$ induces rhythmic bursting activity corresponding to point (A) in panel (i); this bursting is abolished when  either $I_{\rm NaP}$ (A1) or $I_{\rm CAN}$ (A2) is blocked. (B): A sample voltage trace for NE ($I_{\rm CAN}$+$\rm IP_3$)-induced bursting. This bursting persists in the absence of  $I_{\rm NaP}$ (B1), but is eliminated when  $I_{\rm CAN}$ is also blocked (B2).} 
    \label{fig:ts_to_cb}
\end{figure}

We next model the CAN-dependent conditional bursting behavior evoked in tonic spiking neurons as experimentally observed by \cite{VR}. 
In model \cref{eq:pBC_ICa_slow} with $g_{\rm CAN}=0.7$ as the control condition, tonic spiking arises under two distinct parameter regimes: either when calcium conductance ($g_{\rm Ca}$) is very small (less than $0.0001$) (see \cref{fig:activity_patterns}(i), light blue region), or when $g_{\rm Ca}$ is intermediate (e.g., $\sim 0.0002$) (see \cref{fig:ts_to_cb}(i), light blue region). In both cases, $g_{\rm NaP}$ needs to be relatively high to sustain tonic spiking in the voltage dynamics.

We first focus on the tonic spiking regime highlighted in \cref{fig:ts_to_cb}(i), characterized by intermediate $g_{\rm Ca}$ and relatively low $[\rm IP_3]$. 
Our simulations and analyses highlight three important observations: First, neurons in this regime remain tonic spiking as $g_{\rm CAN}$ increases from $0.7$ to $1.4$ (compare \cref{fig:ts_to_cb}(i) and (ii)) and even higher values (data not shown), suggesting that increasing $g_{\rm CAN}$ alone is insufficient to reproduce NE-induced bursting observed experimentally. Second, while increasing $[\rm IP_3]$ alone does induce bursting (\cref{fig:ts_to_cb}A), this activity is not solely dependent on CAN currents. Third, when both $g_{\rm CAN}$ and $[\rm IP_3]$ are increased, the system transitions from tonic spiking to bursting (\cref{fig:ts_to_cb}B), with the bursting behavior solely dependent on CAN currents, consistent with experimental data. These observations suggest that, within this parameter regime, coordinated modulation of both $g_{\rm CAN}$ and $[\rm IP_3]$ is essential to reproduce the experimentally observed NE-induced, CAN-dependent-only bursting from tonic spiking neurons. 

To illustrate this, we show voltage traces from a representative control tonic spiking neuron with $g_{\rm NaP} = 4$, $g_{\rm Ca} = 0.0002$, $[{\rm IP_3}] = 0.1$ and $g_{\rm CAN} = 0.7$ (yellow star in \cref{fig:ts_to_cb}(i)) after it transitions to bursting, either through an increase in $[\rm IP_3]$ alone or under combined increases in both $g_{\rm CAN}$ and $[\rm IP_3]$. 
Importantly, the $\rm IP_3$-induced bursting corresponding to point (A) in \cref{fig:ts_to_cb}(i), with its voltage trace shown in \cref{fig:ts_to_cb}A, depends on both $I_{\rm NaP}$ and $I_{\rm CAN}$. Blocking either current eliminates bursting (see ~\cref{fig:ts_to_cb}A1 and A2), contradicting experimental evidence suggesting that NE-induced bursting is purely CAN-dependent. In contrast, bursting induced by simultaneously increasing $g_{\rm CAN}$ and $[{\rm IP_3}]$, corresponding to point (B) in \cref{fig:ts_to_cb}(ii), with its voltage trace shown in \cref{fig:ts_to_cb}B, is a C-burst. This bursting persists in the absence of $I_{\rm NaP}$ (\cref{fig:ts_to_cb}B1) but is abolished when $I_{\rm CAN}$ is blocked (\cref{fig:ts_to_cb}B2). Below we refer to the bursting in \cref{fig:ts_to_cb}A as $\rm IP_3$-induced bursting, and the bursting in \cref{fig:ts_to_cb}B as NE-induced bursting. 

\begin{figure}[!ht]
    \centering
    \begin{center}
    \begin{tabular}
    {@{}p{0.45\linewidth}@{\quad}p{0.45\linewidth}@{}}
    \subfigimg[width=\linewidth]{\bfseries{\footnotesize{(A)}}}{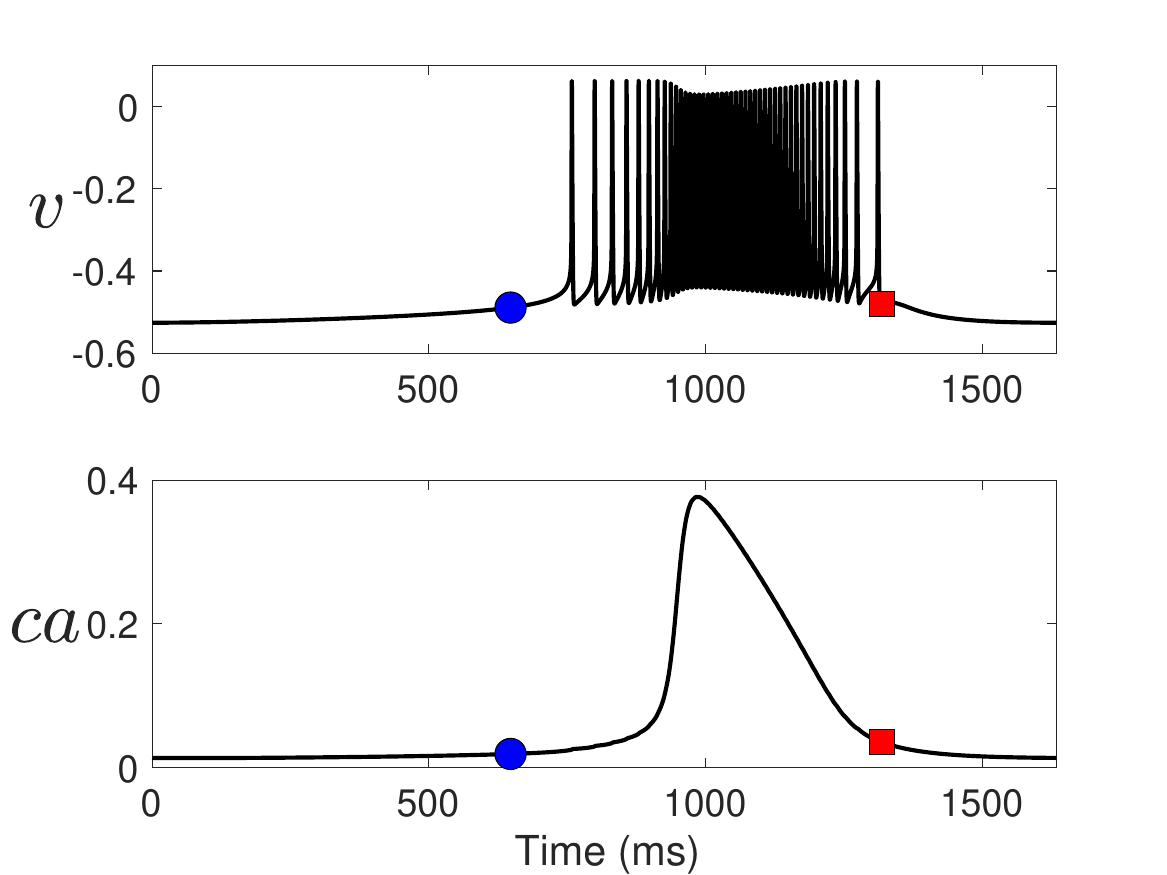} & 
    \subfigimg[width=\linewidth]{\bfseries{\footnotesize{(B)}}}{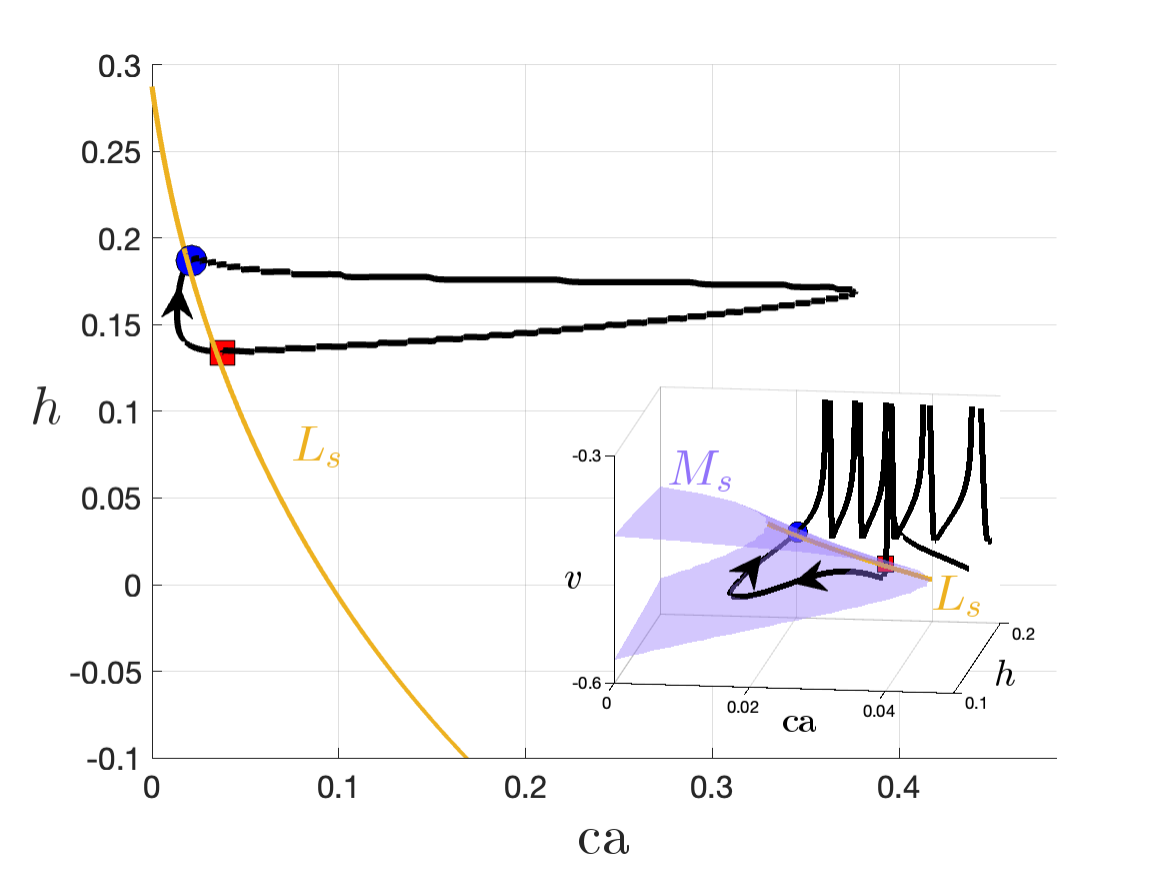}
    \end{tabular}
    \end{center}
    \caption{Simulation of one cycle of NE-induced C-bursting solution from \cref{fig:ts_to_cb}B.  
    The blue circle and red square mark the initiation and termination of the burst.
    (A) Temporal evolution of $v$ and $ca$ over one full cycle. (B) Projection of the black trajectory from panel (A) onto the $(ca, h)$-space, along with the fold $L_s$ (blue curve) of the critical manifold $M_s$. The inset shows the trajectory, $M_s$ (blue surface), and \RED{$L_s$ (yellow curve)} projected onto $(ca, h, v)$-space. The burst is initiated and terminated as the trajectory crosses $L_s$.} 
    \label{fig:ICB_analysis}
\end{figure}

Our GSPT analysis of the NE-induced bursting (\cref{fig:ICB_analysis}) reveals a bursting mechanism \RED{similar to} that of the regular C-burst in \cref{CB}. 
The corresponding temporal evolution of $v$ and $ca$ over one full burst cycle is shown in panel A. The blue circle and red square mark the initiation and termination of the burst, respectively, corresponding to the points where the trajectory projected onto $(ca, h)$- and $(ca,h,v)$-spaces crosses the fold $L_s$ of the critical manifold $M_s$ (see \cref{fig:ICB_analysis}B and its inset). During the silent phase, the trajectory follows the attracting lower branch of $M_s$ until it reaches its fold $L_s$ at the blue circle, triggering burst initiation and entering the active phase. This in turn causes $ca$ to jump up. As $h$ and $ca$ decrease, the trajectory again crosses $L_s$ - now at the red square - which coincides with a homoclinic bifurcation at a saddle-node on invariant circle (SNIC) bifurcation that terminates the periodic branch of the fast layer problem. Thus, the burst ends at the red square. \RED{While both the C-burst analyzed in \cref{fig:CB_burst_mechanism} 
and the NE-induced burst initiate and terminate through a SNIC mechanism, they differ in the temporal ordering of bursting initiation and the $ca$ jump-up. In the C-burst, the $ca$ jump-up occurs first and drives the system across the SNIC to initiate the burst, whereas in the NE-induced burst, this order is reversed. Another distinction lies in the role of the superslow manifold $M_{ss}$. In the NE-induced C-burst, $M_{ss}$ is unstable. As a result, the dynamics transition directly from slow evolution along $M_s$ to the fast $v$ jump upon crossing the SNIC, thereby bypassing the superslow segment exhibited in the C-burst.}
The fact that $ca$-oscillations do not initiate the NE-induced burst does not imply they are not important. The jump-up of $ca$ helps pull the trajectory away from the fold $L_s$ after crossing the blue circle, leading to continuous spiking during the active phase. Moreover, the subsequent decrease of $ca$ is crucial for burst termination - without it, the system would remain in a tonic spiking state. 


\begin{figure}[!ht]
    \begin{center}
    \begin{tabular}    
    {@{}p{0.45\linewidth}@{\quad}p{0.45\linewidth}@{}}
    \subfigimg[width=\linewidth]{\bfseries{\footnotesize{(A)}}}{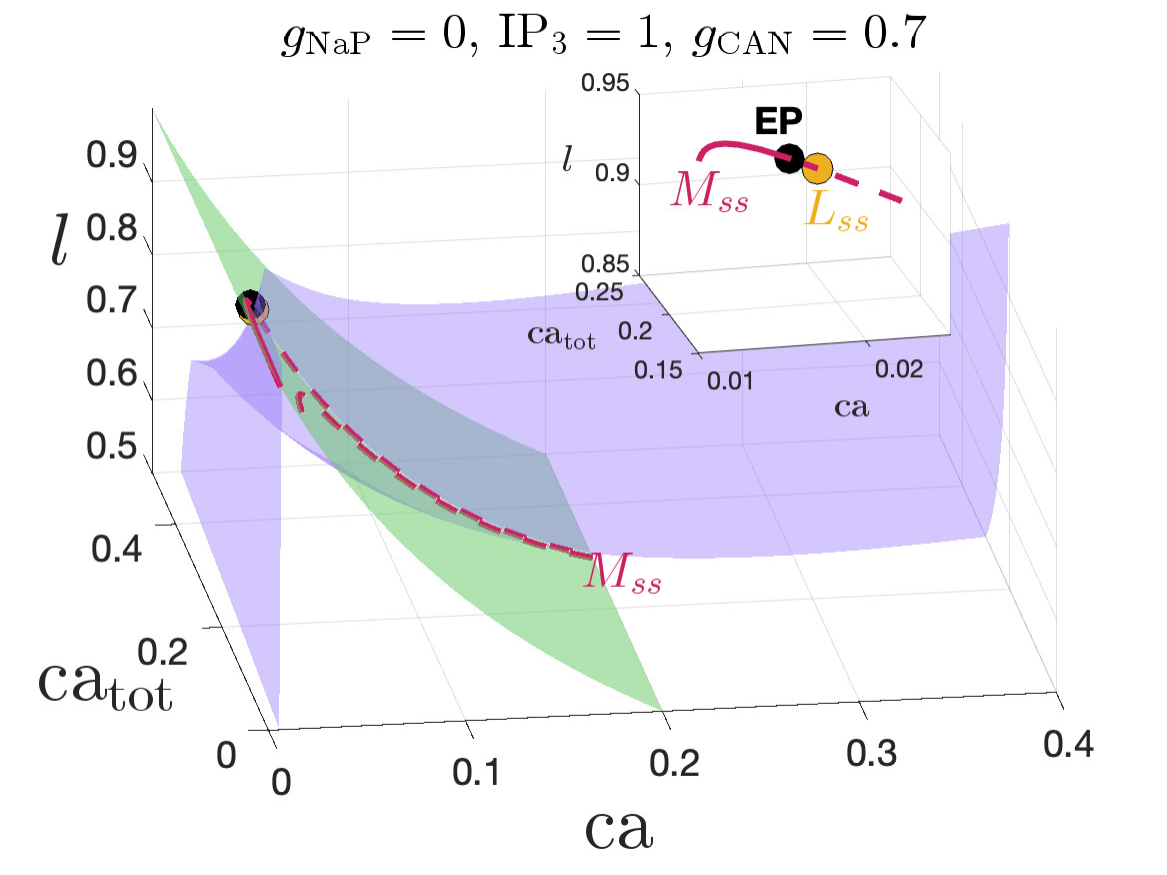} & \subfigimg[width=\linewidth]{\bfseries{\footnotesize{(B)}}}{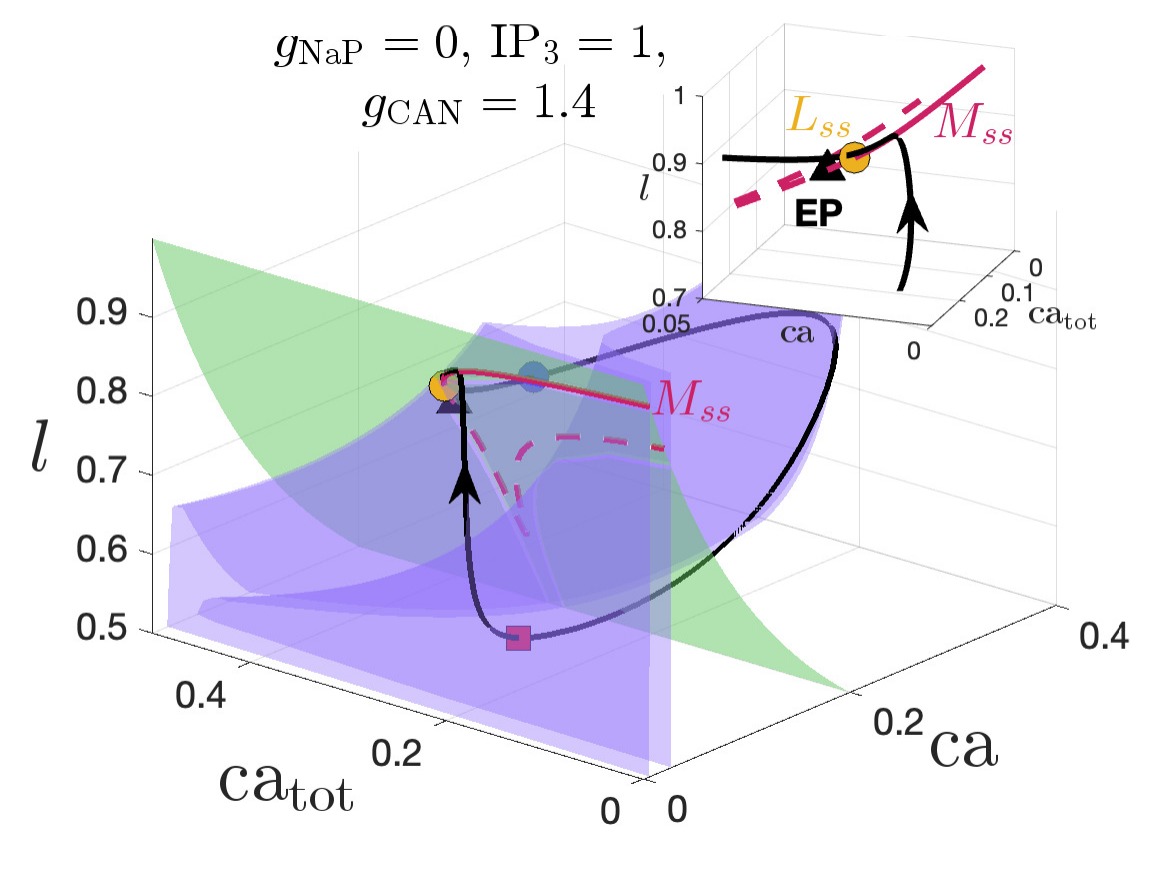} \\
    \subfigimg[width=\linewidth]{\bfseries{\footnotesize{(C)}}}{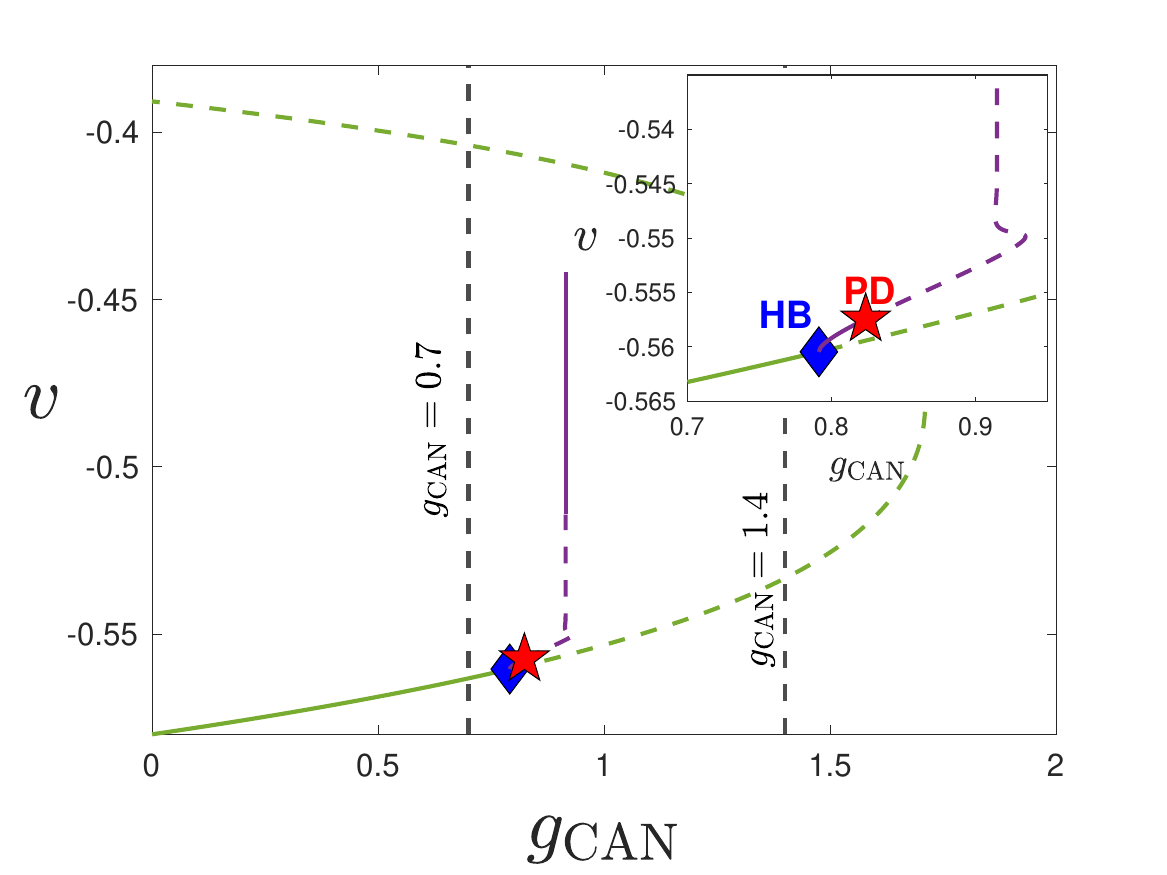} & \subfigimg[width=\linewidth]{\bfseries{\footnotesize{(D)}}}{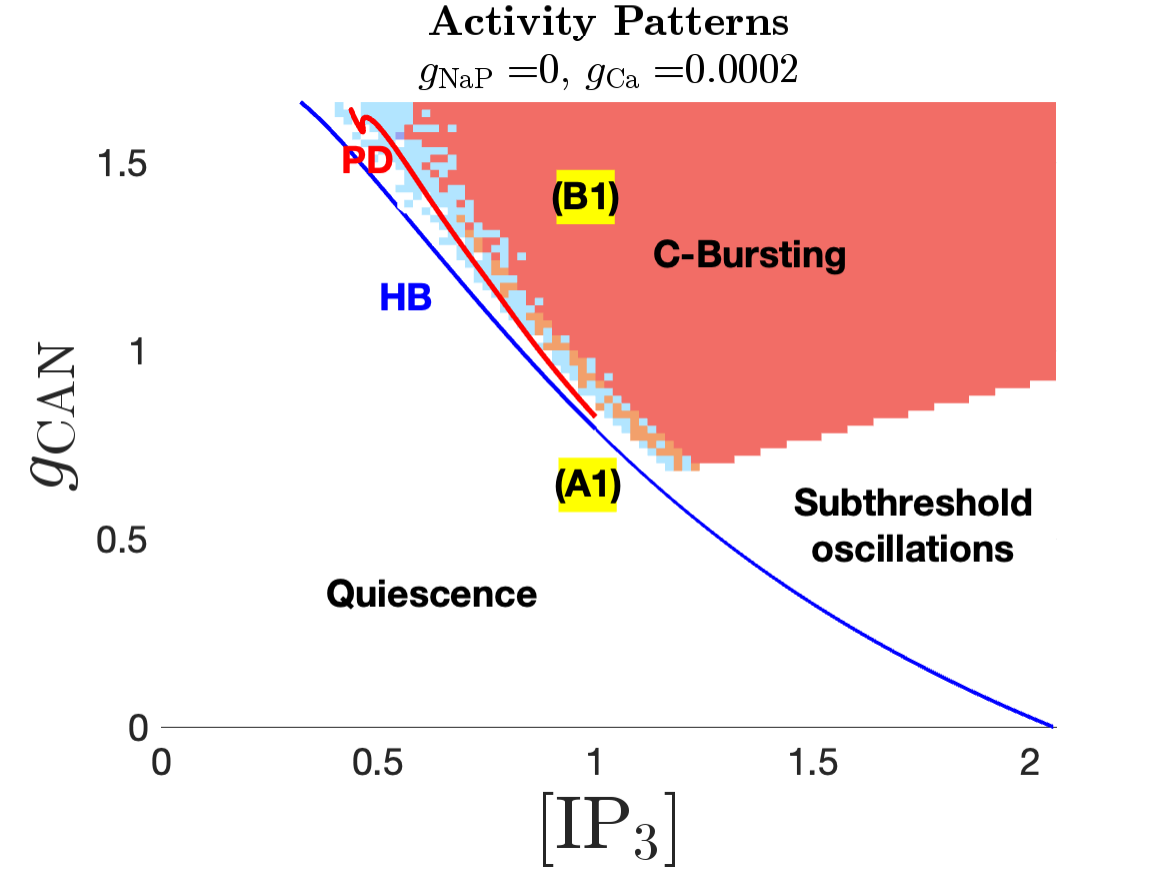}
    \end{tabular}
    \end{center}
    \caption{Projections of the solutions from \cref{fig:ts_to_cb}A1 and B1 onto $(ca,ca_{tot},l)$-space are shown in panels (A) and (B), respectively. Also shown are the $ca$-\RED{nullsurfaces}, $l$-nullsurface and the superslow manifold $M_{ss}$. Full system equilibria are shown by black circle (stable) and black triangle (unstable). The yellow circle denotes the superslow manifold fold $L_{ss}$. Other color coding and symbols are the same as in \cref{fig:CB_burst_mechanism}B.
    (C) Bifurcation diagram of the full system \cref{eq:pBC_ICa_slow} with $g_{\rm CAN}$ as a parameter, for $g_{\rm NaP} = 0,\, [\rm IP_3]=1$ and $g_{\rm Ca}=0.0002$. Green and purple curves represent equilibria and family of periodic orbit solutions. Stable and unstable objects are denoted by solid and dashed lines, respectively. Inset: transitions from quiescence to small-amplitude oscillations to bursting branch via HB (blue diamond) and period-doubling PD (red star) bifurcations. 
    (D) Two-parameter bifurcation diagram in $([\mathrm{IP_3}],g_{\rm CAN}$) plane identifying the boundaries between different patterns based on the HB and PD bifurcations from panel (C), superimposed with the activity patterns for the full system \cref{eq:pBC_ICa_slow}.}
    \label{fig:induced_bursts_analysis}
\end{figure}

To better understand why blocking $g_{\rm NaP}$ eliminates $\rm IP_3$-induced bursting while NE-induced bursting persists, we examine the projections of the dynamics with $g_{\rm NaP}=0$ from \cref{fig:ts_to_cb}A1 and B1 onto $(ca, ca_{tot}, l)$-space, together with the superslow manifold $M_{ss}$ (see \cref{fig:induced_bursts_analysis}A and B). In panel (A), the system settles into a stable equilibrium, corresponding to a silent state. This explains why ${\rm IP_3}$-induced bursting becomes silent in the absence of $I_{\rm NaP}$. In contrast, in panel (B) with a higher $g_{\rm CAN}$ value, the equilibrium is unstable and a bursting solution emerges. The bursting trajectory initially travels along the stable branch of the $ca$-nullsurface, then transitions onto the stable portion of $M_{ss}$, traveling along it until it reaches the fold $L_{ss}$, after which it jumps to higher $ca$ values. This $ca$ jump triggers burst onset at the blue circle. The mechanism is similar to that of the regular C-burst described in \cref{CB} as both are driven by calcium oscillations; however, one is initiated by crossing the fold $L_{ss}$, while the other is triggered by a Hopf bifurcation and experiences a delay. 

\Cref{fig:induced_bursts_analysis}C shows the bifurcation structure of the full system for $g_{\rm NaP}=0$ and $[\rm IP_3]=1$, using $g_{\rm CAN}$ as the bifurcation parameter. As $g_{\rm CAN}$ increases, the equilibrium changes from stable (green solid) to unstable (green dashed) via a supercritical Hopf bifurcation (blue diamond), at which a family of limit cycles emerges (purple curve). This branch loses stability through a period-doubling (PD) bifurcation (red star), \RED{transitioning from a stable small-amplitude periodic orbit (PO) (solid purple line) to an unstable small-amplitude PO branch (dashed purple line) that subsequently regains stability via fold bifurcations (solid purple vertical line). After the PD bifurcation, the full system exhibits bursting dynamics}. \RED{Using this one-parameter diagram, we continue the HB and PD bifurcation curves in $(\mathrm{[IP_3], g_{\rm CAN}})$-space} with $g_{\rm NaP}=0$ to identify regions in which bursting persists in the absence of $I_{\rm NaP}$, i.e., the C-bursting region. \RED{We note that as $[\rm IP_3]$ increases, the PD curve stops at a fold bifurcation at $[\rm IP_3] = 1$ and does not extend to intersect the HB curve. This two-parameter bifurcation diagram} consists of three main distinct dynamic regimes: quiescence, subthreshold oscillations, and C-bursting. Interestingly, \RED{mixed-mode bursting oscillations (MMBOs) also arise in a small region above the HB curve (see the light blue region in panel D)}. 

While intracellular calcium $ca$ oscillations emerge above the HB curve in \cref{fig:induced_bursts_analysis}D, C-bursting occurs only for relatively large $g_{\rm CAN}$ (see the red region). This is because, at low $g_{\rm CAN}$, the amplitudes of $ca$ oscillations are generally too small to sufficiently depolarize the membrane voltage and induce bursting. As a result, the influence of $ca$ dynamics is not effectively transmitted to the voltage compartment through $I_{\rm CAN}$ and cannot sustain bursting in $v$, \RED{resulting in sub-threshold voltage oscillations in the full system (see the white region above the HB curve in \cref{fig:induced_bursts_analysis}D)}. Using this bifurcation analysis (\cref{fig:induced_bursts_analysis}C and D), we can now understand why a combined increase in $\rm [IP_3]$ and $g_{\rm CAN}$ is necessary to model NE-induced C-bursting from the tonic spiking regime associated with low $[\rm IP_3]$ and $g_{\rm CAN}$ values. The $g_{\rm CAN}$ values corresponding to \cref{fig:ts_to_cb}(A1) and (B1) are indicated by the two vertical black lines in \cref{fig:induced_bursts_analysis}C, and by points (A1) and (B1) in \cref{fig:induced_bursts_analysis}D. An increase in $[\rm IP_3]$ alone leads to bursting that transitions to quiescence when $g_{\rm NaP}$ is blocked (point (A1) in \cref{fig:induced_bursts_analysis}D), whereas increasing both $g_{\rm CAN}$ and $\rm [IP_3]$ allows the system to cross the HB bifurcation and reach the C-bursting regime that persists without $I_{\rm NaP}$ (point (B1) in \cref{fig:induced_bursts_analysis}D). Although a further increase in $\rm [IP_3]$ for a fixed low $g_{\rm CAN}$ can also allow the system to cross the HB bifurcation, it results in subthreshold oscillations instead of bursting in the absence of $I_{\rm NaP}$, as discussed previously.

Lastly, it is worth noting that the level of $g_{\rm Ca}$ plays a critical role in determining whether C-bursting can be induced from a tonic spiking state. For example, numerical simulations in the low-$g_{\rm Ca}$ tonic spiking regime (\cref{fig:activity_patterns}(i), light blue region) reveal that when $g_{\rm Ca}$ is too low (e.g., below $0.0001$), increasing $g_{\rm CAN}$ and $[{\rm IP_3}]$ fails to induce C-bursting. Additional analysis of this phenomenon is provided in \RED{Appendix \hyperlink{appF}{F}}.

\subsection{Effect of NE on Silent neurons} \label{Q} 

\begin{figure}[!ht]
    \begin{center}
    \begin{tabular}
    {@{}p{0.4\linewidth}@{\quad}p{0.4\linewidth}@{}}
    \subfigimg[width=\linewidth]{\bfseries{\small{(i)}}}{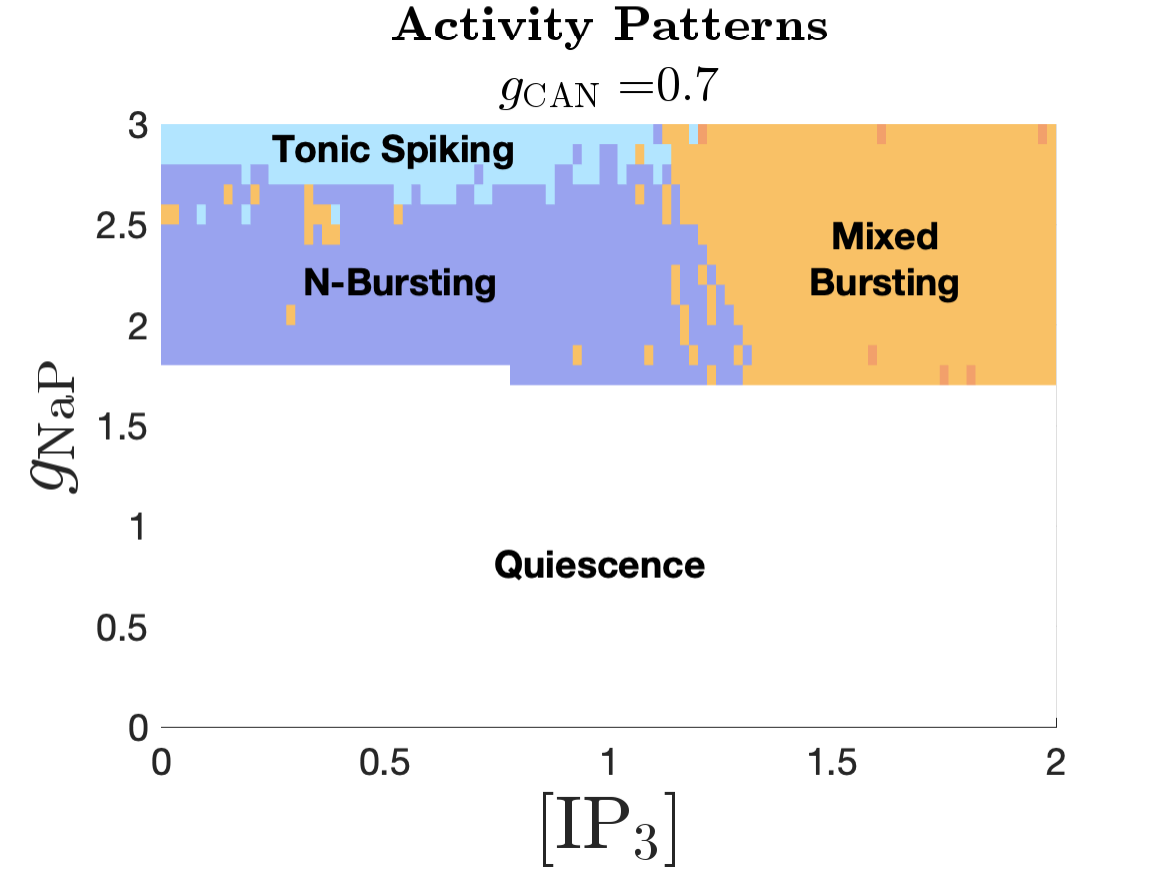} & 
    \subfigimg[width=\linewidth]{\bfseries{\small{(ii)}}}{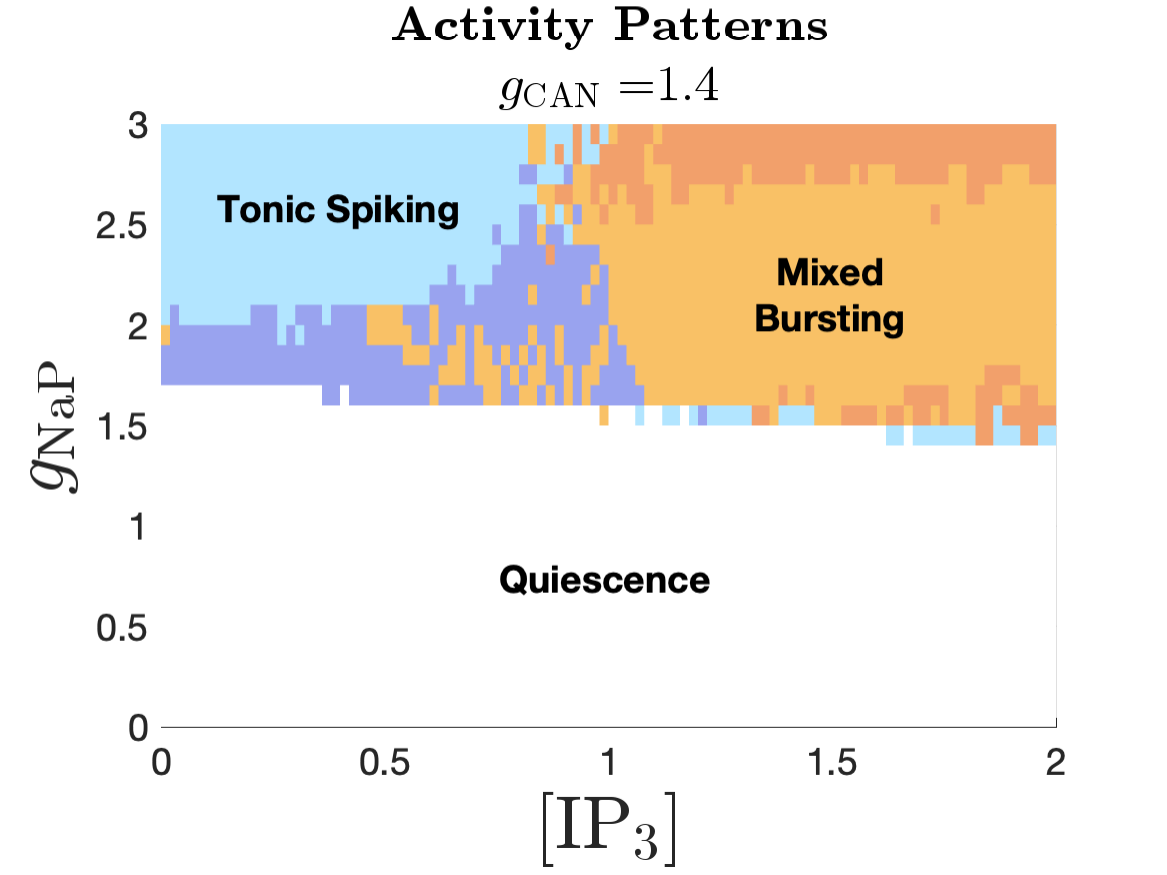}
    \end{tabular}
    \end{center}
    \caption{Activity patterns of model \cref{eq:pBC_ICa_slow} for $g_{\rm Ca} = 5e-5$, shown as a function of $g_{\rm NaP}$ and $[\rm IP_3]$, with (i) $g_{\rm CAN} = 0.7$ and (ii) $g_{\rm CAN} = 1.4$.  Color coding follows that in \cref{fig:activity_patterns}(i).
    In this diagram, the majority of quiescent (silent) neurons remain silent with increase in both $g_{\rm CAN}$ and ${[\rm IP_3]}$, while other activity pattern regions either shrink or expand.}
    \label{fig:Q}
\end{figure}

\RED{Our modeling shows that some quiescent neurons can become active as $g_{\rm CAN}$ increases sufficiently. Specifically, neurons in the white region in \cref{fig:ts_to_cb}(i) can transition to (N)C-bursting in response to increases in $g_{\rm CAN}$ and/or $[\rm IP_3]$ (\cref{fig:ts_to_cb}(ii)).} 
This outcome, however, contradicts experimental observations \RED{in which all} silent neurons remain inactive in the presence of NE. To reconcile this discrepancy, we used our model to identify conditions under which silence is preserved despite NE exposure. Our analysis predicts that neurons with low $g_{\rm Ca}$ values (e.g., $g_{\rm Ca} < 0.0001$) remain quiescent even when both $g_{\rm CAN}$ and $\rm [IP_3]$ are elevated (compare \cref{fig:Q}(i) and (ii), white region). In these neurons, the calcium current is too weak to support intrinsic calcium oscillations, and in the absence of sufficient $I_{\rm NaP}$, neither (N)C- nor C-bursting mechanisms can be activated (see \cref{fig:induced_bursts_gca_ip3} for supporting analysis). Thus, our model predicts \RED{that the level of $g_{\rm Ca}$ can be a critical determinant of the response of silent neurons to NE.}



\section{Discussion} \label{discussion}

Despite previous experimental and computational efforts in understanding the basis for the modulation of norepinephrine (NE) in the preB{\"o}tzinger complex (\pbc{}), there has been limited progress toward understanding the mechanisms by which NE differentially impacts distinct \pbc{} neuronal types. Our paper addresses this gap by proposing a novel mechanism wherein NE enhances neuronal excitability through the concurrent upregulation of calcium-activated non-specific cation conductance ($g_{\rm CAN}$) and intracellular inositol trisphosphate (${\rm IP_3}$) levels. This dual-modulation mechanism engages second messenger-mediated calcium dynamics and successfully captures NE-induced conditional bursting in neurons that in synaptic isolation are tonically spiking, a phenomenon not seen in earlier modeling studies that considered only NE-enhanced $g_{\rm CAN}$ \cite{TB, PR}.  Our \emph{in silico} model also identifies parameter regimes in which neurons remain quiescent despite elevated levels of $g_{\rm CAN}$ and $[{\rm IP_3}]$. In addition, it reveals distinct NE-induced changes in burst frequency and duration between NaP- and CAN-dependent bursters, providing a mechanistic explanation for the experimentally observed heterogeneity in NE responsiveness. Collectively, these findings reconcile discrepancies between previous computational and experimental studies by reproducing key experimental observations and offer novel insights into the diverse modulatory roles of NE within the \pbc{} network. 

To gain deeper insight into the proposed dual-modulation mechanism of NE, we performed GSPT analysis to analyze how this mechanism affects distinct neuronal types. 
This analysis predicts that conditional bursting in tonic spiking neurons requires a concurrent elevation of both [${\rm IP_3}$] and $g_{\rm CAN}$. As illustrated in \cref{TS}, increasing $g_{\rm CAN}$ alone is insufficient to induce bursting; in fact, a sufficient increase in $g_{\rm CAN}$ can even switch a bursting neuron to tonic spiking, opposite to the transition induced by NE. While increasing [${\rm IP_3}$] can induce bursting in some tonic spiking neurons, the resulting bursting is not solely dependent on $I_{\rm CAN}$, as eliminating either $I_{\rm NaP}$ or $I_{\rm CAN}$ abolishes the bursting dynamics. Our analysis predicts that only when both $g_{\rm CAN}$ and $\rm [IP_3]$ are sufficiently elevated does the induced bursting become strictly CAN-dependent, mimicking the NE-induced bursting properties. 

Our analysis of NaP-dependent bursting (N-bursting) neurons in \cref{NB} demonstrates that NE increases N-burst frequency while having minimal impact on burst duration
(\cref{fig:NB_gcan_ip3}), consistent with experimental findings. This effect occurs primarily through the increase of $g_{\rm CAN}$, which both decreases the distance between $I_{\rm NaP}$ inactivation thresholds for burst initiation and termination (i.e., $(h_{\rm SN}-h_{\rm HC})$ in \cref{fig:NB_analysis}C) and accelerates the increasing rate of $h_{\rm NaP}$ during the silent phase, together leading to a shortened interburst interval and thus a higher burst frequency. In contrast, during the active phase, the effects of $g_{\rm CAN}$ on $(h_{\rm SN}-h_{\rm HC})$ and the rate of $h_{\rm NaP}$ are opposing, effectively counterbalancing each other to maintain a nearly constant burst duration. Interestingly, our simulations suggest that the increase in frequency occurs in discrete steps as the model reaches certain thresholds in $g_{\rm CAN}$ values. With higher $[{\rm IP_3}]$, these thresholds shift slightly to higher $g_{\rm CAN}$ values. Our analysis suggests that each threshold corresponds to a saddle-node bifurcation of the bursting branch in the full system (see \cref{fig:NB_freq_analysis}). Between bifurcations, the burst frequency remains nearly constant, but experiences a sudden increase upon crossing each bifurcation threshold, resulting in a phasic pattern of burst frequency in the $([\mathrm{IP_3}], g_{\rm CAN})$ parameter space. This pattern of discrete changes is also reflected in the N-burst duration (see \cref{fig:NB_gcan_ip3}(ii)), though we only observe a sudden increase in burst duration along the bifurcation threshold curve, while the N-burst duration otherwise remains nearly constant. Understanding the mechanisms that maintain homeostatic properties between bifurcation thresholds, and the abrupt changes across them,  remain an interesting direction for future research. Based on this, we predict that the baseline conditions and the conditions where NE concentrations change correspond to different homeostatic regions in the $([\mathrm{IP_3}], g_{\rm CAN})$ parameter space. Whether a similar phasic pattern also exists in experimental data remains a question, but can be tested through dose-response experiments with NE, as existing literature primarily focuses on single-dose responses. 
 
Our analysis shows that CAN-dependent bursting (C-bursting) is driven by calcium oscillations via $I_{\rm CAN}$ activation, with burst onset mediated through a delayed HB on the superslow manifold $M_{ss}$. As explained in \cref{CB}, increasing $g_{\rm CAN}$ alone increases burst frequency, whereas an initial increase in $[\rm IP_3]$ slows down the frequency, followed by acceleration with further increases in $[\rm IP_3]$. As a result, there exist regions in the $([\mathrm{IP_3}], g_{\rm CAN})$-parameter space where the C-burst frequency remains nearly constant despite increases in both parameters. This highlights the importance of incorporating both mechanisms to reproduce the experimental observation that the C-burst frequency remains nearly unchanged under NE application. Nonetheless, this unchanged frequency is accompanied by a nearly constant burst duration (\cref{fig:CB_gcan_ip3}C-G). There are also regions where increasing $[\rm IP_3]$ and $g_{\rm CAN}$ leads to increases in both burst frequency and duration (\cref{fig:CB_gcan_ip3}G-H). In the first case, while C-burst frequency remains similar, the lack of an increase in burst duration contradicts experimental observations from \cite{VR}.
In the second case, although the increase in burst duration aligns with the experimental results, it comes at the cost of an increase in burst frequency, which again differs from experimental findings.

One possible explanation for this discrepancy is that the relatively high (20~$\mu$M) dose of NE used in the experimental study by \cite{VR} may have depolarized the CAN-dependent bursters sufficiently to the point where burst frequency does not further increase with more excitability. Such a burst frequency saturation effect has been shown in C-bursters where sequentially increasing depolarizing steps of current injections initially increase burst frequency, but subsequent larger depolarizations no longer increase it \cite{thoby2001identification}. Thus, it is possible that NE at 20~$\mu$M quickly saturated the frequency response of the C-bursters. Another possibility is that NE could activate more than one adrenergic receptor subtype, resulting in activities not captured by our modeling.

$g_{\rm CAN}$-induced increases in C-burst frequency were not observed in the earlier models \cite{TB, PR}, where $ca$-oscillations acted as an independent relaxation oscillator, unaffected by membrane voltage due to the absence of $I_{\rm Ca}$. This decoupling was a consequence of the closed-cell assumption, which excluded calcium influx from the extracellular space and limited the responsiveness of C-burst frequency (i.e., the frequency of calcium oscillations) to membrane depolarization. In contrast, our model incorporates $I_{\rm Ca}$ to relax the closed-cell assumption \cite{jasinski, phillips2019biophysical, wang2020complex}, leading to more complex effects of $g_{\rm CAN}$ on C-bursting dynamics through the voltage dependence of the calcium subsystem, as demonstrated by our analysis in \cref{CB}. A more detailed GSPT analysis, incorporating additional subsystems and projections, may be necessary to further clarify these effects and determine how to carefully balance $g_{\rm CAN}$ and $[\rm IP_3]$ to achieve an increase in burst duration while maintaining a constant burst frequency in an open cell model.  
Thus, although our model does not fully reproduce all experimental observations regarding the effects of NE on C-bursters, it provides specific, testable predictions about the interplay between $g_{\rm CAN}$ and $\rm [IP_3]$.

While we do not vary $g_{\rm Ca}$ when modeling NE, its baseline level varies across neuronal types and can have an important impact on the effects of NE. For example, $g_{\rm Ca}$ remains relatively low in N-bursting neurons but is much higher in C-bursting neurons (see \cref{fig:activity_patterns}), which partly contributes to the different effects of NE on these two bursting dynamics as discussed above. Moreover, \cref{TS} shows that NE induces C-bursting in tonic spiking neurons with intermediate $g_{\rm Ca}$, but not in those with low $g_{\rm Ca}$. The role of $g_{\rm Ca}$ in modeling NE-induced C-bursting is further analyzed in \RED{Appendix \hyperlink{appF}{F}}. 
Similarly, our model predicts that the effect of NE on silent neurons also critically depends on the level of $g_{\rm Ca}$. \RED{In experiments, silent \pbc{} neurons depolarize with NE but do not transition into spiking or bursting \cite{VR}. In contrast, our model produces NE-induced bursting in silent neurons when $g_{\rm Ca}$ is sufficiently high, whereas with low $g_{\rm Ca}$, the majority of silent neurons in the model remain quiescent despite increases in both $g_{\rm CAN}$ and $[\rm IP_3]$ (see \cref{fig:Q}). 
Thus, our analysis predicts that the response of a silent neuron may depend sensitively on its intrinsic calcium conductance. 
Whether similar $g_{\rm Ca}$-dependent differential effects occur in \pbc{} neurons \emph{in vitro} remains unclear and will require further experimental investigation.}

Our modeling also revealed a variety of interesting respiratory bursting dynamics. In modeling the effects of NE, we observed both mixed bursting (see \cref{fig:activity_patterns}D in \cref{activity_patterns} and \RED{Appendix \hyperlink{appE}{E}}) and mixed-mode bursting oscillations (see \cref{fig:CB_gcan_ip3}A,B in \cref{CB}). We show that mixed bursting behavior arises during the transition from a tonic spiking neuron to bursting, suggesting that an initial application of NE might induce irregular activity before stabilizing into regular bursting with continued NE application. Since existing experiments have typically used a fixed dosage of NE, testing this hypothesis would require further experimental investigation. Beyond their biological relevance, these mixed bursting and mixed-mode bursting oscillations (MMBOs) are also of interest from a dynamical systems perspective. 
In particular, the small-amplitude oscillations can arise either from a delayed Hopf bifurcation (DHB) on the superslow manifold $M_{ss}$ or through a canard mechanism associated with folded node singularities along the fold of $M_s$ \cite{Desroches2012}. A key singularity known as canard-delayed-Hopf singularity naturally arises in three-timescale models such as our model \cref{eq:pBC_ICa_slow}, and may facilitate the coexistence and interaction of the DHB and canard mechanisms to co-modulate local oscillator behaviors \cite{phan2024mixed}. 
Uncovering the mechanisms underlying these mixed bursting and MMBO dynamics in three-timescale settings would require a more detailed GSPT analysis, as carried out in previous studies (e.g.,  \cite{bertram2008phantom,Desroches2013,koksal2020canard,Desroches2022,wang2016jcompneur,wang2017timescales,wang2020complex,phan2024mixed}), and represents an intriguing direction for future research.

Beyond its single-neuron effects, neuromodulation plays a crucial role in shaping rhythmogenesis at the network level. Recent studies have shown that network sensitivity to opioid-based neuromodulation is critically determined by network configuration and the intrinsic cellular properties of neurons embedded in the network \cite{baertsch2021dual,burgraff2021dynamic,chou2024modeling}. Our findings highlight the complex ways in which NE differentially modulates distinct neuronal subtypes, supporting the view that neurons play distinct roles in shaping network-level responses based on their intrinsic states. Extending our modeling framework to the network level will be crucial for understanding how NE-induced, cell-specific changes influence coordinated activity across the \pbc{} network, and how these effects interact with network architecture. 

More broadly, the role of neuromodulation in rhythmogenesis is known to depend on physiological context. While NE can stabilize inspiratory activity in healthy networks \cite{VR}, it may destabilize inspiratory network rhythms under pathological conditions \cite{zanella2014norepinephrine}. For example, under acute intermittent hypoxia, NE has been shown to promote variability in respiratory rhythms and impair transmission to motor outputs—effects implicated in disorders such as sleep apnea and apneas of prematurity \cite{zanella2014norepinephrine, garcia2016chronic, browe2023gasotransmitter}.
Although these disruptions were originally attributed to changes in inhibitory synaptic drive, our findings raise an alternative, testable hypothesis: that variability in intrinsic NE responsiveness among preB\"{o}tC neurons may also contribute to such divergent outcomes. Our study lays the groundwork for future investigations into how cell-specific modulation may shape network-level effects under both healthy and disease conditions.

\RED{\section*{Codes and Data  Availability}

Simulation code for the full model is publicly available on GitHub at \url{https://github.com/sreshta-venkat/pBC_Single_Neuron_Model_for_NE}.
    Parameter sweep and bifurcation analysis scripts (XPP/AUTO continuation files and associated analysis code) are available from the authors upon reasonable request.}

\begin{appendix}
\section{Mathematical Preliminaries} \hypertarget{appA}{}

In 1952, Hodgkin and Huxley \cite{hodgkin1952quantitative} developed a mathematical model of the squid giant axon using a system of differential equations. 
In this paper, we model the individual preB\"{o}tC neuron as a single-compartment unit with Hodgkin-Huxley conductances adapted from previously described models \cite{PR, wang2020complex}. 
It is well established that neural processes can evolve over very different time scales. Identifying relevant time scales is a useful step when modeling neural systems. In mathematical analysis, the time scales in a system of differential equations are often grouped into a small number of classes to facilitate simplification of analysis via limiting processes. This enables the application of a method known as \emph{Geometric Singular Perturbation Theory} (GSPT), as we review in Subsection \ref{gspt}. 
GSPT leverages insights from the reduced ``limit systems", which are obtained by taking the singular limit at which certain variables evolve infinitely fast or slow, to gain information about the behavior of the original system. Examples of this approach are ubiquitous in the literature, with ideas applicable to systems in which time scales can be grouped into two classes being particularly well developed. Typical simple models of membrane potential oscillations have at least two time scales. In our model, which combines membrane potential and intracellular calcium dynamics, dimensional analysis in Appendix \hyperlink{appB}{B} reveals three distinct timescales. This motivates the use of a three-timescale extension of the GSPT framework \cite{Nan2015} to decomposition the full system into fast, slow and superslow subsystems (see Appendix \hyperlink{appC}{C} for their derivations). The aim of GSPT is to combine information of these reduced systems to infer dynamics in the full system. 


In addition to GSPT, we also make extensive use of bifurcation analysis to analyze the system's dynamic behaviors.  
Many interesting transitions in activity arise at bifurcation points, locations where the stability, number, and/or type of equilibria and periodic orbits change. To carry out this analysis, we construct \emph{one-parameter bifurcation diagrams}, which track how equilibria and periodic orbits evolve as a single \emph{bifurcation parameter} varies. These parameters may be biologically meaningful system parameters, or, in systems with multiple timescales, a slow variable can be treated as a bifurcation parameter for the fast subsystem in combination with GSPT. We use this method in our analysis of the N-bursting solution in \cref{NB}. Such bifurcation techniques have been widely used in the study of neuronal bursting dynamics \cite{rinzel1987formal, bertram1995topological, izhikevich2000neural, golubitsky2001unfolding}, including in models of respiratory neurons \cite{PR, wang2016jcompneur, wang2017timescales, wang2020complex}.  
Further, since biological systems are often influenced by multiple factors that can change simultaneously, we also make use of  \emph{2-parameter bifurcation diagrams}, which track how bifurcation curves change as a function of two parameters. We apply this analysis in Sections~\ref{NB} and~\ref{TS} to study our model dynamics, with examples shown in Figures~\ref{fig:NB_analysis}C and~\ref{fig:induced_bursts_analysis}D.


In Appendices \hyperlink{appB}{B} and \hyperlink{appC}{C} below, we detail other mathematical procedures such as nondimensionalization of the model and preliminaries of GSPT necessary for the analysis of our model.

\section{Nondimensionalization of the Model} \hypertarget{appB}{}

In order to verify whether our model evolves in multiple timescales, we need to nondimensionalize our model. As a first step towards that, we rescale the variables so that the different timescales can be clearly identified, and to do so, we define new dimensionless variables $\tau, v, ca, ca_{tot}$ such that
\begin{equation}
    t = Q_t \tau, \;\; V = Q_v v, \;\; [{\rm Ca}] = Q_{ca} ca, \;\; [{\rm Ca}]_{\rm Tot} = Q_{ca_{tot}} ca_{tot},
\end{equation}
where $Q_t, Q_v, Q_{ca} \; \text{and} \; Q_{ca_{tot}}$ are time, voltage, calcium and total intracellular calcium scales respectively. We note that $n$, $h$ and $l$ are already dimensionless variables.

Next, from numerical simulations, we find that the membrane potential $V$ lies between -60 mV and 10 mV. Correspondingly, we define $T_x := \max\limits_{V \in [-60, 10]} (1/\tau_x(V))$, for $x \in \{n,h\}$. Then, define $t_x(V) := T_x \tau_x(V)$, a rescaled version of $\tau_x$. We also define $g_{\rm max} := \max \{g_{\rm L}, \allowbreak g_{\rm K}, \allowbreak g_{\rm Na}, \allowbreak g_{\rm NaP}, \allowbreak g_{\rm CAN}, \allowbreak g_{\rm Ca}\}$. Further, we let $G([{\rm Ca}]) := \frac{[{\rm IP_3}][{\rm Ca}]}{([{\rm IP_3}] + K_I)([{\rm Ca}] + K_a)}$ and $g_{\rm SERCA}([\rm Ca]) := V_{\rm SERCA} \frac{[\rm Ca]}{K_{\rm SERCA}^2 + [\rm Ca]^2}$. Then, ${\rm J_{ER_{in}}} = (L_{\rm IP_3} + P_{\rm IP_3}G^3([{\rm Ca}])l^3)([\rm Ca]_{\rm ER} - [\rm Ca])$ and ${\rm J_{ER_{out}}} = g_{\rm SERCA}([\rm Ca])[\rm Ca]$. Substituting these in equation \cref{eq:pBC_Ica} and rearranging, we get the following system:

\begin{subequations}
    \begin{align}
        \frac{C_m}{Q_t g_{\rm max}} \frac{dv}{d\tau} &= -\bar{I}_{\rm L} - \bar{I}_{\rm K} - \bar{I}_{\rm Na} - \bar{I}_{\rm NaP} - \bar{I}_{\rm CAN} - \bar{I}_{\rm Ca} \label{eq:nondim1v} \\
        \frac{1}{Q_t T_n} \frac{dn}{d\tau} &= \frac{n_{\infty}(v)-n}{t_n(v)} \label{eq:nondim1n} \\
        \frac{1}{Q_t T_h} \frac{dh}{d\tau} &= \frac{h_{\infty}(v)-h}{t_h(v)} \label{eq:nondim1h} \\
        \frac{d[\rm Ca]}{dt} &= f_i (\rm{J_{ER_{in}}} - \rm{J_{ER_{out}}}) - \alpha_{\rm Ca}I_{\rm Ca} - ([\rm Ca] - [\rm Ca]_{min})/\tau_{\rm Ca} \label{eq:nondim1ca} \\
        \frac{d[\rm Ca]_{\rm Tot}}{dt} &= -\alpha_{\rm Ca}I_{\rm Ca} - ([\rm Ca] - [\rm Ca]_{min})/\tau_{\rm Ca} \label{eq:nondim1catot} \\
        \frac{dl}{dt} &= AK_d(1-l) - A[{\rm Ca}]l, \label{eq:nondim1l} 
    \end{align}
    \label{eq:nondim1}
\end{subequations}
with dimensionless currents $\bar{I}_x = \frac{I_x}{g_{\rm max} Q_v}$.

Thus, we have nondimensionalized the voltage subsystem \cref{eq:nondim1v} - \cref{eq:nondim1h}. Next, we focus on the calcium subsystem \cref{eq:nondim1ca} - \cref{eq:nondim1l}. Again, from numerical simulations, we know that $[\rm Ca] \in [0, 2] \; \mu$M and $[\rm Ca]_{\rm Tot} \in [0, 5] \; \mu$M. Therefore, we define $G_c := \max (G^3([\rm Ca]))$ and $G_S := \max (g_{\rm SERCA}([\rm Ca]))$ over the range $[\rm Ca] \in [0, 2]$. Then, define \RED{$P_{\rm max} = \max \{ L_{\rm IP_3}, P_{\rm IP_3}G_c, \sigma G_S\}$}. Making the necessary substitutions in and reducing the equations from \cref{eq:nondim1}, we then obtain the following system of equations:

\begin{subequations}
    \begin{align}
        \frac{C_m}{Q_t g_{\rm max}} \frac{dv}{d\tau} &= -\bar{I}_{\rm L} - \bar{I}_{\rm K} - \bar{I}_{\rm Na} - \bar{I}_{\rm NaP} - \bar{I}_{\rm CAN} - \bar{I}_{\rm Ca} =: f_1(v,n,h,ca)\label{eq:nondim2v} \\
        \frac{1}{Q_t T_n} \frac{dn}{d\tau} &= \frac{n_{\infty}(v)-n}{t_n(v)} \label{eq:nondim2n} \\
        \frac{1}{Q_t T_h} \frac{dh}{d\tau} &= \frac{h_{\infty}(v)-h}{t_h(v)} \label{eq:nondim2h} \\
        \RED{\frac{Q_{ca} K_{ca}}{Q_t} \frac{d ca}{d\tau}} &= \RED{ (\bar{J}_{\rm ER_{in}} - \bar{J}_{\rm ER_{out}}) - \alpha_{\rm Ca} K_{ca} I_{\rm Ca} - K_{ca} \frac{Q_{ca} ca - [\rm Ca]_{min}}{\tau_{\rm Ca}} =: g(v,ca,ca_{tot},l)} \label{eq:nondim2ca} \\
        \RED{\frac{Q_{ca_{tot}} \tau_{\rm Ca}}{Q_t K_{ca_{tot}}} \frac{d ca_{tot}}{d\tau} } &= \RED{ -\Tilde{I}_{\rm Ca} - \frac{Q_{ca} ca - [\rm Ca]_{min}}{K_{ca_{tot}}} =: h(v,ca) } \label{eq:nondim2catot} \\
        \frac{1}{Q_t Q_{ca} A} \frac{dl}{d\tau} &= \Bar{K}_d (1-l) - ca \cdot l, \label{eq:nondim2l}
    \end{align}
    \label{eq:nondim2}
\end{subequations}
where $\Tilde{I}_{\rm Ca} = \frac{I_{\rm Ca}}{g_{\rm Ca} Q_v}$, \RED{$K_{ca} = \frac{\sigma}{f_i P_{\rm max}}$}, \RED{$K_{ca_{tot}} = \tau_{\rm Ca} \alpha_{\rm Ca} g_{\rm Ca} Q_v$}, \RED{$\bar{J}_{\rm ER_{in}} = (\frac{1}{P_{\rm max}}) J_{\rm ER_{in}}$}, \RED{$\bar{J}_{\rm ER_{out}} = (\frac{\sigma}{P_{\rm max}}) J_{\rm ER_{out}}$} and $\Bar{K}_d = \frac{K_d}{Q_{ca}}$.

From expected ranges for $V$, $[\rm Ca]$ and $[\rm Ca]_{Tot}$ mentioned a priori, suitable choices for the voltage and calcium scales are $Q_v = 100$ mV, $Q_{ca} = 2 \; \mu$M and $Q_{ca_{tot}} = 5 \; \mu$M, respectively. We note that the values of $n_{\infty}, h_{\infty}, n, h, l \in [0, 1]$. Moreover, for the values specified in \cref{tab:par_val}, the maximum of the conductances is $g_{\rm Na} = 28$ nS. Further, from numerical evaluations of $1/\tau_n(V)$ and $1/\tau_h(V)$, for $V \in [-60, 10]$, we find that $T_n \approx 6.5491 \; {\rm ms}^{-1}$ and $T_h \approx 0.0165 \; {\rm ms}^{-1}$. We also compute $G_c \approx 0.0214$ and $G_S \approx 1000 \; {\rm pL \cdot ms}^{-1}$, so we then have that \RED{$P_{\rm max} \approx 664.4376 \; {\rm pL \cdot ms}^{-1}$}. \RED{Therefore,} we find that this leads to \RED{$K_{ca} \approx 11.1372$} and \RED{$K_{ca_{tot}} \approx 0.025$ for $g_{\rm Ca} = 0.00002$ (N-bursting), $K_{ca_{tot}} \approx 0.25$ for $g_{\rm Ca} = 0.0002$ (NE-induced bursting), and $K_{ca_{tot}} \approx 0.625$ for $g_{\rm Ca} = 0.0005$ (C-bursting)}.  




\RED{Using these values, we observe that all quantities on the right-hand side of equations \cref{eq:nondim2v} - \cref{eq:nondim2l} are bounded by 1 in absolute values and are O(1). Hence, the coefficients of the derivatives on the left-hand side of all the equations in \cref{eq:nondim2} convey the relative rates of evolution of the variables. Since $C_m/g_{\rm max} \approx 0.75 = O(1)$ ms, $1/T_n \approx 0.1527 = O(0.1)$ ms, $1/T_h \approx 60.5505 = O(100)$ ms, $Q_{ca} K_{ca} \approx 22.2744 = O(10)$ ms, $\frac{Q_{ca_{tot}} \tau_{\rm Ca}}{K_{ca_{tot}}} = 10^5 = O(10^5)$ ms (for N-bursting), $\frac{Q_{ca_{tot}} \tau_{\rm Ca}}{K_{ca_{tot}}} = 10^4 = O(10^4)$ ms (for NE-induced bursting), $\frac{Q_{ca_{tot}} \tau_{\rm Ca}}{K_{ca_{tot}}} = 4000 = O(10^4)$ ms (for C-bursting), and $1/(Q_{ca} A) = 100 = O(100)$ ms, we can conclude that $v$ and $n$ evolve on a fast timescale, $ca$ evolves on an intermediate slow timescale, $h$ and $l$ evolve on a slow timescale, while $ca_{tot}$ evolves on a superslow timescale. Although $ca$ is faster than $h$ and $l$, it still evolves on a significantly slower timescale than $v$ and $n$. Therefore, instead of working with four distinct timescales, in this work, we choose to group $ca$ along with $h$ and $l$ as a slow variable for the ease of the geometric singular perturbation analysis as presented in the next section. We choose the slow timescale as our reference time by letting $Q_t = 100$ ms, and set}:

\begin{subequations}
\begin{align}
    \varepsilon &:= \frac{C_m}{Q_t g_{\rm max}} = 0.0075 = O(0.01) \\
    R_h &:= Q_t T_h = O(1)\\
    \RED{R_{ca}} &:= \RED{\frac{Q_t}{Q_{ca} K_{ca}}=O(10)} \\
    R_l &:= Q_t Q_{ca} A = O(1)\\
    \RED{\delta} &:= \RED{ \frac{Q_t K_{ca_{tot}}}{Q_{ca_{tot}} \tau_{\rm Ca}} = O(0.001-0.001)}.
\end{align}
\label{nondimpar}
\end{subequations}

Substituting the above expressions \cref{nondimpar} in \cref{eq:nondim2} and rearranging the necessary terms, we obtain:

\begin{subequations}
    \begin{align}
        \varepsilon \frac{dv}{d\tau} &= -\bar{I}_{\rm L} - \bar{I}_{\rm K} - \bar{I}_{\rm Na} - \bar{I}_{\rm NaP} - \bar{I}_{\rm CAN} - \bar{I}_{\rm Ca} =: f_1 (v, n, h, ca) \label{eq:nondimv} \\
        \varepsilon \frac{dn}{d\tau} &= \frac{1}{r_{12}} \frac{n_{\infty}(v)-n}{t_n(v)} =: f_2 (v, n) \label{eq:nondimn} \\
        \frac{dh}{d\tau} &= R_h \frac{h_{\infty}(v)-h}{t_h(v)} =: g_1 (v,h) \label{eq:nondimh} \\
        \RED{\frac{d ca}{d\tau}} &= \RED{ R_{ca}  g (v, ca, ca_{tot}, l)} \label{eq:nondimca} \\
        \RED{\frac{d ca_{tot}}{d\tau}} &= \RED{ \delta h (v, ca) } \label{eq:nondimcatot} \\
        \frac{dl}{d\tau} &= R_l \left[ \bar{K}_d (1-l) - ca \cdot l \right] =: g_2 (ca, l), \label{eq:nondiml}
    \end{align}
    \label{eq:nondim}
\end{subequations}
where $\varepsilon, \delta \ll 1$, $r_{12} = \frac{g_{\rm max}}{T_n C_m} \approx 0.2036$, 
and all functions $f_1,\, f_2,\,g_1,\,g,\,h,\, g_2$ on the right-hand side of \cref{eq:nondimv} - \cref{eq:nondiml} are about $O(1)$. \\

\section{Singular limits in the Nondimensionalized Model}\hypertarget{appC}{}

In this section, we perform GSPT analysis on \cref{eq:nondim} by treating $\varepsilon$ and $\delta$ as two independent singular perturbation parameters. \RED{For simplicity, we group $ca$ as the slow variable and denote $\tilde{g}=R_{ca}\cdot g$.} Since the primary focus of this work is to explore the effects of norepinephrine on the preB{\"o}tC neuron, we only provide a brief overview of the GSPT analysis and the derivation of key subsystems for the sake of completeness. Readers interested in the full details are referred to \cite{Nan2015,phan2024mixed}. \\

Choose $\tau=t_s$ to be the reference time, we call the dimensionless system \cref{eq:nondim} that evolves over the \emph{slow time $t_s$} the \emph{slow system}. Introducing a superslow time $t_{ss} = \delta t_s$ yields the following equivalent descriptions of dynamics:
\begin{equation}
    \begin{aligned}
        \varepsilon \delta \frac{dv}{dt_{ss}} &= f_1 (v, n, h, ca) \\
        \varepsilon \delta \frac{dn}{dt_{ss}} &= f_2 (v, n) \\
        \delta \frac{dh}{dt_{ss}} &= g_1 (v, h) \\
        \RED{\delta \frac{d ca}{dt_{ss}}} &= \RED{\tilde{g} (v, ca, ca_{tot}, l)} \\
        \frac{d ca_{tot}}{dt_{ss}} &= h(v, ca) \\
        \delta \frac{dl}{dt_{ss}} &= g_2 (ca,l),
    \end{aligned}
    \label{eq:pBC_ICa_superslow}
\end{equation}
which evolves on the \emph{superslow timescale} and is called the \emph{superslow system}. Alternatively, defining a fast time $t_f = t_s / \varepsilon$ leads to another equivalent \emph{fast system}:
\begin{equation}
    \begin{aligned}
        \frac{dv}{dt_f} &= f_1 (v, n, h, ca) \\
        \frac{dn}{dt_f} &= f_2 (v, n) \\
        \frac{dh}{dt_f} &= \varepsilon g_1 (v, h) \\
        \RED{\frac{d ca}{dt_f}} &= \RED{\varepsilon \tilde{g} (v, ca, ca_{tot}, l)} \\
        \frac{d ca_{tot}}{dt_f} &= \varepsilon \delta h(v, ca) \\
        \frac{dl}{dt_f} &= \varepsilon g_2 (ca,l),
    \end{aligned}
    \label{eq:pBC_ICa_fast}
\end{equation}
which evolves on the \emph{fast timescale}. 

The existence of two independent singular perturbation parameters, $\varepsilon$ and $\delta$, implies there are various ways to implement GSPT, each yielding distinct singular limit predictions. 

\textbf{Singular Limit as $\varepsilon \to 0$} \\
Fixing $\delta > 0$ and letting $\varepsilon \to 0$ in the fast system \cref{eq:pBC_ICa_fast} yields the \RED{two-dimensional (2D)} \textit{fast layer problem}, a system that describes the dynamics of the fast variables $v$ and $n$ for fixed values of $ca, h, l$ and $ca_{tot}$:
\RED{
\begin{equation}
    \begin{aligned}
        \frac{dv}{dt_f} &= f_1 (v, n, h, ca) \\
        \frac{dn}{dt_f} &= f_2 (v, n). \\
    \end{aligned}
    \label{eq:pBC_ICa_fastlayer}
\end{equation}}
The set of equilibrium points of the \emph{fast layer problem} is called the \textit{critical manifold}, denoted by $M_s$:
\RED{
\begin{equation}
    M_s := \{(v,n,h,ca,ca_{tot},l) : f_1 (v,n,h,ca)  = f_2 (v,n) = 0\}.
    \label{eq:pBC_ICa_Ms_app}
\end{equation}}
For sufficiently small $\varepsilon>0$, normally hyperbolic parts of $M_s$ perturb to a locally invariant manifold called a \emph{slow manifold} \cite{fenichel1979}. In this paper, for simplicity, we simply use $M_s$ as a convenient numerical approximation of these slow manifolds.

$M_s$ is folded with a set of saddle-node bifurcations of the fast subsystem 
\begin{equation}
    L_s := \{(v,n,h,ca,ca_{tot},l)\in M_s: \det(J)=0 \},
    \label{eq:pBC_ICa_fold_app}
\end{equation}
where $\det(J)$ denotes the determinant of the \RED{$2 \times 2$} Jacobian matrix of the fast subsystem \cref{eq:pBC_ICa_fastlayer}. The fold curves $L_s$ separate the attracting and repelling branches of the critical manifold $M_s$. 

Taking the same limit $\varepsilon \to 0$ with $\delta>0$ in the slow system \cref{eq:pBC_ICa_slow} yields the \RED{4D} \textit{slow reduced problem}, which describes the dynamics of the slow and the superslow variables along the critical manifold $M_s$ \cref{eq:pBC_ICa_Ms}:  
\RED{
\begin{equation}
    \begin{aligned}
        \frac{dh}{dt_s} &= g_1 (v, h) \\
        \frac{d ca}{dt_s} &= \tilde{g} (v, ca, ca_{tot}, l) \\
        \frac{d ca_{tot}}{dt_s} &= \delta h(v, ca) \\
        \frac{dl}{dt_s} &= g_2 (ca,l),
    \end{aligned}
    \label{eq:pBC_ICa_slowreduced}
\end{equation}}
where $f_1=f_2=0$. 

\textbf{Singular Limit as $\delta \to 0$} \\
Alternatively, letting $\delta \to 0$ and fixing $\varepsilon > 0$ in the slow system \cref{eq:pBC_ICa_slow} yields the 5D \textit{slow layer problem}:
\begin{equation}
    \begin{aligned}
        \varepsilon \frac{dv}{dt_s} &= f_1 (v, n, h, ca) \\
        \varepsilon \frac{dn}{dt_s} &= f_2 (v, n) \\
        \frac{dh}{dt_s} &= g_1 (v, h) \\
        \RED{\frac{d ca}{dt_s}} &= \RED{\tilde{g} (v, ca, ca_{tot}, l)} \\
        \frac{dl}{dt_s} &= g_2 (ca,l),
    \end{aligned}
    \label{eq:pBC_ICa_slowlayer}
\end{equation}
where the superslow variable $ca_{tot}$ is a constant parameter.

The set of equilibrium points of the slow layer problem \cref{eq:pBC_ICa_slowlayer} is a one-dimensional subset of $M_s$ called the \textit{superslow manifold} and is denoted by $M_{ss}$:
\RED{
\begin{equation}
    M_{ss} := \{(v,n,h,ca,ca_{tot},l)\in M_s : g(v, ca, ca_{tot}, l) =  
    g_1 (v,h)  = g_2 (ca,l) = 0\}.
    \label{eq:pBC_ICa_Mss_app}
\end{equation}}
Again, similar to $M_s$, for $0 < \delta \ll 1$, the normally hyperbolic parts of $M_{ss}$ perturb to locally invariant manifolds, which we shall simply refer to as $M_{ss}$. Interesting dynamics are expected to occur near nonhyperbolic points on $M_{ss}$ where Fenichel's theory (GSPT) breaks down. These include Hopf or saddle-node fold bifurcations of the slow layer problem \cref{eq:pBC_ICa_slowlayer}.

Taking the same limit, i.e., $\delta \to 0$ with $\varepsilon>0$, in the superslow system \cref{eq:pBC_ICa_superslow} leads to the \textit{superslow reduced problem} representing the dynamics of the superslow variable $ca_{tot}$ along the $M_{ss}$ \cref{eq:pBC_ICa_Mss}:
\begin{equation}
    \frac{d ca_{tot}}{dt_{ss}} = h(v, ca).
    \label{eq:pBC_ICa_superslowreduced}
\end{equation} 
where $f_1=f_2=g=g_1=g_2=0$. The superslow dynamics of \cref{eq:pBC_ICa_superslowreduced} are slaved to $M_{ss}$ until nonhyperbolic points are reached. 

\textbf{Double Singular Limits as $\varepsilon \to 0$ and $\delta \to 0$} \\
Further, since both the slow reduced problem \cref{eq:pBC_ICa_slowreduced} and the slow layer problem \cref{eq:pBC_ICa_slowlayer} still evolve on two distinct timescales, taking the limit $\delta \to 0$ in \cref{eq:pBC_ICa_slowreduced} or $\varepsilon \to 0$ in \cref{eq:pBC_ICa_slowlayer} yields the same \emph{slow reduced layer problem}:
\RED{
\begin{equation}
    \begin{aligned}
        \frac{dh}{dt_s} &= g_1 (v, h) \\
        \frac{d ca}{dt_s} &= \tilde{g} (v, ca, ca_{tot}, l) \\
        \frac{dl}{dt_s} &= g_2 (ca,l),
    \end{aligned}
    \label{eq:pBC_ICa_slowreducedlayer}
\end{equation}}
which describes the dynamics of the slow variables $ca, \; h$ and $l$ along $M_s$ and the superslow variable $ca_{tot}$ is a constant.

\RED{
\section{Effect of $[\rm IP_3]$ on C-burst frequency} \hypertarget{appD}{}

In \cref{CB}, we explained that the non-monotonic effects of $[\rm IP_3]$ on the frequency of C-bursts arises from its differential impacts on the $l$ values at the burst initiation and termination, based on a comparison of the system dynamics at three representative $[\rm IP_3]$ values. To verify that this mechanism is not specific to those particular values, we numerically measure the $l$-values of the full system solution at both burst initiation and termination over a range of $[\rm IP_3]$ values. The length of the trajectory traversed by the slow variable $l$ in the silent phase is given by the difference of the $l$-values at burst initiation and termination, that is, \[l_{SP} = l_{\rm burst_{init}} - l_{\rm burst_{term}}.\] and is depicted in \cref{fig:CB_lval_SP}. The three representative $[\rm IP_3]$ values used in our earlier analysis (\cref{CB}) are indicated by the magenta dashed lines. From the bottom panel in \cref{fig:CB_lval_SP}, we note that the length of $l$ traversed in silent phase initially increases as $[\rm IP_3]$ increases, but begins to decrease for $[\rm IP3] > 0.4$ (approximately). This is consistent with the initial increase in the burst period and its subsequent decrease as $[\rm IP_3]$ increases (see \cref{fig:CB_gcan_ip3}(i)), thereby confirming our analysis in \cref{CB}.   

\begin{figure}[!ht]
    \begin{center}
    \subfigimg[width=0.6\linewidth]{}{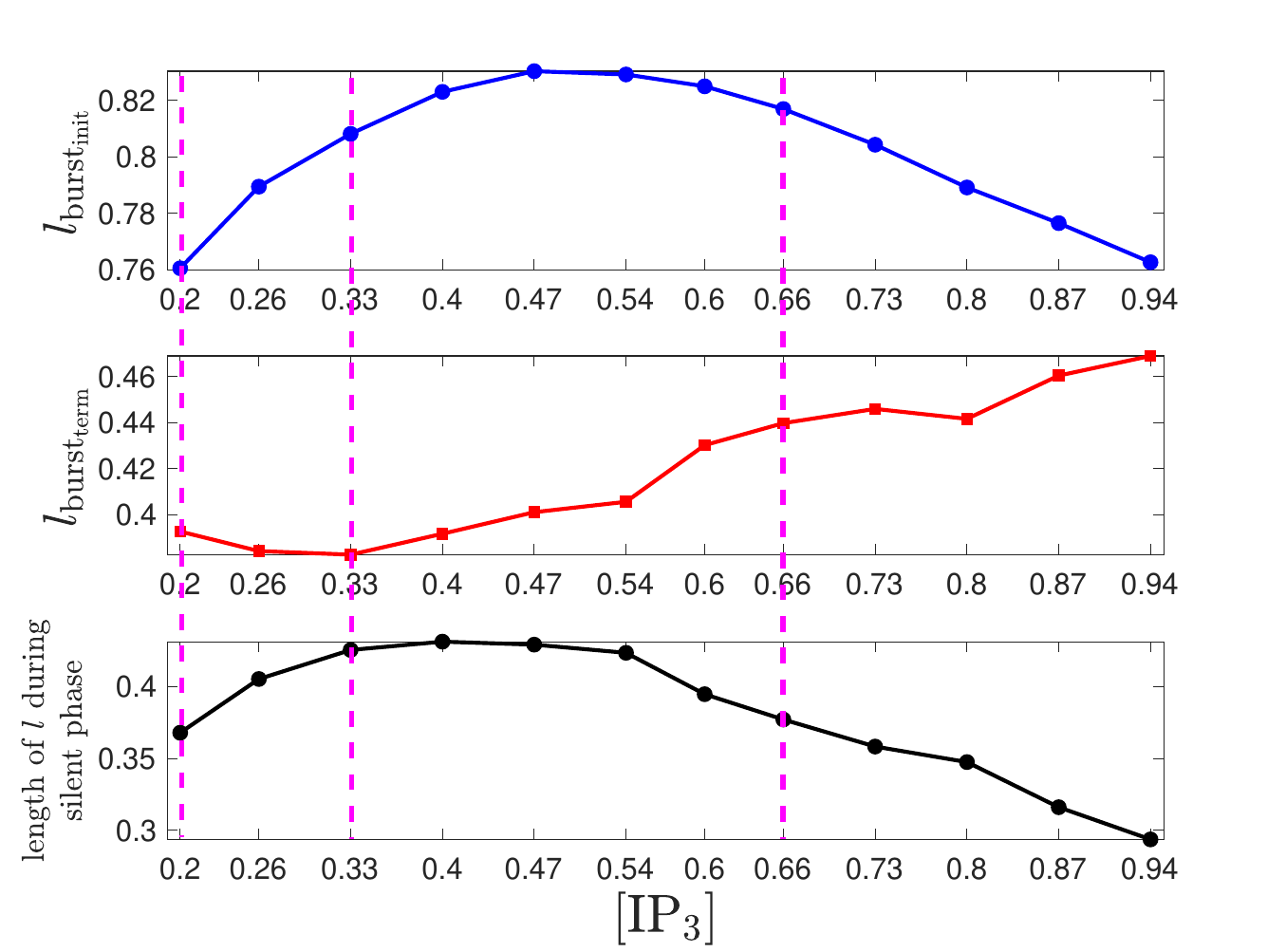}
    \end{center}
    \caption{Numerical computation of $l$-values at (\emph{top panel}) burst initiation, (\emph{middle panel}) burst termination, and (\emph{bottom panel}) the length of the trajectory traversed in the silent phase by $l$, as a function of increasing $[\rm IP_3]$, for C-bursting with parameters: $g_{\rm Ca} = 0.0005$, $g_{\rm CAN} = 0.7$, and $g_{\rm NaP} = 0$. Also marked by dashed lines in magenta are the $[\rm IP_3]$ reference values used for our qualitative analysis in \cref{fig:CB_analysis_ip3}. }
    \label{fig:CB_lval_SP}
\end{figure}

}

\section{Mixed Bursting Dynamics} \hypertarget{appE}{}

In \cref{TS}, when a tonic spiking neuron switches to a bursting behavior, there exists an intermediate parameter region along the transitional region in the $([{\rm IP_3}], g_{\rm NaP}$)-space (see \cref{fig:ts_to_cb}(i)) where the model exhibits some interesting \textit{mixed bursting} dynamics. We use the term mixed bursting to describe all such dynamics where a period may include two or more bursts, with each of these bursts having similar or different burst widths. These mixed bursts persist with an increase in $g_{\rm CAN}$ (\cref{fig:ts_to_cb}(ii)) as well as with small variations in $g_{\rm Ca}$ (\cref{fig:activity_patterns}(i)).

\begin{figure}[!ht]
    \begin{center}
    \subfigimg[width=0.9\linewidth]{}{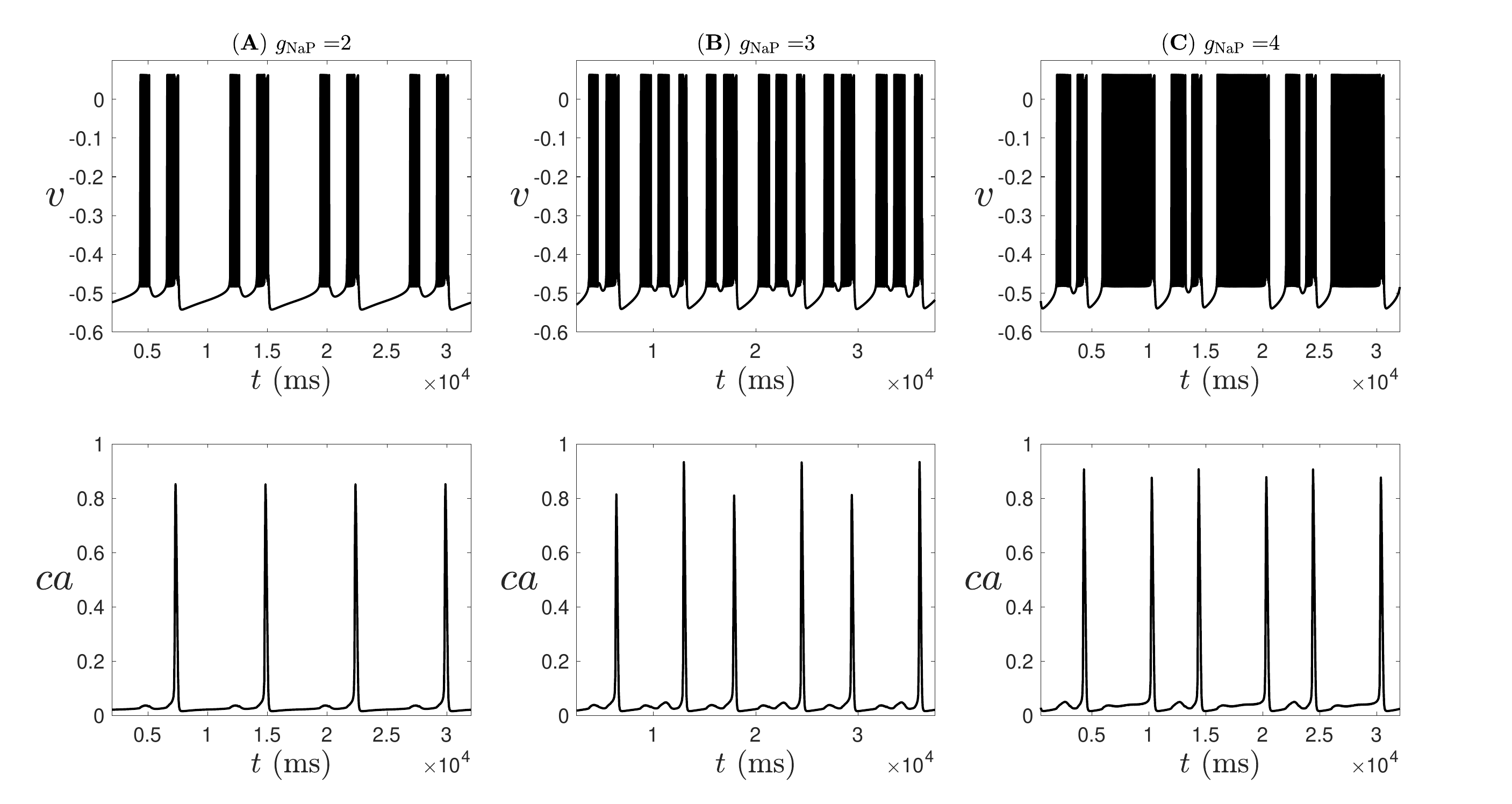}
    \end{center}
    \caption{Some example voltage and calcium temporal traces from \cref{eq:pBC_ICa_slow} exhibiting mixed bursting, occurring during the transition from tonic spiking to induced (N)C-bursting (see \cref{fig:ts_to_cb}), with parameters: $g_{\rm Ca} = 0.0002$, $g_{\rm CAN} = 0.7$, [${\rm IP_3}] = 0.4$, and (A) $g_{\rm NaP} = 2$, (B) $g_{\rm NaP} = 3$, (C) $g_{\rm NaP} = 4$. }
    \label{fig:MB}
\end{figure}

In our model, this mixed bursting behavior is a subclass of Type 1 (N)C-bursters in which the bursting mechanism depends on both the NaP and CAN mechanisms in such a way that in the absence of either, the bursts are lost. Such a mixed bursting behavior arises for $g_{\rm Ca} \approx 0.0002$, which is just sufficient to produce independent $ca$-oscillations. These kinds of bursts are especially interesting since, from \cref{fig:MB}, looking at the $ca$-dynamics, we note multiple small-amplitude oscillations that occur before a full oscillation in the $ca$-subsystem. This suggests the presence of canard-like dynamics in our model. Moreover, the fact that such bursts occur over a relatively large range of $g_{\rm NaP}$, as well as $g_{\rm Ca}$ and [${\rm IP_3}$] to a reasonable extent, makes them a compelling behavior in need of further analytical exploration. 

These bursts are also interesting from the perspective of understanding the effect of NE in our model. Because they occur for an intermediate range of parameter values observed during the transition of a tonic spiking neuron to an induced (N)C-burster, this might suggest that the concentration of NE applied may play an important role in seeing a successful transition. Further, the concentration could also be a factor of consideration at the network level in studying the network synchrony.

\section{Role of $g_{\rm Ca}$ in modeling NE-induced C-bursting}\hypertarget{appF}{}

\begin{figure}[!ht]
    \begin{center}
    \begin{tabular}    
    {@{}p{0.45\linewidth}@{\quad}p{0.45\linewidth}@{}}
    \subfigimg[width=\linewidth]{\bfseries{\footnotesize{(A)}}}{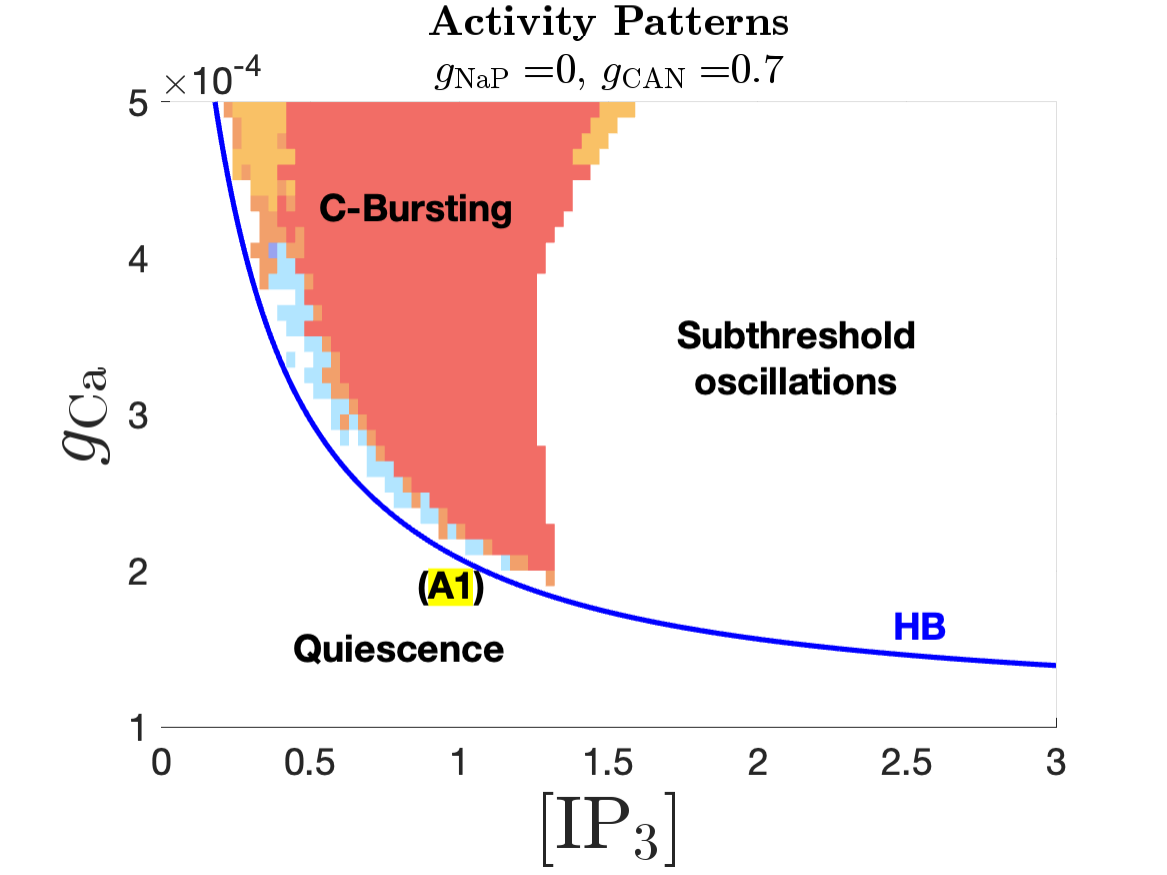} & \subfigimg[width=\linewidth]{\bfseries{\footnotesize{(B)}}}{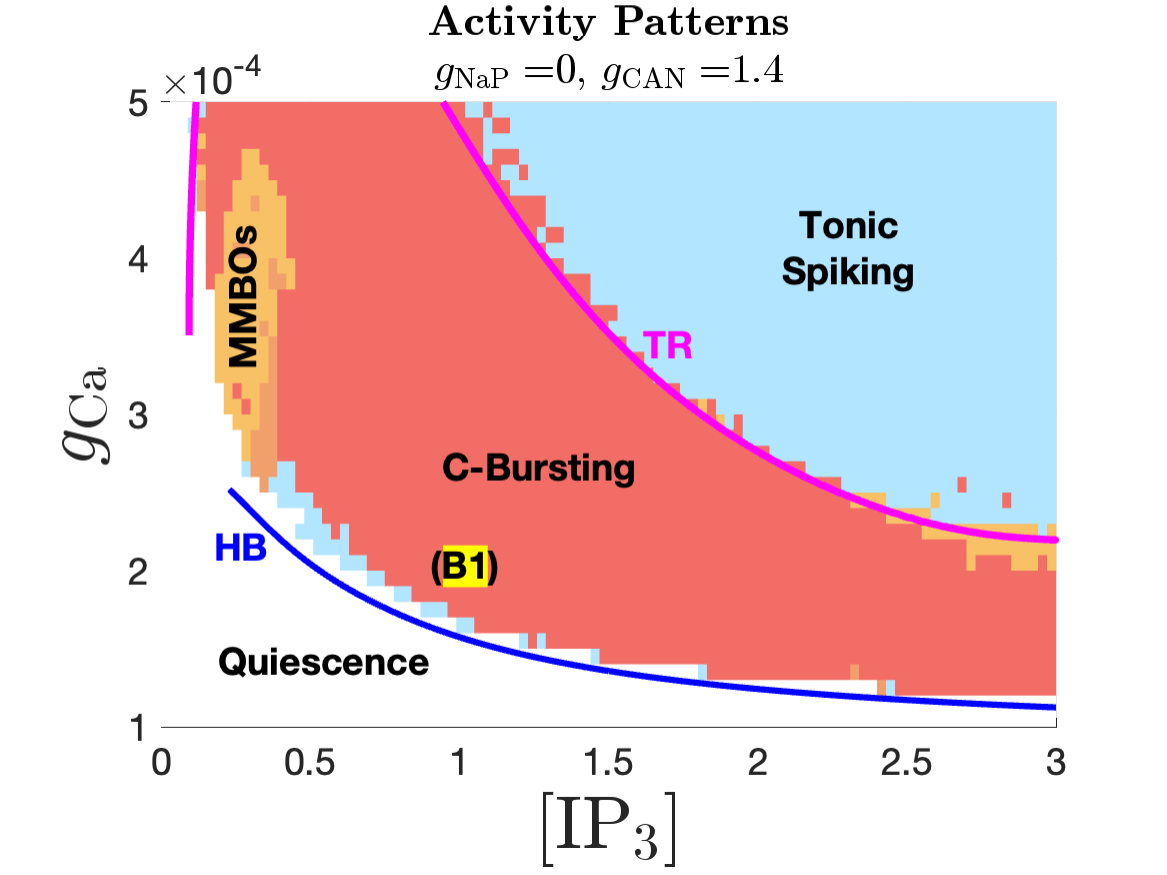} 
    \end{tabular}
    \end{center}
    \caption{Two-parameter bifurcation diagrams of the full system \cref{eq:pBC_ICa_slow} in the $([{\rm IP_3}], g_{\rm Ca})$-space, with $g_{\rm NaP} = 0$. (A) $g_{\rm CAN} = 0.7$; (B) $g_{\rm CAN} = 1.4$. The Hopf bifurcation curve (HB) is shown in blue, and the torus bifurcation curve (TR) is shown in magenta. Activity patterns of the full system \cref{eq:pBC_ICa_slow} are superimposed.} 
    \label{fig:induced_bursts_gca_ip3}
\end{figure}

Recall from \cref{TS} that in model \cref{eq:pBC_ICa_slow}, tonic spiking arises under two distinct parameter regimes: when calcium conductance ($g_{\rm Ca}$) is low (less than $0.0001$) (see \cref{fig:activity_patterns}(i), light blue region), or when $g_{\rm Ca}$ is intermediate (e.g., $\sim 0.0002$) (see \cref{fig:ts_to_cb}(i), light blue region). In both cases, $g_{\rm NaP}$ needs to be relatively high to sustain tonic spiking in the voltage dynamics. As shown in \cref{TS}, while NE can induce C-bursting from the tonic spiking regime with intermediate $g_{\rm Ca}$, it fails to do so when $g_{\rm Ca}$ is too low.  

To understand this, we compute the bifurcation structure of the full system \cref{eq:pBC_ICa_slow} with respect to $[\rm IP_3]$ and $g_{\rm Ca}$, again setting $g_{\rm NaP} = 0$ (see \cref{fig:induced_bursts_gca_ip3}). It is important to note that only the quiescent region with low or intermediate $g_{\rm Ca}$ and low $[\rm IP_3]$ in \cref{fig:induced_bursts_gca_ip3} corresponds to the tonic spiking regime in the full system when $g_{\rm NaP}$ is high. To examine the condition on $g_{\rm Ca}$ under which C-bursting can be induced, we restrict our attention to the range $g_{\rm Ca}\leq 0.0002$, ensuring the system begins in a tonic spiking state at high $g_{\rm NaP}$ and low $[\rm IP_3]$. 
\cref{fig:induced_bursts_gca_ip3} shows that increasing $g_{\rm CAN}$ from 0.7 (panel (A)) to $1.4$ (panel (B)) shifts the HB curve to lower $g_{\rm Ca}$, thereby expanding the C-bursting region. Moreover, subthreshold oscillations above the HB curve are eliminated with increased $g_{\rm CAN}$, while tonic spiking solutions are observed in the light blue region above the TR bifurcation curve (magenta). It follows that when $g_{\rm Ca}$ is below the horizontal asymptote of the HB curve at high $g_{\rm CAN}$ (e.g., $g_{\rm Ca} < 0.00012$), increasing $g_{\rm CAN}$ and $\rm [IP_3]$ is insufficient to cross the HB curve and induce a transition to C-bursting. 
In contrast, for intermediate $g_{\rm Ca}$ (e.g., $g_{\rm Ca}=0.0002$ at points (A1) and (B1)), increasing $g_{\rm CAN}$ and $\rm [IP_3]$ can drive a neuron —initially silent in the absence of $g_{\rm NaP}$ but corresponding to tonic spiking at high $g_{\rm Nap}$— across the HB bifurcation and transition to a C-bursting state that persists without $I_{\rm NaP}$.

\end{appendix}

\section*{Acknowledgments}
We thank Dr.~Jean-Charles Viemari for helpful comments on this work. 
This work was supported by NIH/NIDA R01DA057767, as part of the Collaborative Research in Computational Neuroscience Program. A.J.~Garcia 3rd is also supported by NIH/NHLBI R01HL163965 and NIH/NIDA R01DA061412.

\bibliographystyle{siamplain}
\bibliography{references}
\end{document}

%% file: commands.tex
\usepackage{amsmath}
\usepackage{amsfonts}
\usepackage{amssymb}
\usepackage{comment}
\usepackage{subfig}
\usepackage{wrapfig}
\usepackage{enumerate}
\usepackage{float}
\usepackage{booktabs} 
\usepackage{graphicx}
\usepackage{multirow}
\usepackage{xcolor}
\usepackage[normalem]{ulem}

\usepackage{tikz}
\usepackage[utf8]{inputenc}
\usepackage{pgfplots} 
\usepackage{pgfgantt}
\usepackage{pdflscape}
\pgfplotsset{compat=newest} 
\pgfplotsset{plot coordinates/math parser=false}

\newlength\fwidth
	
\usepackage{color}
\usepackage{latexsym}
\usepackage{lscape}

\newcommand{\RED}[1]{{\color{black}{#1}}}

\newcommand{\pbc}{preB\"{o}tC}

\newcommand{\subfigimg}[3][,]{%
  \setbox1=\hbox{\includegraphics[#1]{#3}}
  \leavevmode\rlap{\usebox1}
  \rlap{\hspace*{-5pt}\raisebox{\dimexpr\ht1-2\baselineskip}{#2}}
  \phantom{\usebox1}
}